\documentclass{jfm}

\usepackage{graphicx}
\usepackage{natbib}

\usepackage{float}
\usepackage{amsmath}

\ifCUPmtlplainloaded \else
  \checkfont{eurm10}
  \iffontfound
    \IfFileExists{upmath.sty}
      {\typeout{^^JFound AMS Euler Roman fonts on the system,
                   using the 'upmath' package.^^J}%
       \usepackage{upmath}}
      {\typeout{^^JFound AMS Euler Roman fonts on the system, but you
                   dont seem to have the}%
       \typeout{'upmath' package installed. JFM.cls can take advantage
                 of these fonts,^^Jif you use 'upmath' package.^^J}%
      }
  \else
  \fi
\fi

\ifCUPmtlplainloaded \else
  \checkfont{msam10}
  \iffontfound
    \IfFileExists{amssymb.sty}
      {\typeout{^^JFound AMS Symbol fonts on the system, using the
                'amssymb' package.^^J}%
       \usepackage{amssymb}%
         
       \let\ge=\geqslant  
      }{}
  \fi
\fi

\ifCUPmtlplainloaded \else
  \IfFileExists{amsbsy.sty}
    {\typeout{^^JFound the 'amsbsy' package on the system, using it.^^J}%
     \usepackage{amsbsy}}
    {}
\fi

\newsavebox{\astrutbox}
\sbox{\astrutbox}{\rule[-5pt]{0pt}{20pt}}

\newcommand\p{\ensuremath{\partial}}

\title[SSG turbulence]{A study of surface semi-geostrophic turbulence: freely decaying dynamics}

\author[F. Ragone and G. Badin]%
{Francesco Ragone$^{1,2}$%
  \thanks{Email address for correspondence: francesco.ragone@ens-lyon.fr},\ns
Gualtiero Badin$^1$}

\affiliation{$^1$Institute of Oceanography, University of Hamburg, Hamburg \\
$^2$Laboratoire de Physique, Ecole Normale Sup\'{e}rieure de Lyon, Lyon, France 
}

\pubyear{}
\volume{}
\pagerange{}
\date{}

\begin{document}

\maketitle

\begin{abstract}
In this study we give a characterization of semi-geostrophic turbulence by performing freely decaying simulations 
for the case of constant uniform potential vorticity, a set of equations known as surface semi-geostrophic approximation. The equations are formulated as conservation laws for potential temperature and potential vorticity, with a nonlinear Monge-Amp\'{e}re type inversion equation for the streamfunction, expressed in a transformed coordinate system that follows the geostrophic flow. We perform model studies of turbulent surface semi-geostrophic flows in a doubly-periodic domain in the horizontal limited in the vertical by two rigid lids, allowing for variations of potential temperature at one of the boundaries, and we compare them with the corresponding surface quasi-geostrophic case. Results show that, while surface quasi-geostrophic  dynamics is dominated by a symmetric population of cyclones-anticyclones, surface semi-geostrophic  dynamics features a more prominent role of fronts and filaments. The resulting distribution of potential temperature is strongly skewed and peaked at non-zero values at and close to the active boundary, while symmetry is restored in the interior of the domain, where small-scale frontal structures do not penetrate. In surface semi-geostrophic turbulence energy spectra are less steep than in the surface quasi-geostrophic case, with more energy concentrated at small scales for increasing Rossby number. Energy connected to frontal structures, lateral strain rate and vertical velocities are largest close to the active boundary. These results show that the semi-geostrophic model could be of interest for studying the lateral mixing of properties in geophysical flows.
\end{abstract}


\section{Introduction}\label{sec:introduction}

Geophysical flows are characterized by a wide spectrum of spatial and temporal scales. Given the multiscale nature of the system, different dynamical theories are needed in order to understand the properties of processes occurring at different subranges of this spectrum. A particular class of models that has proved very useful in theoretical and qualitative studies of geophysical fluid dynamics is the one of the balanced models, simplified versions of the primitive equations obtained by scale analysis filtering out the high-frequency inertial-gravity waves. Among the balanced models, the quasi-geostrophic (QG) and semi-geostrophic (SG) approximations have been extensively employed to study the properties of dynamics at scales larger than the Rossby deformation radius. 

The classical QG approximation is obtained assuming that the Rossby number is much smaller than one. The SG approximation is obtained by the less restrictive condition that the Lagrangian time scale, that is the time scale of change of the momentum following the motion of a particle, is much longer than $f^{-1}$, or equivalently that the correspondingly defined Lagrangian Rossby number is much smaller than one.  The condition of small Lagrangian Rossby number is much less stringent than the condition of small traditional Rossby number. Consequently, the SG approximation has proved to be more realistic than the QG approximation in the representation of large-scale geophysical flows, at the price of showing a substantially more complex mathematical structure.


Given its simple formal structure, the QG approximation has become a standard model for studying the qualitative properties of large-scale geophysical flows. A model based on the QG approximation that has proved particularly useful for theoretical studies 
is the surface quasi-geostrophic (SQG) model \citep{Blumen1978, Held&al1995}. SQG is realized imposing constant potential vorticity in the interior of the domain and allowing advection of a conserved scalar (potential temperature or buoyancy) on a boundary. The SQG dynamics is thus effectively two-dimensional, where the structure of the flow in the interior of the domain is determined uniquely by the values of the scalar at the boundary through a linear elliptic inversion equation. In this model the conserved scalar takes the role taken by vorticity in the Euler equations. SQG  is characterized by an inverse cascade of total energy at low wavenumbers and a forward cascade of potential temperature variance at high wavenumbers. Classic SQG is unbounded at the bottom boundary and shows a -5/3 kinetic energy spectrum  \citep{Blumen1978, Held&al1995}. Finite-depth SQG is obtained introducing a bottom boundary with potential temperature set to zero \citep{Tulloch&Smith2006}. Finite-depth SQG features a critical scale, with a transition from a QG/2D-like kinetic energy spectrum with a -3 slope at large scales to a SQG-like kinetic energy spectrum with a -5/3 slope at small scales. 

In atmospheric dynamics, the SQG approximation has been used to study the dynamical properties of potential temperature anomalies on the tropopause \citep{Juckes1994,smithbernard13}, the asymmetry between cyclones and anticyclones \citep{Hakim&al2002} and the shape of the tropopause spectra \citep{Tulloch&Smith2009}. In ocean dynamics, the SQG approximation has been used  to infer the interior dynamics of the ocean from the knowledge of sea surface temperature anomalies \citep{LaCasce&Mahadevan2006,Lapeyre&Klein2006,Wang2013,liuetal14}. Results show, however, that SQG  underestimates the buoyancy anomaly and current fields at depth. Employment of exponentially decaying stratification \citep{LaCasce2012} retains depth buoyancy anomaly values smaller than the observations. The projection of the SQG modes into normal modes has been studied by \cite{Lapeyre2009} and \cite{smithvanneste13}. The relationship between SQG and QG dynamics in a limited number of vertical layers and the emerging geostrophic turbulence have been studied by \cite{tullochsmith09b} and \cite{Badin2014}.


The SG approximation was originally developed by \cite{Eliassen1948} in a three-dimensional formulation, and subsequently applied by \cite{Hoskins&Bretherton1972} in two dimensions to study frontogenesis in the atmosphere. \cite{Hoskins1975,Hoskins1976} further developed the theory in the shallow water approximation. The SG approximation is obtained from the primitive equations taking the geostrophic momentum approximation, where the ageostrophic motion is retained in the advective velocity. The equations are naturally formulated in geostrophic coordinates, a modified coordinate system that follows the geostrophic flow \citep{Eliassen1948,Fjortoft1962}. In geostrophic coordinates, SG features a nonlinear, mixed-type Monge-Amp\'{e}re partial differential equation for a modified streamfunction of the flow. 

The SG approximation has been subject to a fairly large number of studies in the past decades. The case with potential vorticity constant in the interior has been studied for the Eady wave by \cite{Hoskins1976}, \cite{Juckes1998} and \cite{Davies&Mueller1988}. The SG model has been successfully employed for the study of the geostrophic adjustment problem \citep{Plougonven&Zeitlin2005}.  A comparison between the behavior of baroclinic waves in SG dynamics and primitive equations has been studied by \cite{Snyder&al1991}, while a comparison with observations was presented by \cite{Blumen1979b}. The linear and nonlinear stability of SG flows was studied, amongst others, in a series of articles by \cite{Kushner&Shepherd1995a,Kushner&Shepherd1995b}, \cite{Kushner1995} and in a series of articles by Ren \citep{Ren1998,Ren1999,Ren2000,Ren2000b,Ren2005}. The Hamiltonian structure of the SG equations was introduced by \citet{salmon83}, \citet{salmon85} and \citet{salmon88}, see also \citet{Purser93,Oliver06}, \citet{Oliver14} and \citet{BlenderBadin15}. The geometry was instead studied by \citet{roubtsov_roulstone97,roubtsov_roulstone01} and \citet{delahaies_roulstone10} (see also \citet{McIntyreRoulstone02}). \citet{nagai2006two} and \citet{badinetal09} applied the SG approximation to study the stability of ocean fronts. See \cite{CullenBook} and references therein for a complete overview on SG theory, including results on the existence of solutions of the SG equations expressed as a mass transportation problem.


Despite the rich literature on the subject, there are no studies aimed at characterizing the properties of SG turbulence. In this paper we want to give a characterization of SG turbulence in a simple, idealized setup, performing numerical simulations of a fully turbulent, freely decaying flow in finite-depth surface semi-geostrophic (SSG) approximation. SSG is obtained imposing on the SG equations the same boundary conditions as in SQG \citep{Badin2013}. In general, SG dynamics can better represent the formation of fronts and filaments, which in turn can generate ageostrophic instabilities. From a quantitative point of view the SG approximation is not able to represent accurately instabilities at scales smaller than the Rossby radius of deformation because of the lack of a proper vortex dynamics \citep{Malardel_et_al97}. However, because of the representation even in a crude form of these dynamics, SSG turbulence is expected to differ substantially from SQG turbulence. For example, in the ocean, surface fronto- and filamentogenesis induce restratification due to the asymmetry in the divergence field (absent in SQG) and in the structure of upward and downward velocities associated with filaments and fronts. This has been observed in primitive equations simulations  \citep{Lapeyre&al2006,Klein&al2008} and explained in the context of the SQG$^{+1}$ model \citep{Hakim&al2002}, a first order correction in Rossby number to the SQG equations, which accounts for the effects of ageostrophic advection. 

In this paper we investigate the properties of SSG turbulence, testing whether its dynamics can represent at a qualitative level some of the features of the observed ocean dynamics that are not captured by the SQG model. We show how the inclusion of the nonlinear term of the Monge-Amp\'{e}re equation induces a distinctive qualitative difference in the statistics of potential temperature, allowing for the emergence of filaments as a prominent part of the dynamics. The coordinate transformation on the other hand increases the amount of energy at small scales, flattening the kinetic energy spectra, and induces the net surface cooling. 
Vertical velocities, horizontal divergence and lateral strain are generally enhanced in amplitude and penetrate further in depth for increasing Rossby number in SSG. These results suggest that SSG could be proposed as a theoretical laboratory to study at a qualitative level certain aspects of submesoscale dynamics.

In this paper we investigate the properties of SSG turbulence. We aim to answer the following questions: how do SQG and SSG turbulence differ? What is the role of the inversion of coordinates on the emerging turbulence? The nonlinear terms of the inversion equation that are present in SSG have been typically neglected in the literature, with very few exceptions  \citep{Snyder&al1991}. As a result, the difference between QG and SG has been limited to the presence of a coordinate transformation between physical and geostrophic coordinates accounting for the divergent nature of the SG flow. What is the effect of these nonlinear terms for example on the cyclone/anticyclone asymmetry and on the turbulent spectra? How do vertical velocities differ in SQG and SSG?


The paper is structured as follows. In Section \ref{sec: theory} we summarize the basics of SG theory following classic papers like \cite{Hoskins1975} and \cite{Hoskins&West1979}, and we derive the SSG model in nondimensional coordinates, following \cite{Hoskins&West1979} and \cite{Badin2013}. In Section \ref{sec: methods} we show how we solve the inversion equation present in SSG. In Section \ref{sec:SSG_turbulence} we describe the details of the numerical model that is used to perform simulations of SSG turbulence. We perform a sensitivity analysis varying the Rossby number, comparing the resulting turbulence with what obtained with the SQG model. We analyze fields both at the active boundary and in the interior of the domain, showing how the characteristic properties of SSG vary with depth. In Section \ref{sec: conclusions} we present our conclusions and final discussions. 

\section{Semi-geostrophic approximation}\label{sec:semi_geostrophic_approximation} \label{sec: theory}

\subsection{General equations}\label{sec: SG}

We start from the SG equations in the $f$-plane in Boussinesq form. For a complete analysis of the properties of this system of equations and a general overview on SG theory the reader can refer to \cite{CullenBook}. Introducing the velocity field $(u,v,w)$, the streamfunction (rescaled pressure) $\phi$, the potential temperature anomaly $\theta$ and background $\theta_0$, and the gravitational acceleration $g$, the SG equations in dimensional form are
\begin{equation}\label{eq:SG_Boussinesq}
\begin{array}{r}  
\displaystyle \frac{D u_g}{D t} - f v + \frac{\p \phi}{\p x} = 0, \\[8 pt]
\displaystyle \frac{D v_g}{D t} + f u + \frac{\p \phi}{\p y} = 0, \\[8 pt]
\displaystyle g\frac{\theta}{\theta_0} - \frac{\p \phi}{\p z} =0, \\[8 pt]
\displaystyle \frac{D \theta}{D t} = 0, \\[8 pt]
\displaystyle \frac{\p u}{\p x} + \frac{\p v}{\p y} + \frac{\p w}{\p z} = 0 ,
\end{array}
\end{equation}
where
\begin{equation}
\displaystyle \frac{D}{D t} = \frac{\p}{\p t} + u \frac{\p}{\p x} + v \frac{\p}{\p y},
\end{equation}
and the geostrophic velocity field $(u_g,v_g)$ is defined by the streamfunction as
\begin{equation}
\begin{array}{l}
\displaystyle u_g = - \frac{1}{f} \frac{\p \phi}{ \p y},\\[8 pt]
\displaystyle v_g  = + \frac{1}{f} \frac{\p \phi}{\p x}.
\end{array}
\end{equation}
The model conserves the potential temperature $\theta$ and the SG potential vorticity
\begin{equation}
\displaystyle Q_{sg}= {{g}\over{f \theta_0}} \zeta_{sg} \cdot \nabla \theta,
\end{equation}
where $\zeta_{sg}$ is the SG absolute vorticity
\begin{equation}\label{eq:vort_sg}
\displaystyle \zeta_{sg} = \left( -\frac{\p v_g}{\p z}, \frac{\p u_g}{\p z}, f + \frac{\p v_g}{\p x}-\frac{\p u_g}{\p y}\right) +\left(\frac{1}{f} J_{yz} (u_g,v_g),\frac{1}{f} J_{zx} (u_g,v_g), \frac{1}{f}J_{xy} (u_g,v_g) \right),
\end{equation}
with $J_{ab}$ the Jacobian operator with respect to the variables $a$ and $b$. The definition of SG vorticity differs from the QG case for the presence of the nonlinear terms in \eqref{eq:vort_sg}. The model conserves also the energy integral over the entire domain
\begin{equation}\label{eq:energy_sg}
\displaystyle E_{sg} = \int_V \left[ \frac{1}{2}\left( u_g^2 + v_g^2 \right) - \frac{g}{\theta_0}z\theta \right]dV,
\end{equation}
where the horizontal kinetic energy involves only the geostrophic velocities and is identical to the QG case.

The basic difference between QG and SG is that the materially conserved quantities in QG are advected by the geostrophic velocity field $(u_g,v_g)$, while in SG they are advected by the full (geostrophic and ageostrophic) horizontal velocity field $(u,v)$. However, $u$ and $v$ are implicit in the SG equations, therefore \eqref{eq:SG_Boussinesq} as it is can not be written as a system of conservation equations with an inversion equation for the streamfunction. \cite{Hoskins&Bretherton1972} and \cite{Hoskins1975}, inspired by previous work by \cite{Eliassen1948} and \cite{Fjortoft1962}, have shown that introducing  the geostrophic coordinates
\begin{equation}\label{eq:transform}
\begin{array}{l}
\displaystyle X=x + f^{-1} v_g, \\[8 pt]
\displaystyle Y=y -  f^{-1} u_g, \\[8 pt]
\displaystyle Z=z,
\end{array}
\end{equation} 
the horizontal advection operator expressed in the new coordinates includes explicitly only the geostrophic velocities
\begin{equation}
\displaystyle \frac{D}{D t} = \frac{\p}{\p t} + u \frac{\p}{\p x} + v \frac{\p}{\p y} = \frac{\p}{\p t} + u_g \frac{\p}{\p X} + v_g \frac{\p}{\p Y}.
\end{equation}
The Bernoulli function
\begin{equation} \label{eq:Bernoulli}
\displaystyle \Phi=\phi+\frac{1}{2}\left(u_g^2 + v_g^2 \right)
\end{equation}
acts then as a streamfunction in the new coordinate system 
\begin{equation}\label{eq:property_GT}
\displaystyle \left( \frac{\p \Phi}{\p X} , \frac{\p \Phi}{\p Y} , \frac{\p \Phi}{\p Z} \right) = \left( \frac{\p \phi}{\p x} , \frac{\p \phi}{\p y} , \frac{\p \phi}{\p z} \right) = \left( f v_g , - f u_g , g \frac{\theta}{\theta_0}  \right).
\end{equation}
Equations \eqref{eq:transform} and \eqref{eq:Bernoulli} define a Legendre or contact transformation, and their mathematical properties have been studied by \cite{Blumen1981} and \cite{Purser93,Purser1999}. In the geostrophic space $(X,Y,Z)$ the dynamics can be expressed in terms of conservation equations for the potential temperature $\theta$ and the potential vorticity $Q_{sg}$, plus the inversion equation for the Bernoulli function 
\begin{equation}\label{eq:inversion}
\frac{1}{Q_{sg}}\frac{\p^2 \Phi}{\p Z^2} + \frac{1}{f^2} \left( \frac{\p^2 \Phi}{\p X^2} + \frac{\p^2 \Phi}{\p Y^2}\right)  -  \frac{1}{f^4}\left[\frac{\p^2 \Phi}{\p X^2} \frac{\p^2 \Phi}{\p Y^2} -\left( \frac{\p^2 \Phi}{\p X \p Y} \right)^2 \right] =  1.
\end{equation}
The SG problem in this form can in principle be approached as the QG problem, with two crucial differences: 1) the inversion equation for the streamfunction is a nonlinear Monge-Amp\'{e}re type equation instead of a linear elliptic equation, and 2) the model is formulated in geostrophic coordinates, which include implicitly the advection by the geostrophic velocity field.

\subsection{Boundary conditions and finite-depth surface semi-geostrophic equations}\label{sec: SSG}

Finite-depth SSG approximation is obtained defining the domain and the boundary conditions as for finite-depth SQG, and setting constant potential vorticity in the interior of the domain \citep{Badin2013}. The final equations are consistent with, e.g., \cite{Hoskins&West1979} and \cite{Snyder&al1991}, with a few differences. We have chosen the finite-depth version of SSG instead than the version defined on a semi-infinite domain because in the former case the solution procedure is much easier to treat numerically, as discussed in the following. We consider a domain doubly periodic in the horizontal directions  and bounded in the vertical by two horizontal surfaces at $Z=0$ and $Z=H$. At the boundaries we impose the rigid lid condition $w=0$. Setting constant stratification and potential vorticity $Q_{sg}$ uniform in the interior of the domain, the time-dependent problem occurs only at the boundaries as horizontal advection of $\theta$. Fixing $\theta=0$ at $Z=H$ \citep{Tulloch&Smith2006}, the state of the system is determined by the evolution of the two-dimensional dynamics of $\theta$ at $Z=0$, which provides the boundary condition to \eqref{eq:inversion} to determine the full three-dimensional structure of the flow. As in SQG, in SSG the surface potential temperature takes the same role taken by vorticity in the Euler equations, as the active tracer advected by the dynamics \citep{Held&al1995}. Note that here we have chosen the active boundary at the bottom of the domain, consistently with the the formalism of the SQG$^{+1}$ model of \cite{Hakim&al2002} (which is defined in a semi-infinite domain, as classic SQG) and of the finite-depth SQG model \cite{Tulloch&Smith2006}. This choice is appropriate for atmospheric applications. If one wants to consider an oceanic application, the active boundary has to be taken at the top of the domain, which results in an inversion of the vertical coordinates. From the structure of the equations it is immediate that the results of our paper hold also in this case, simply changing sign to the potential temperature.

Note that, in order to compare the results of SSG and SQG turbulence, as well as with the results from SQG$^{+1}$ turbulence that are present in the literature, we have adopted here the formulation of SG in geostrophic and height coordinates. Although this is the most commonly used form of the equations, the natural formulation of SG in terms of potential vorticity advection and inversion is, instead, in geostrophic and isentropic coordinates. The proofs that SG is well posed construct a mapping from geostrophic and isentropic coordinates to physical space, but can not require that the boundary of the region where $Q_{sg}$ is constant in geostrophic and isentropic space maps to the physical boundary \citep{CullenBook}. As a result, the boundary conditions (\ref{eq:Monge-Ampere b.c.}) could result in an overdetermined problem and there is the possibility that the results are affected by a lack of well- posedness in the problem. Future work will have to clarify these issues.

In order to formulate the model in nondimensional form, we define as typical vertical length scale the depth of the domain $H$. The typical horizontal length scale in geostrophic space is taken as the Rossby deformation radius $L=L_R=N H /f$, where $N$ is the Brunt-V\"{a}is\"{a}l\"{a} frequency, here taken as constant. Note that, introducing the length scale in geostrophic space, $L$ represents the typical distance between lines of absolute momentum \citep{Craig1993}. Denoting the horizontal velocity scale as $U$, we introduce the geostrophic space Rossby number $\epsilon=U/Lf<<1$, and we consider a time scale $T=1/\epsilon f$ larger than the inertial time scale. Setting $Q_{sg}=N^2>0$  and with a suitable rescaling of streamfunction and potential temperature \citep{Hoskins&West1979}, the inversion equation in nondimensional form can be written as
\begin{equation}\label{eq:Monge-Ampere}
 \frac{\p^2 \Phi}{\p X^2} + \frac{\p^2 \Phi}{\p Y^2} + \frac{\p^2 \Phi}{\p Z^2} -  \epsilon \left[\frac{\p^2 \Phi}{\p X^2} \frac{\p^2 \Phi}{\p Y^2} - \left( \frac{\p^2 \Phi}{\p X \p Y} \right)^2 \right] =  0,
\end{equation}
with boundary conditions
\begin{equation}\label{eq:Monge-Ampere b.c.}
\begin{array}{l}
\displaystyle  \frac{\p \Phi}{\p Z}|_{Z=0} = \theta, \\[8pt]
\displaystyle  \frac{\p \Phi}{\p Z}|_{Z=1} = 0.
\end{array}
\end{equation}
Note that the rescaling transforms (\ref{eq:Monge-Ampere b.c.}) into a homogeneous equation.
The time evolution at $Z=0$ is given by
\begin{equation}\label{eq:Monge-Ampere t.e.}
\displaystyle  \frac{D \theta}{D t} = \left( \frac{\p}{\p t} + u_g \frac{\p}{\p X} +  v_g \frac{\p}{\p Y} \right) \theta = 0,
\end{equation}
with geostrophic velocities
\begin{equation}
\begin{array}{l}
\displaystyle u_g = -\frac{\p \Phi}{\p Y} ,\\[8 pt]
\displaystyle v_g = +\frac{\p \Phi}{\p X}.
\end{array}
\end{equation} 
The nondimensional geostrophic coordinates and the Bernoulli function are connected to the nondimensional physical coordinates and to the streamfunction by
\begin{equation}\label{eq: nondimensional transform}
\begin{array}{l}
\displaystyle X=x + \epsilon v_g, \\[8 pt]
\displaystyle Y=y -  \epsilon u_g, \\[8 pt]
\displaystyle Z=z , \\[8 pt]
\displaystyle \Phi=\phi+\frac{\epsilon}{2}\left(u_g^2 + v_g^2 \right).
\end{array}
\end{equation} 
Equations \eqref{eq:Monge-Ampere} - \eqref{eq:Monge-Ampere t.e.} are the same as in \cite{Hoskins&West1979}, with two differences: 1) we keep the full nonlinear form of equation \eqref{eq:Monge-Ampere} instead of neglecting the nonlinear terms, and 2) we take homogenous boundary conditions at the bottom as in the finite-depth SQG model \citep{Tulloch&Smith2006}. 

Note that for $\epsilon$ that goes to zero the finite-depth SSG equations tend to the finite-depth SQG equations, as (\ref{eq:Monge-Ampere}) tend to a Laplace equation and (\ref{eq: nondimensional transform})  to the identity. Finite-depth SQG features a transition scale corresponding to a critical wavenumber $k_t=f/NH$ \citep{Tulloch&Smith2006}, such that the kinetic energy spectrum follows a $k^{-3}$ law for $k<<k_t$ and a $k^{-5/3}$ law for $k>>k_t$. When the length scale $H$ is taken as the whole depth of the domain (as in our case), the transition scale corresponds to the Rossby radius of deformation, that is the horizontal unit length. In this case the finite-depth SQG spectrum is expected to follow essentially a $k^{-5/3}$ law, since the $k^{-3}$ regime is not represented in the power law regime of the spectrum (the power law scaling emerges for scales smaller than the length scale of the vortices dominating the dynamics,  that is the Rossby radius of deformation). In the current work we set the system in these conditions, as we are interested in the dynamics emerging in the frontal regime. It must be noted that when dealing with finite-depth SSG, we do not know in general what will be the effect on the transition scale due to 1) the presence of the nonlinear term and 2) the coordinate transformation. We reserve to explore this point in future studies.

The invertibility of the coordinate transformation requires the Jacobian of \eqref{eq: nondimensional transform} to be positive. Following \cite{Hoskins1975}, the Jacobian  of \eqref{eq: nondimensional transform} is equal to the vertical component of the absolute SG vorticity \eqref{eq:vort_sg}. In nondimensional geostrophic coordinates one obtains that
\begin{equation}\label{eq:Jacobian}
J^{-1} =  1 -  \epsilon \left( \frac{\p^2 \Phi}{\p X^2} + \frac{\p^2 \Phi}{\p Y^2}\right) +  \epsilon^2 \left[\frac{\p^2 \Phi}{\p X^2} \frac{\p^2 \Phi}{\p Y^2} - \left( \frac{\p^2 \Phi}{\p X \p Y} \right)^2 \right],
\end{equation}
When the invertibility condition breaks down, the model produces singular solutions. From a physical point of view the invertibility condition limits the values of the relative SG vorticity to be smaller than $f$, thus filtering out inertial instabilities.  

In general, if $Q_{sg}$ changes sign within the domain, \eqref{eq:inversion} is a mixed type non-homogeneous partial differential equation \citep{Tricomi1923} of Monge-Amp\'{e}re type (elliptic for $Q_{sg}>0$, hyperbolic for $Q_{sg}<0$). For example, in the ocean the potential vorticity can assume positive and negative values, e.g. due to the action of down-front winds which act to destroy potential vorticity \citep{Dasaro&al2011, Thomas2005}. However, for $Q_{sg}<0$ the assumptions behind the SG equations break down, therefore, given the qualitative, theoretical nature of our study, we limit our analysis to the case $Q_{sg}>0$. 

The SSG inversion equation \eqref{eq:Monge-Ampere} with $Q_{sg}>0$ is elliptic and homogeneous, and substantially easier to treat than the more general form \eqref{eq:inversion}. However, solving \eqref{eq:Monge-Ampere} is still much less straightforward than solving the Laplace equation arising in SQG \citep{Held&al1995,Tulloch&Smith2006}. Although other methods of solution of the SG equations have been developed in the past (see \cite{CullenBook} and references therein), in the vast majority of studies present in the literature numerical solutions of the SG equations have been obtained by approximating equations \eqref{eq:Monge-Ampere} - \eqref{eq:Monge-Ampere t.e.}. With one notable exception \citep{Snyder&al1991}, when treating the three-dimensional problem the nonlinear terms of equations \eqref{eq:inversion} and \eqref{eq:Monge-Ampere} have always been neglected, advocating the fact that they were small compared to the linear terms \citep{Hoskins1975,Hoskins1976,Hoskins&West1979}. Alternatively, the dynamics has been limited to a two-dimensional vertical plane, thus automatically eliminating the nonlinearity \citep{Hoskins&Bretherton1972,Badin2013}. In both cases the difference between QG and SG dynamics was therefore limited to the use of geostrophic coordinates in the latter: see, e.g., the early comment in  \cite{Hoskins1975} about SG theory providing merely a distortion of the QG solutions, its utility being thus essentially limited to justify the use of the formally and practically much simpler QG theory outside its strict range of applicability. However, \cite{Snyder&al1991} showed that including the nonlinear terms of  \eqref{eq:Monge-Ampere} does have a substantial effect on the development of cyclones. It is therefore arguable that these terms could play a non-negligible role in determining the properties of SSG turbulence.

\section{Solution procedure}\label{sec: methods}

In SQG, the inversion equation can be solved analytically, providing an explicit formula for the streamfunction as a function of the surface potential temperature in Fourier space \citep{Held&al1995,Tulloch&Smith2006}. On the contrary, \eqref{eq:Monge-Ampere} has to be solved numerically. Here we employ, adapting it to our specific problem, an iterative Poisson solver similar to the one introduced by \cite{Benamou&al2010} for the two-dimensional elliptic Monge-Amp\'{e}re equation. This is essentially the same method already successfully used in a three-dimensional case by \cite{Snyder&al1991}, but with a different way of solving the Poisson problem at each iteration. Let us define the operator $D$ as
\begin{equation}\label{eq: D term}
 \displaystyle  D \Phi  = \frac{\p^2 \Phi}{\p X^2}  \frac{\p^2 \Phi}{\p Y^2}  -  \left( \frac{\p^2 \Phi}{\p X \p Y} \right)^2 ,
 \end{equation}
Note that $D \Phi$ is the Jacobian of the velocity gradient matrix, and it corresponds to the Okubo-Weiss parameter \citep{okubo_1970,weiss_1991} up to a multiplicative constant. Positive values of $- \epsilon D \Phi$ in (\ref{eq:Monge-Ampere}) correspond thus to vorticity dominated regions, while negative values of $- \epsilon D \Phi$ correspond to strain dominated regions. From this point of view, the term $- \epsilon D \Phi$ can be seen, in geostrophic coordinates, as a forcing term, acting differently for different flow regimes, and representing thus a further difference between the SQG and SSG dynamics. 

Let us also define the operator $T_{\epsilon}$ as
\begin{equation}\label{eq: operator}
 \displaystyle  T_{\epsilon} \Phi =   \epsilon \Delta^{-1} D \Phi,
\end{equation}
where $\Delta$ represents the three dimensional Laplacian and $T_{\epsilon}$ incorporates the boundary conditions \eqref{eq:Monge-Ampere b.c.}. The solution of \eqref{eq:Monge-Ampere} can be obtained by iteration of the application of the operator $T_{\epsilon}$
\begin{equation}\label{eq: iterative problem} 
\displaystyle \Phi=\lim_{n \to +\infty} \Phi^{(n)} = \lim_{n \to +\infty} T_{\epsilon}^n\Phi^{(0)}.
 \end{equation}
In terms of a regular perturbation expansion, (\ref{eq: iterative problem}) corresponds to
\begin{equation}\label{eq: iterative problem new} 
\displaystyle \Phi^{(n)} = \sum_{j=0}^{n} \epsilon^{j} \Phi_{j}.
 \end{equation} 
Each step of the iteration requires solving a Poisson problem for $\Phi^{(n)}$
\begin{equation}\label{eq: Poisson problem} 
 \displaystyle \Delta \Phi^{(n)}  = \epsilon D \Phi^{(n-1)},
 \end{equation}
and inhomogeneous Neumann boundary conditions as in \eqref{eq:Monge-Ampere b.c.}. As starting point of the iteration it is natural to consider $\Phi^{(0)}$ such that  $\Delta \Phi^{(0)}=0$. Taking Fourier transforms in the horizontal directions, each step of the iteration becomes for each horizontal wavenumber $\vec{k}$ an Helmotz problem on the interval $[0,1]$
\begin{equation}\label{eq: Helmoltz problem}
 \displaystyle \frac{\p^2 \hat{\Phi}^{(n)}}{\p Z^2} - k^2 \hat{\Phi}^{(n)}  = \epsilon \hat{D} \Phi^{(n-1)},
\end{equation}
where $\hat{D}$ is defined such that  $\hat{D} \Phi$ is equal to the Fourier transform of $D \Phi$,  $k=|\vec{k}|$ and the inhomogeneous Neumann boundary conditions are
\begin{equation}\label{eq: Helmoltz problem b.c.}
\begin{array}{l}
 \displaystyle \frac{\p \hat{\Phi}^{(n)}}{\p Z}|_{Z=0} = \hat{\theta}, \\ [8pt]
 \displaystyle \frac{\p \hat{\Phi}^{(n)}}{\p Z}|_{Z=1} = 0.
 \end{array}
\end{equation}
\cite{Snyder&al1991} solved \eqref{eq: Helmoltz problem} via finite differences and numerical inversion of the resulting linear system. However, \eqref{eq: Helmoltz problem} with \eqref{eq: Helmoltz problem b.c.} has the formal solution
\begin{equation}\label{eq: solution with Green function}
 \displaystyle \hat{\Phi}^{(n)} (\vec{k},Z)= \hat{\theta}(\vec{k},0) G_k(Z,0)   + \epsilon \int_{0}^1 G_k(Z,Z') \hat{D} \Phi^{(n-1)}(\vec{k},Z')   dZ' ,
\end{equation}
where $G_k(Z,Z')$ is the Green's function of \eqref{eq: Helmoltz problem} (see e.g. \cite{Tulloch&Smith2006})
\begin{equation} \label{eq: Green function verbose} 
 \displaystyle G_k(Z,Z')=
\left \{
\begin{array}{l}
 \displaystyle -\frac{1}{k}\frac{cosh(k Z) cosh(k (Z'-1))}{sinh(k)} ,\,\,\,\,\,\, Z<Z',  \\ [8pt]
 \displaystyle -\frac{1}{k}\frac{cosh(k (Z-1)) cosh(k Z')}{sinh(k)} ,\,\,\,\,\,\, Z \ge Z' .
\end{array}
\right .
\end{equation}
Equation \eqref{eq: Helmoltz problem} can therefore be solved directly by computing \eqref{eq: solution with Green function}. In this way, each step of the iterative procedure requires for each wavenumber $\vec{k}$ the numerical evaluation of the integral in the second term of \eqref{eq: solution with Green function}. Note that this is greatly facilitated by the fact that we have chosen a domain limited in the vertical. The case with a semi-infinite domain would be much more complicated to solve numerically.

The Green's function depends only on the geometry of the system, therefore it can be computed exactly for each value of $k$ and $Z$ at the beginning of the integration. Our tests have shown that solving \eqref{eq: Helmoltz problem} with \eqref{eq: solution with Green function} is computationally much more convenient than the approach of \cite{Snyder&al1991}. Moreover, since the source term of \eqref{eq: Helmoltz problem} scales with $\epsilon$, the output of the iteration appears as a series in increasing powers of $\epsilon$, whose first terms can be computed explicitly using \eqref{eq: solution with Green function}. The zeroth order solution $\hat{\Phi}^{(0)}$ is the starting point of the iteration and is given only by the boundary term
\begin{equation}\label{eq:solution zeroth order}
 \displaystyle \hat{\Phi}^{(0)} =  -\frac{\hat{\theta}(\vec{k},0)}{k}\frac{cosh(k (Z-1))}{sinh(k)} \equiv \hat{\Phi}_0.
\end{equation}
Note that this is the is finite-depth SQG and SSG solution of \citet{Tulloch&Smith2006} and \citet{Badin2013}. After one application of $T_{\epsilon}$ we have the first order correct solution
\begin{equation}\label{eq:solution first order}
 \displaystyle \hat{\Phi}^{(1)} = -\frac{\hat{\theta}(\vec{k},0)}{k}\frac{cosh(k (Z-1))}{sinh(k)} + \epsilon \int_{0}^1 G_k(Z,Z')  \hat{D}\Phi^{(0)}(\vec{k},Z')  dZ' \equiv \hat{\Phi}_0 +\epsilon \hat{\Phi}_1.
\end{equation}
Further applications of $T_{\epsilon}$ will add terms order $O(\epsilon^2)$ and higher, with a more involved formal structure. However, if we consider small $\epsilon$ we can stop at first order with a reasonable degree of accuracy. In the simulations performed in this work we have used small values of $\epsilon$, so that we can take $\hat{\Phi}^{(1)}$ as solution of the problem. Our tests have shown that the method converges relatively fast for any value of $\epsilon$ (maximum 10-15 iterations), and that for the  values of $\epsilon$ used in the paper considering second-order correct solutions does not change the results. Note that as $\epsilon$ decreases the SSG and SQG solutions converge, since $\Phi$ converges to $\Phi_0$ and the coordinate transformation to the identity. 

In most works on SG approximation the analysis of the properties of the solutions was limited to the geostrophic space, and the transformation back to physical space performed only for visualization purposes, with the aid of graphical methods. In oder to perform a quantitative analysis also in physical space, \cite{Hoskins1975} proposed a simple method for performing the inverse coordinate transformation. \cite{Schaer&Davis1990} developed a more complex (and rigorous) iterative algorithm, that for small values of $\epsilon$ reduces to the method proposed by \cite{Hoskins1975}  (which corresponds to the first step of the iterative algorithm of  \cite{Schaer&Davis1990}). In this work we consider small values of $\epsilon$. We have therefore used the simple method of \cite{Hoskins1975}, consistently with limiting the iteration of the solution procedure of the Monge-Amp\'{e}re equation to first order. For more details on the iteration of the iterative procedure to find the solution of the Monge-Amp\'{e}re equation and on the transformation of coordinates, see Appendix \ref{sec: A} and \ref{sec: B} respectively.

\section{SSG turbulence}\label{sec:SSG_turbulence}

\subsection{Model description}\label{sec: model}

The numerical integration of the SSG model is performed in geostrophic space and it involves two steps. First, starting from an initial condition in geostrophic coordinates $(X,Y,Z)$ satisfying (\ref{eq:Monge-Ampere}), the potential temperature $\theta$ at $Z=0$ is advected with velocities given by the streamfunction $\Phi$ at $Z=0$. Then, the streamfunction $\Phi$ is computed in the whole domain solving (\ref{eq:Monge-Ampere}) with the new potential temperature field as boundary condition at $Z=0$. The solution of (\ref{eq:Monge-Ampere}) is given by (\ref{eq:solution first order}) at first order in $\epsilon$. This gives a new streamfunction field at $Z=0$ that is used to advect the potential temperature, and so on. The potential temperature in the whole domain can be reconstructed at each time-step by differentiating the vertical profile of the streamfunction, but since the values of $\theta$ in the interior are not involved in the dynamics, this is done only in post-processing. All the fields are then transformed in physical space in post-processing.

The advection of potential temperature at $Z=0$ is performed with the semi-spectral method employed by \cite{Constantin&al2012} to study the formation of singular solutions in the SQG equations. Taking advantage of the periodic boundary conditions in the horizontal directions, the potential temperature field at $Z=0$ and time $t$ is approximated as 
\begin{equation}\label{eq:Fourier zeta}
\theta(\vec{X},t) = \sum_{k_X,k_Y=-N/2}^{N/2-1} \hat{\theta}(\vec{k},t)e^{i \vec{k}\cdot \vec{X}}
\end{equation}
where $\vec{X}=(X,Y)$ at $Z=0$, $\vec{k}=(k_X,k_Y)$ and $N$ is the number of grid-points in the horizontal directions. Time evolution is then performed integrating with a fourth order Runge-Kutta scheme the prognostic equation for the Fourier transform of the potential temperature $\hat{\theta}(\vec{k},t)$ at $Z=0$. The Jacobian is computed in geostrophic space, after a Fast Fourier Transform, using the Arakawa discretization \citep{Arakawa1966}, which guarantees conservation of the energy and enstrophy invariants. In order to remove numerical instabilities, it is common practice to introduce a dissipation operator, typically in the form of a hyperdiffusion. Following \cite{Constantin&al2012}, we employ instead an exponential filter introduced by \citep{Hou&Li2007}, multiplying each Fourier mode $\hat{\theta}(\vec{k},t)$ by $\rho(2k_X/N)\rho(2k_Y/N)$, where $\rho(x) = \exp{ \left[ -\alpha x^m \right]}$. We take $\alpha=36$ and $m=19$, as in \cite{Constantin&al2012}. This high order exponential spectral filter has been developed by \citet{Hou&Li2007} specifically for the discretization of systems of equations developing singular or nearly-singular solutions (as the semi-geostrophic equations). With this choice of the parameters, the application of the filter keeps the 2/3 lower wavenumber modes unchanged and suppresses the 1/3 higher wavenumber modes. Note that this filter is a less sharp version of the classical Orszag 2/3 method, where the highest 1/3 of the wavenumbers coefficient is simply put to zero. \cite{Hou&Li2007} showed that the use of this high order exponential filter captures up to $15\%$ more effective Fourier modes than the 2/3 Orszag method, producing more accurate approximations of the solutions. We have performed experiments with different choices of the parameters of the filter, in particular using $\alpha=512$ and $m=24$, such that the filter was sharper and thus more effective in removing small scale features, without observing differences in the results.

The streamfunction is computed by solving the nonlinear Monge-Amp\'{e}re equation with the method described in the previous section. We limit our investigation to small Rossby numbers, so that we consider only the first term of the expansion resulting from the iterative procedure. The solution of the problem is computed at each time step by numerically computing \eqref{eq:solution first order}. While in SQG the streamfunction can be computed from the potential temperature at the top boundary through a simple inversion in Fourier space, effectively reducing the problem to two-dimensional, in SSG the presence of the nonlinear term in the Monge-Amp\'{e}re equation requires taking explicitly into account the full three-dimensional structure of the streamfunction and including a discretization of the vertical coordinate. However, like in SQG, also in SSG the vertical integration is made from a diagnostic equation and does not need time integration at each vertical layer. We have performed simulations on a $512 \times 512$ horizontal square grid and 20 vertical levels. As one can see from equation \eqref{eq:solution first order}, high wavenumber modes decay faster with depth, so that high vertical resolution is needed only close to the surface. The vertical levels are exponentially spaced, with the layers depth varying from $\Delta Z=0.004$ to $\Delta Z=0.18$  from the top to the bottom.

The initial condition is defined as a random field of surface $\theta$ in geostrophic space, as common practice in the literature on SG approximation. The random field of surface $\theta$ is defined as in \cite{Hakim&al2002}, such that the corresponding zeroth order surface streamfunction follows
\begin{equation} \label{eq: IC} 
\displaystyle \hat{\Phi}_0(\vec{k},0) \propto \frac{k^{m/4-1}}{(k+k_0)^{m/2}},
\end{equation}
with $m=25$ and $k_0$=14. A random phase is given to each mode, in order to have a random initial condition with a prescribed kinetic energy spectrum. After several tests aimed at avoiding cases with too many points at which the invertibility condition was violated, we have selected the cases with surface kinetic energy normalized at ${KE}_{sg}=5$. Note that here as well as in the following we consider the kinetic energy associated with the geostrophic velocity field $(u_g,v_g)$, as this is the quantity that enters in the conserved energy integral \eqref{eq:energy_sg}. The time step is taken as $dt=0.005$, and the simulations are performed up to $T=100$, corresponding to about 300 eddy turnover times, computed according to \cite{Kerr1990}. We performed a sensitivity analysis on the Rossby number, performing simulations for $\epsilon$=(0.02,~0.05,~0.1,~0.15,~0.2). The range of values of $\epsilon$ is in agreement with the range chosen by \citet{Snyder&al1991}, that consider a maximum value $\epsilon = 0.3$.  We compare the results also with runs of the standard finite-depth SQG.

Note that our experiments do not include forcing and large scale dissipation. Among other works that have performed freely decaying simulations with SQG and SQG-like models, \cite{Hakim&al2002} limited the analysis at long times, while \cite{Capet&al2008} limited the analysis at early times during which nonlinear interactions are rather strong. Both note that for long times the system shows kinetic energy spectra much steeper than what predicted by the theory. For shorter times the spectra of \cite{Capet&al2008} are closer to the theoretical prediction. However, \cite{Capet&al2008} study the transfer of surface kinetic energy in SQG flows, and their analysis focuses on a stage of the evolution that is far from approximating a stationary state of the SQG equations (they have almost not vortices formed yet). We have therefore chosen the approach of \cite{Hakim&al2002}. We have restricted our analysis  to the last 10 units time (corresponding to about 30 eddy turnover times), where we have tested that the statistics of the system does not change with time in a significant way. In this regime energy is reduced to about 30\% of its value at time zero, but further decays by less then 5\% within the time window where the analysis is performed. Still, our freely decaying flow does not reach a real stationary state, and we expect to observe SQG kinetic energy spectra deviating from the theoretical prediction, as in \cite{Hakim&al2002}.

Transformation from geostrophic to Cartesian coordinates is performed in post-processing, checking if the integration produces singularities \citep{Schaer&Davis1990}. We have checked that in the full turbulent state the invertibility conditions is violated in the worst cases in a few percents of the total points of the domain only, in the upper layers of the model and for the largest values of the Rossby number. The local formation of singularities is unavoidable in numerical simulations of the SG equations, and the problem occurs also when employing different methods of solution that do not involve the coordinate transformation (see \cite{CullenBook} and reference therein). However, in the numerical model these singularities are systematically eliminated by the spectral filter, so that they are not problematic from the stability point of view. We have performed a large number of tests varying the total kinetic energy of the initial condition (see above), and selected those simulations for which the points at which the invertibility condition was not satisfied in the fully turbulent phase of the flow remained few and isolated. Note that the problem of violation of the invertibility condition at the initial condition could be avoided formulating the model in isentropic coordinates. In this way, defining $Z=g\theta/\theta_0$, SSG would be determined by imposing the condition that $Q_{sg}$ is constant on a region periodic in $(X,Y)$ with $Z_0(X,Y,t)<Z<Z_1$, where $Z_1$ is constant. The evolution equation would then apply as an equation for $Z_0$ instead than for $\theta$ (although with different boundary conditions). In this case one could use any random field $Z_0$ as initial condition without violating the invertibility condition, thus avoiding the initial dissipation. However, one should note that local violations of the invertibility condition and dissipation would anyway emerge during the evolution of the flow due to numerical effects. Further, this would not allow for comparison of the results of this study with classical studies of SQG turbulence.

\begin{figure}
  \begin{center}
   \includegraphics[height=6.5cm,width=6.5cm]{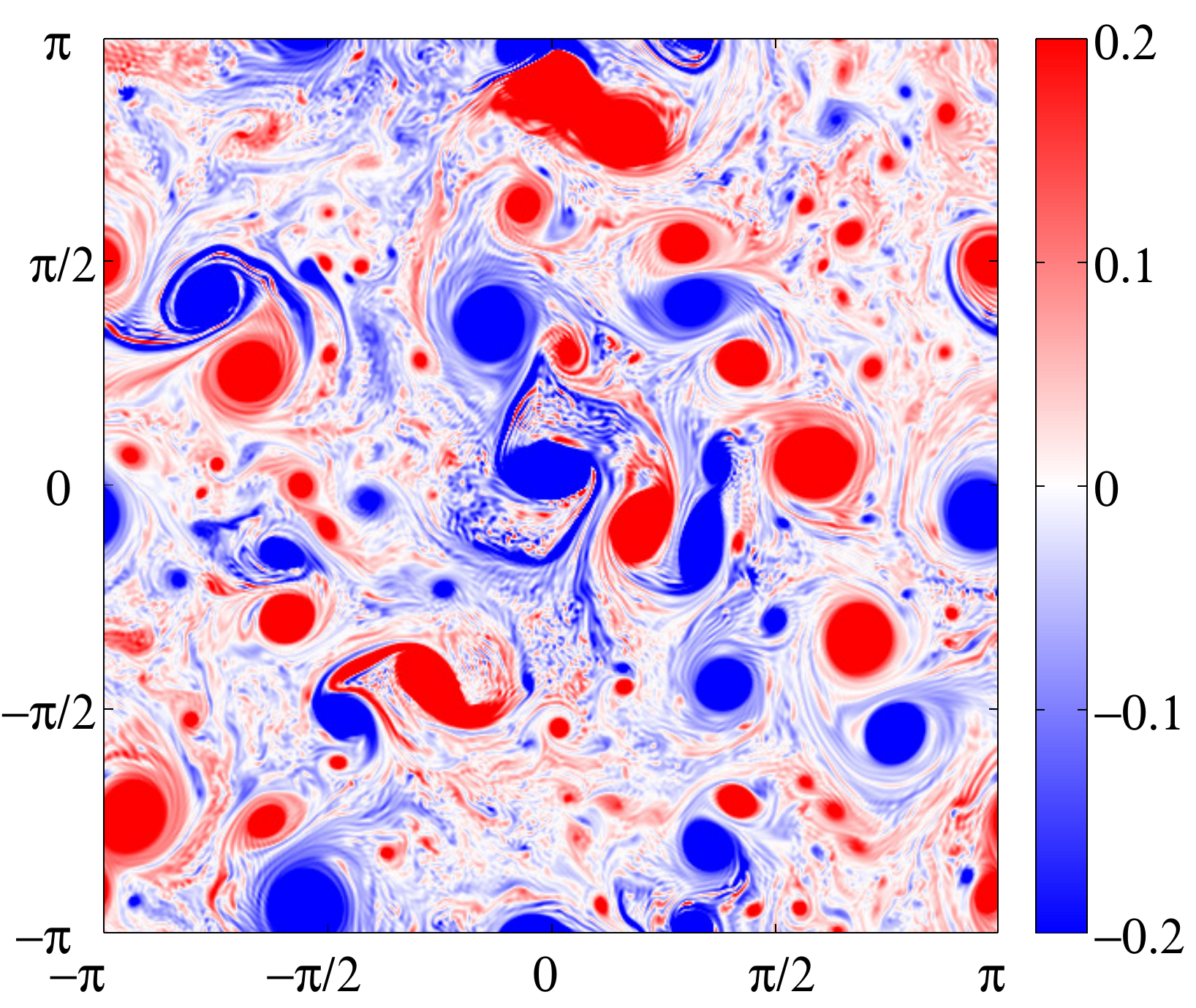}
   \includegraphics[height=6.5cm,width=6.5cm]{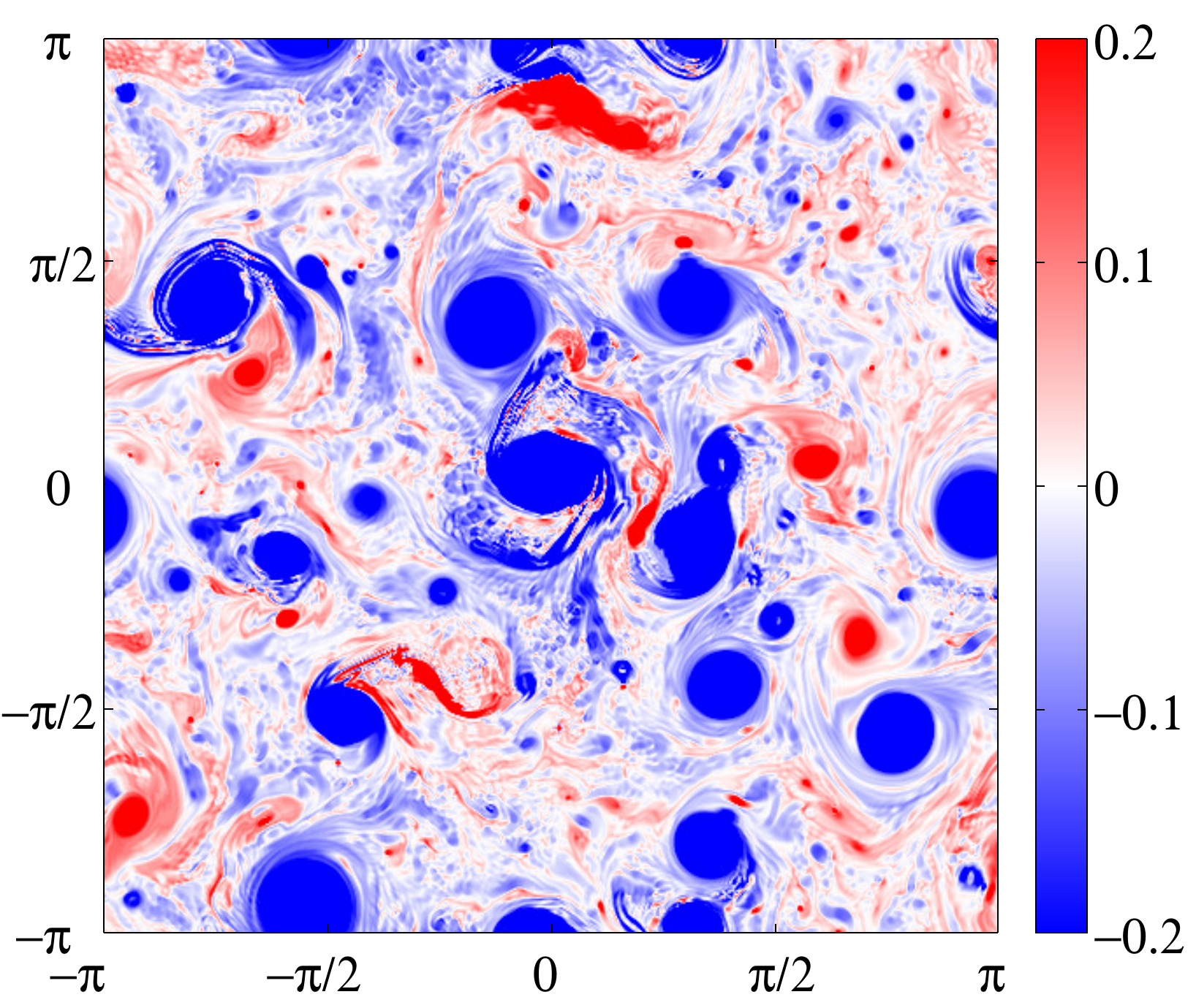}
   \includegraphics[height=6.5cm,width=6.5cm]{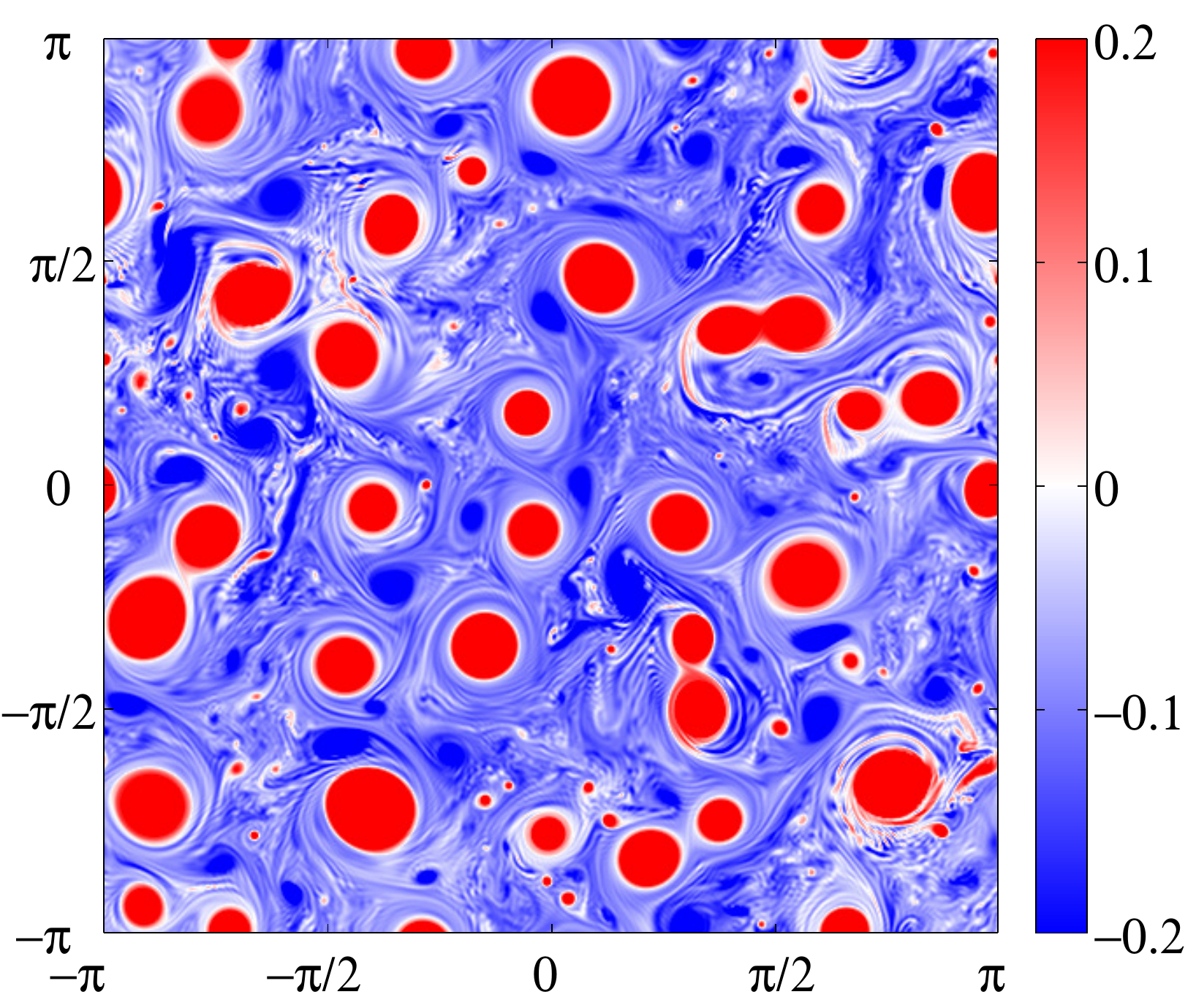} 
   \includegraphics[height=6.5cm,width=6.5cm]{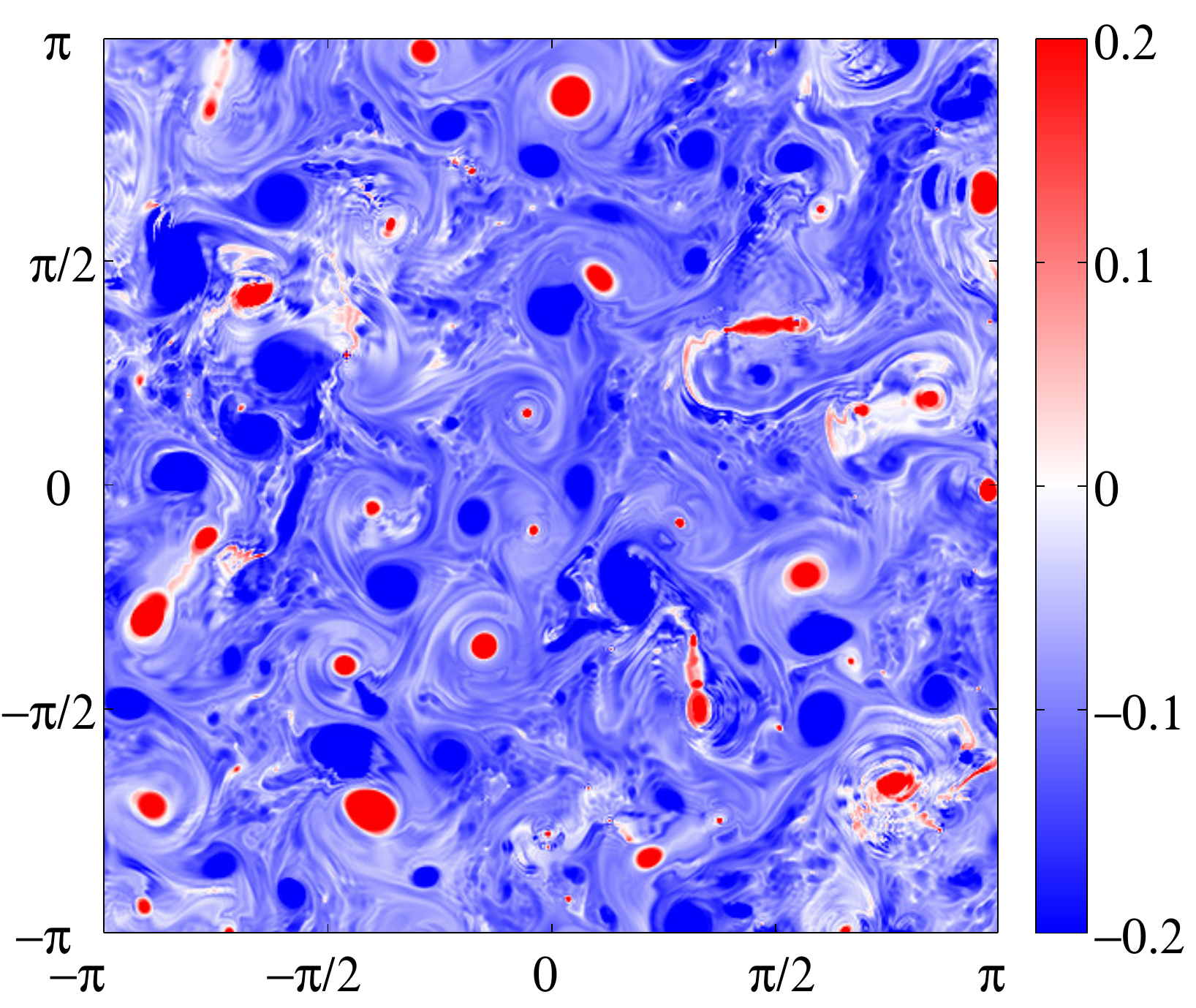}  
 \end{center}   
 \caption{Snapshots at T=100 of $\theta$ at $z$=0 for SQG (top-left), SQG under coordinate transformation with $\epsilon=0.2$ (top-right), SSG for $\epsilon$=0.2 in geostrophic coordinates (bottom-left) and SSG for $\epsilon$=0.2 in physical coordinates (bottom-right).}
 \label{fig:snapshots surface}
\end{figure}

\subsection{Surface statistics}\label{sec: surface}

\begin{figure}
  \begin{center}
  \includegraphics[height=6.5cm,width=6.5cm]{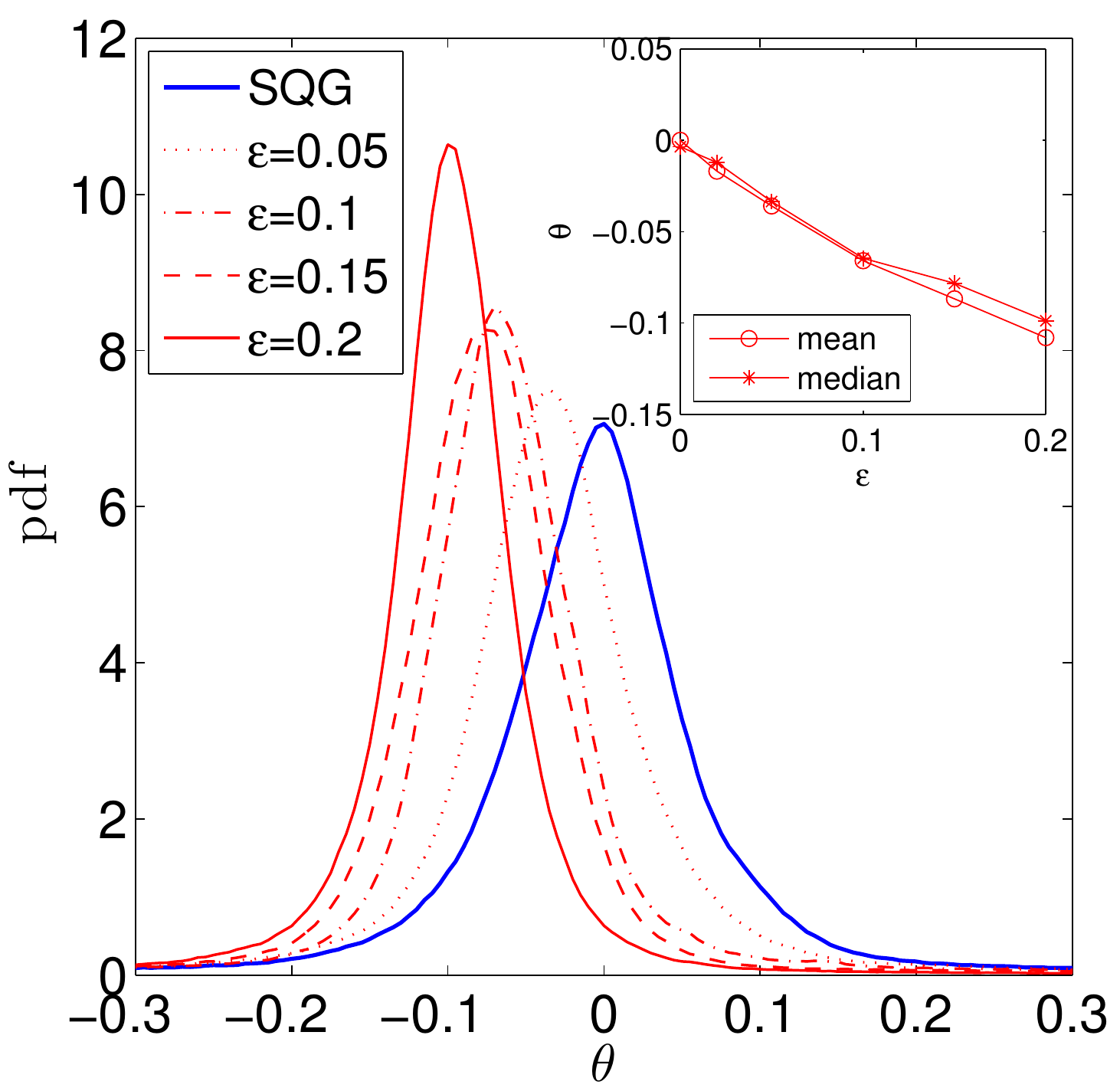}
  \includegraphics[height=6.5cm,width=6.5cm]{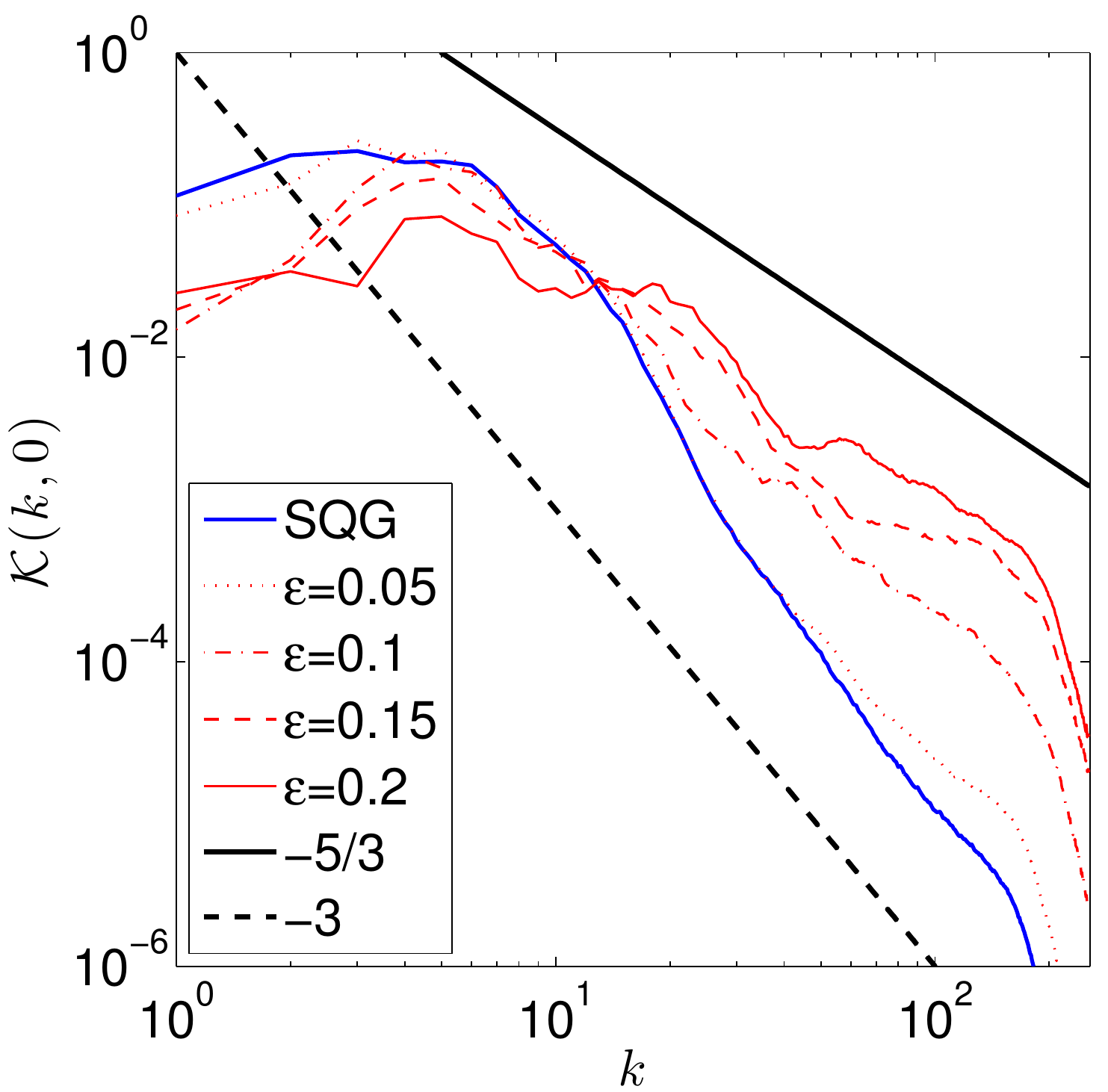}
  \end{center}   
  \caption{(Left: PDFs of surface $\theta$ for SQG (blue line) and SSG for different values of Rossby number in physical coordinates (red lines). In the inset we show as a function of the Rossby number the mean and median of the distributions. The SQG case corresponds to the origin. Right: surface kinetic energy spectra for SQG (blue line) and SSG for different values of Rossby number in physical coordinates (red lines).}
\label{fig:statistics surface}
\end{figure}

Figure \ref{fig:snapshots surface} shows snapshots of surface $\theta$ at $T$=100 for SQG (top-left), SQG under coordinate transformation with $\epsilon=0.2$ (top-right, corresponding to a SSG simulation as traditionally proposed in the literature where only the coordinate transformation is considered), SSG for $\epsilon$=0.2 in geostrophic coordinates (bottom-left, where only the effect of the inclusion of the nonlinearity of the inversion equation is visible) and SSG for $\epsilon$=0.2 in physical coordinates (bottom-right, the full SSG case). While the range of variability is one order of magnitude larger, the scale of $\theta$ is limited to the range [-0.2,0.2], in order to highlight structures with lower intensity rather than the vortices. Note that $\theta$ is the active conserved scalar in SQG and SSG, so that it takes the role normally taken by vorticity. At z = 0, SQG (top-left panel) produces mostly localized coherent structures with filaments formed by secondary instabilities in between. In comparison, SSG turbulence (bottom-right panel) is a mixture of local structures, represented by the coherent vortices, and nonlocal structures, represented by fronts and filaments dominating the dynamics.  The filaments are not produced by secondary instabilities but rather by the organization in features with skewed potential temperature, induced by the joint action of the nonlinear term and the transformation of coordinates. Cyclones (region with positive $\theta$ anomaly) are smaller than anticyclones and isolated. The appearance of an asymmetry in the distribution and size of cyclonic and anticyclonic regions is an expected result of the coordinate transformation. As can be seen comparing the SQG snapshot (top-left panel) and the SQG snapshot under coordinate transformation (top-right panel), the effect of the coordinate transformation is that positive relative vorticity (potential temperature) is increased in magnitude and the areas where it occurs are compressed, while negative relative vorticity (potential temperature) is decreased in magnitude and the areas where it occurs are expanded \citep{Hoskins1975}. On the other hand, the different vortex dynamics induced by the nonlinear term is clear comparing the SQG snapshot (top-left panel) and the SSG snapshot in geostrophic coordinates (bottom-left panel), which show how the nonlinear terms act to enhance the negative values of $\theta$, corresponding to the strain dominated regions of the flow. 

The left panel of Figure \ref{fig:statistics surface} shows the PDFs of surface $\theta$ for SQG (blue line) and SSG for different values of Rossby number in physical coordinates (red lines). While SQG produces a non-Gaussian but zero-centered PDF, SSG  produces instead skewed PDFs that are centered around non-zero negative values. Since the isolated coherent structures control the tails of the distribution, it is the filaments that are responsible for this asymmetry in the bulk of the distribution. The increase in magnitude of the mean and median of the PDFs with $\epsilon$ shows an almost linear trend (in the inset), as a signature of a continuous shift towards negative values of the potential temperature anomaly as $\epsilon$ increases. This effect has been observed also in \citep{Hakim&al2002}, and it is linked to a net cooling of the surface. The net cooling is due to results from the increase in surface kinetic energy that accompanies the forward cascade of buoyancy variance, which implies a decrease in available potential energy, which lowers the center of gravity of the fluid. In an oceanic case, where the temperature anomaly has the opposite sign, this effect is associated to the restratification effect associated with filamento- and frontogenesis.   

The right panel of Figure \ref{fig:statistics surface} shows the surface horizontal kinetic energy spectra for SQG (blue line) and SSG for different values of Rossby number and physical coordinates (red lines). In general the spectra at level $Z$ have been computed taking the radial average of the kinetic energy spectral density field
\begin{equation}\label{eq:spectra}
\displaystyle \mathcal{K}(k,Z) = \frac{1}{4\pi}\int_0^{2\pi}k^2\left|\hat{\phi}(\vec{k},Z)\right|^2d\omega,
\end{equation}
\citet{AndrewsHoskins78} predicted a semi-geostrophic $k^{-8/3}$ power law for both the kinetic and available potential energy of a one dimensional front reaching an infinite value for vorticity. Following the behaviour of the front past the singularity, \citet{Boyd92} corrected the value to $k^{-2}$. Both these values refer however to the case of an isolated one dimensional front.  

Finite-depth SQG spectra are supposed, in the chosen setup, to follow a $k^{-5/3}$ law, as the transition between the $k^{-3}$ and $k^{-5/3}$ occurs at $k=6$, so that the regime $k^{-3}$ is not represented. However, our SQG spectra are steeper than $k^{-5/3}$, due to the fact that our free decaying flows does not reach a real steady state, as the energy dissipation at small scales by the spectral filter is not compensated by injection of energy by an external forcing. Without a forcing, the flow does not stabilize in a scale invariant stationary state, but tends to develop coherent structures that prevent the spectra to attain the theoretical slope. The same effect is observed and discussed in \cite{Hakim&al2002} and \cite{Capet&al2008}, that have performed similar freely decaying simulations of the SQG and SQG$^{+1}$ models. SSG spectra are less steep and tend to $k^{-5/3}$ as $\epsilon$ increases, in agreement with the emerging role of filaments which implies that more energy is stored at smaller scales for larger values of $\epsilon$. SSG spectra show also the presence of "bumps", which might be an indication of the enhanced energy present in coherent structures  which are not symmetric around $\overline{\theta}=0$, i.e. associated to the fact that in SSG regions with positive and negative potential temperature are not symmetric.

\subsubsection{Role of the nonlinear term and of the transformation of coordinates}

\begin{figure}
  \begin{center}
  \includegraphics[height=6.5cm,width=6.5cm]{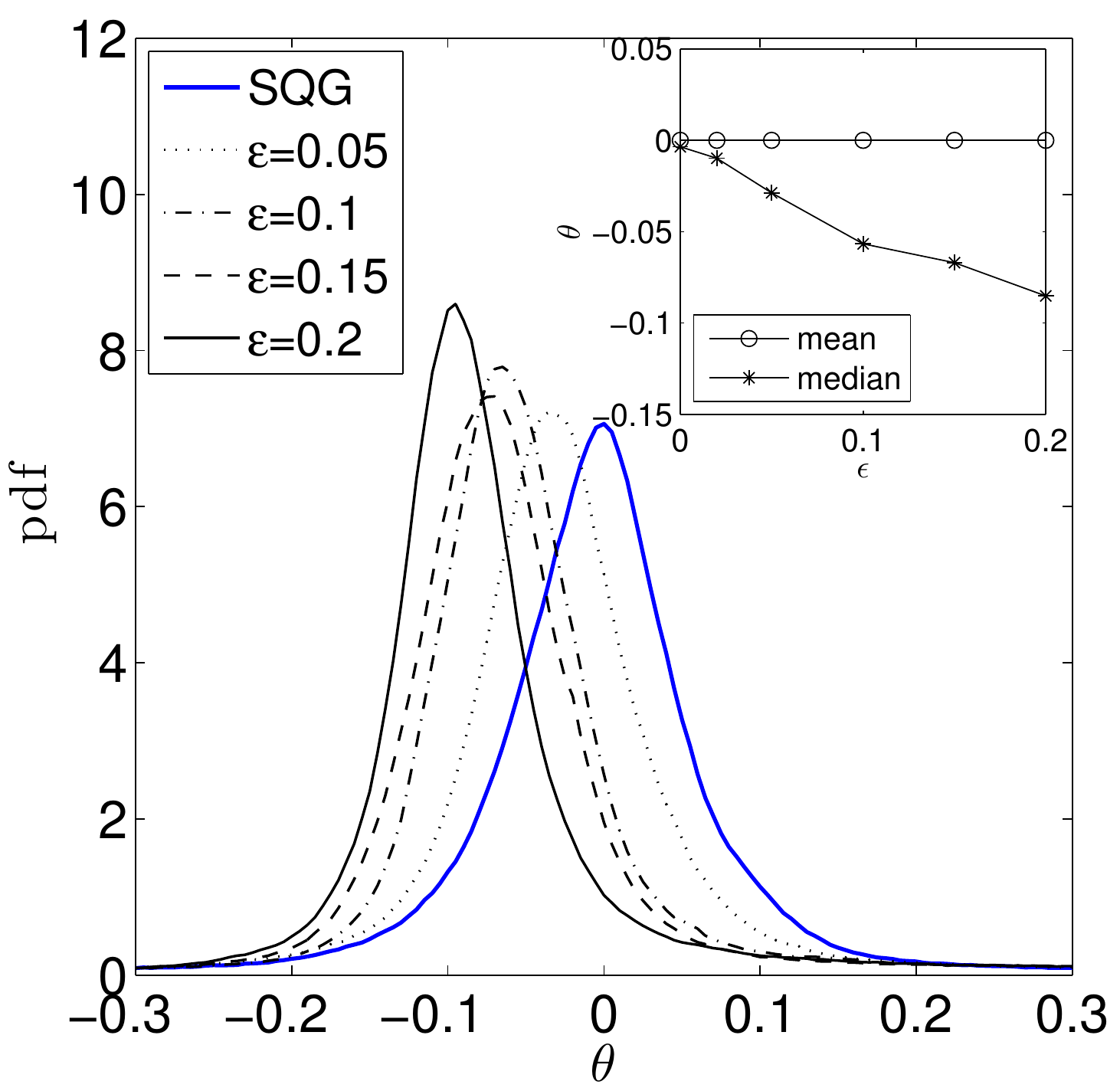}
 \includegraphics[height=6.5cm,width=6.5cm]{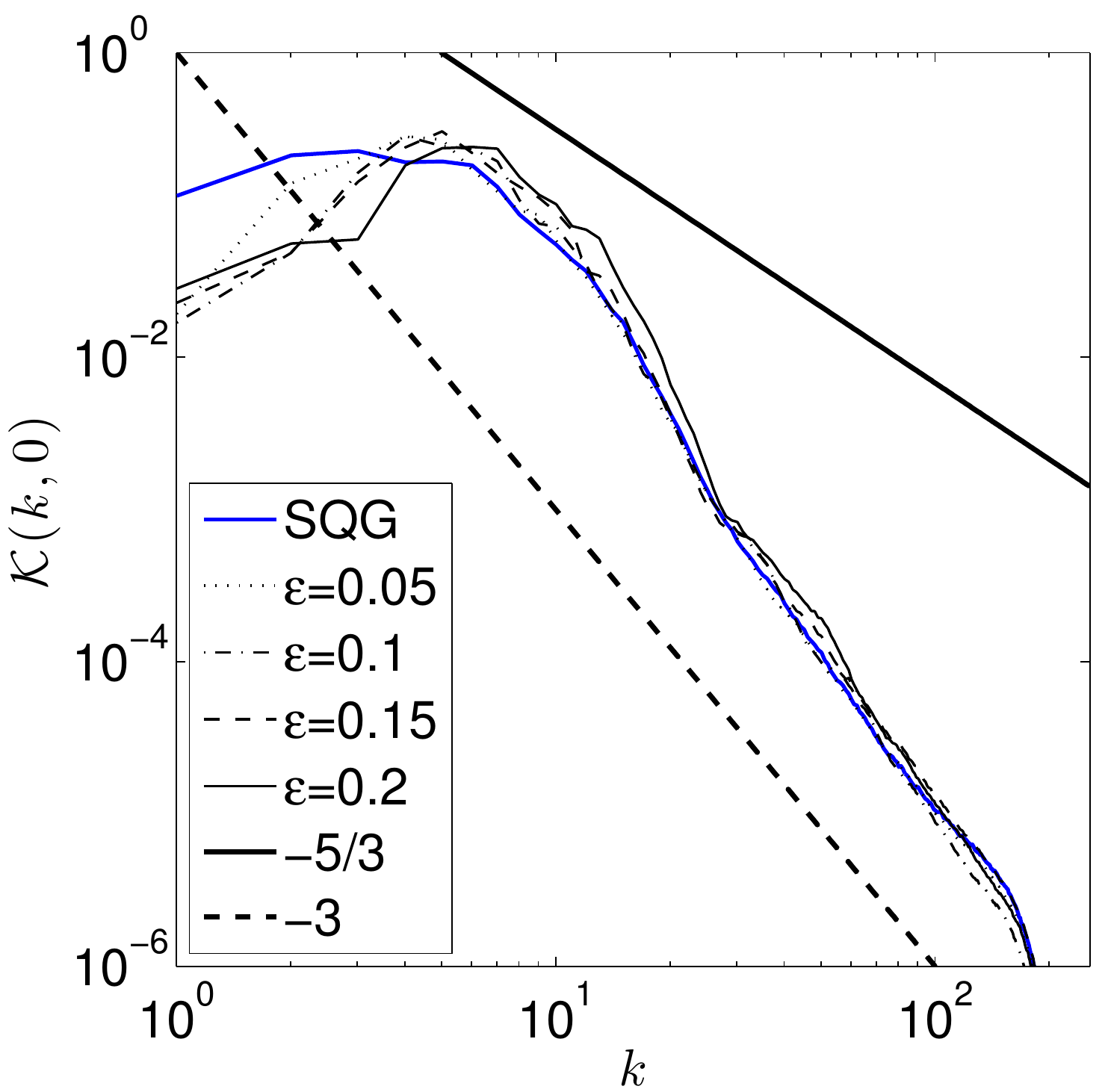}
   \includegraphics[height=6.5cm,width=6.5cm]{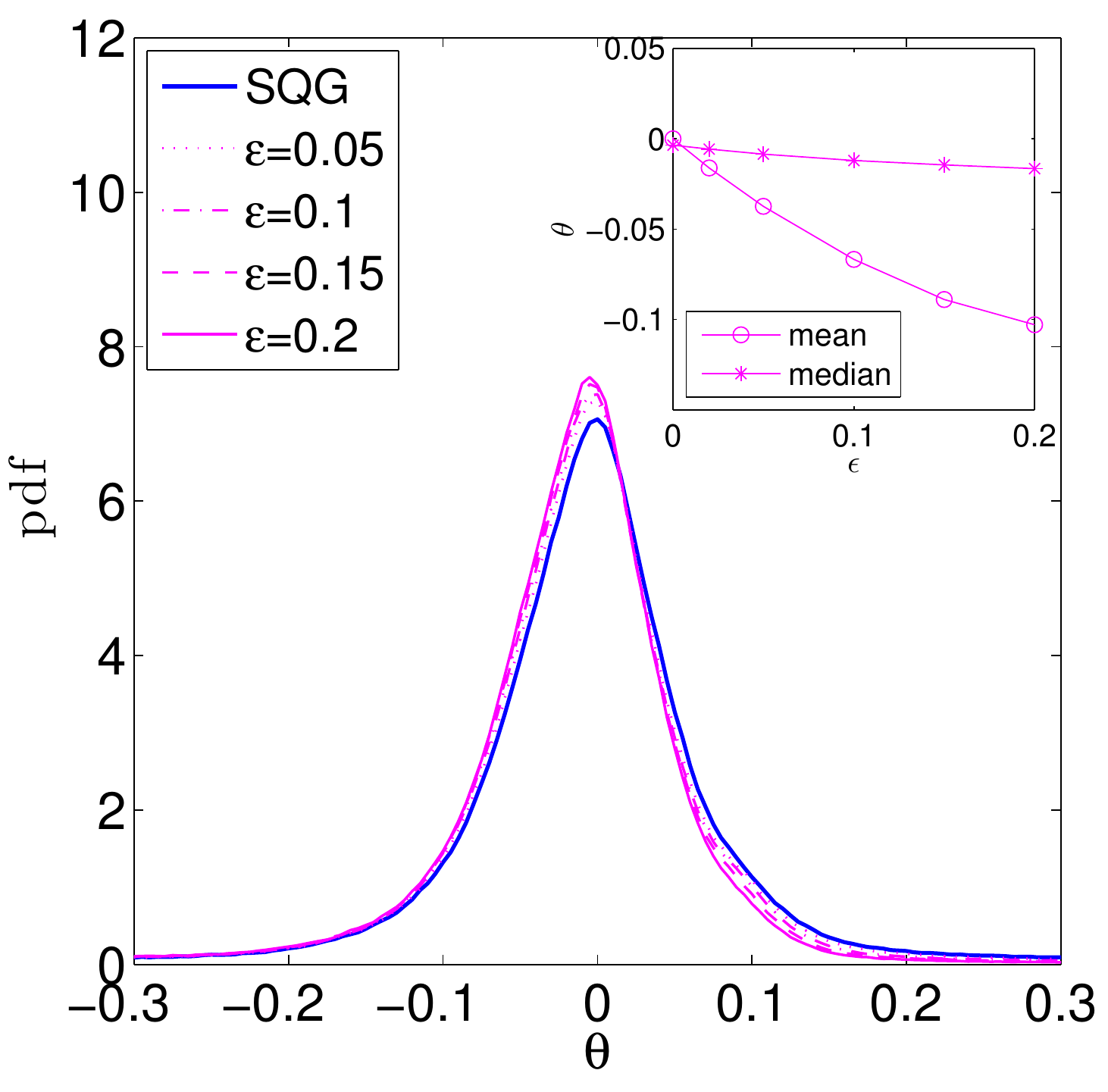}
 \includegraphics[height=6.5cm,width=6.5cm]{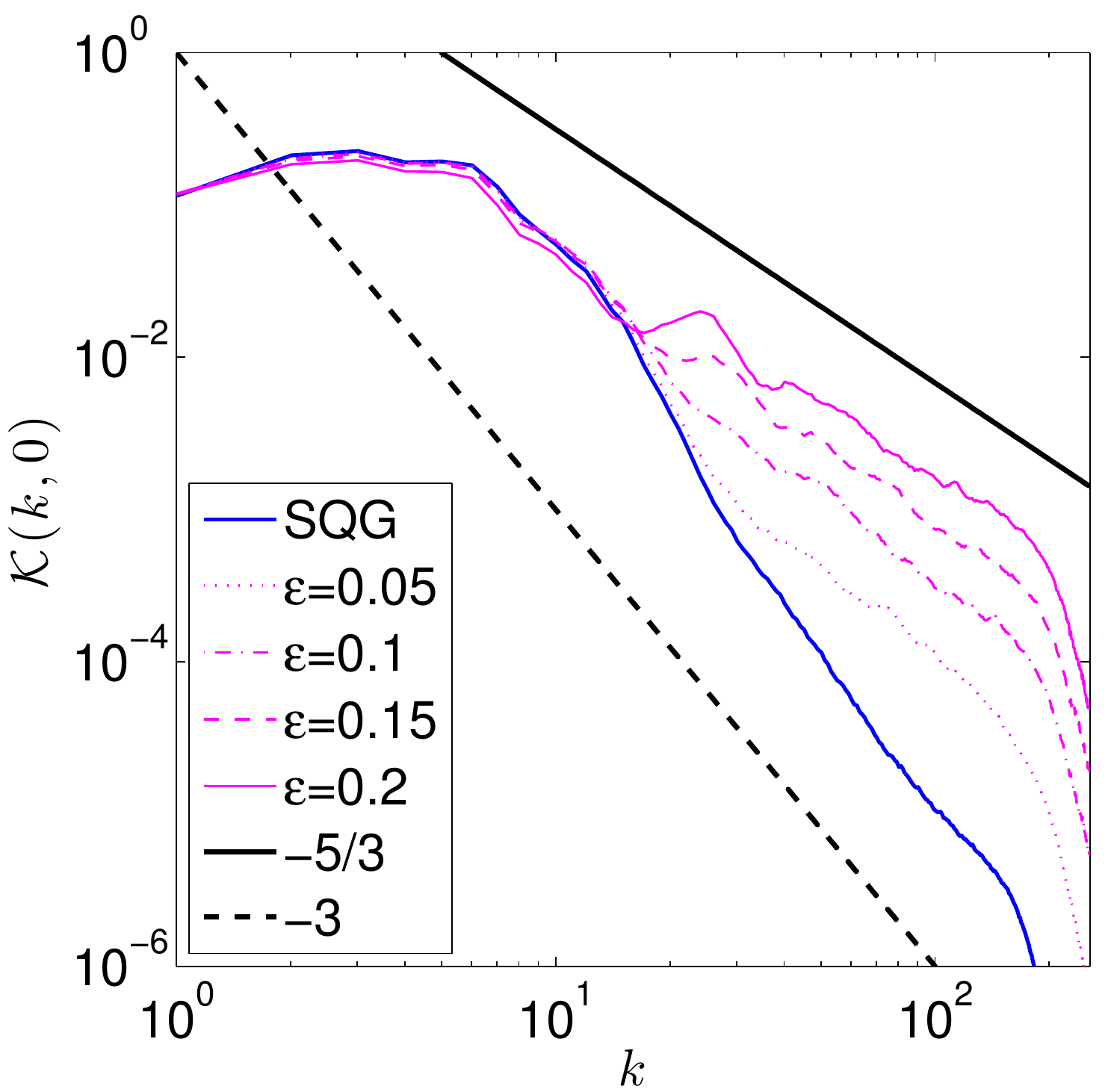}
  \end{center}   
  \caption{Top-left: PDFs of surface $\theta$ for SQG (blue line) and SSG in geostrophic coordinates(i.e. without the transformation of coordinates) for different values of Rossby number (black lines), with mean and median in the inset. Top-right: surface kinetic energy spectra for SQG (blue line) and SSG in  geostrophic coordinates (i.e. without the transformation of coordinates) for different values of Rossby number (black lines). Bottom-left: PDFs of surface $\theta$ for SQG (blue line) and SQG under the application of the inverse coordinate transformation (i.e. without the nonlinear term) for different values of the Rossby number (magenta lines), with mean and median in the inset. Bottom-right: surface kinetic energy spectra for SQG (blue line) and SQG under the application of the inverse coordinate transformation (i.e. without the nonlinear term) for different values of the Rossby number (magenta lines).}
  \label{fig:separation effects}
\end{figure}

SSG differs from SQG because of 1) the presence of the nonlinear term in \eqref{eq:Monge-Ampere} and 2) the application of the coordinate transformation. We try to disentangle the role of these two elements by estimating their individual impact on PDFs and spectra. In the top row of Figure \ref{fig:separation effects} we show the PDFs of surface $\theta$ (left panel) and the kinetic energy spectra (right panel) for SQG (blue lines) and SSG in geostrophic coordinates for different values of the Rossby number (black lines). In this way we retain only the effects due to the presence of the nonlinear term in \eqref{eq:Monge-Ampere}. Horizontally averaging the surface temperature equation we have
\begin{equation}\label{eq: averaged temperature}
 \displaystyle \frac{d \overline{\theta}}{ dt} = - \overline{\theta \nabla \cdot u},
\end{equation}
For horizontally non-divergent dynamics, the r.h.s. must vanish and the mean of the potential temperature must be zero. Since the SSG dynamics in geostrophic coordinates is non-divergent (as it is determined by the streamfunction $\Phi$), the presence of the nonlinearity in the inversion equation can not be responsible for the net cooling. Remarkably however, the PDFs of $\theta$ in geostrophic coordinates maintain the shift of the peak to negative values. This is not in contradiction with what said: the inset of the top-left panel of Figure 3 shows that the mean surface temperature is indeed zero for any value of $\epsilon$, while it is only the median of the distributions that departs from zero for increasing values of $\epsilon$. The shift of the center of the PDFs is compensated by their increasing skewness.

As discussed in Section \ref{sec: methods}, the presence of the nonlinear term acts as a forcing, enhancing the values of negative potential temperature in the strain dominated regions between the coherent structures, resulting thus in a shift of the median of the PDF. This effect is seen also in Figure 1, which shows that in absence of the transformation of coordinates the effect of the nonlinear term is to enhance the regions of negative potential temperature in the stirring dominated regions, resulting in an effective disappearance of the region of zero potential temperature in between coherent structures that is instead visible in the simulations without the inclusion of the nonlinearity in the inversion equation. The kinetic energy spectra are however not affected by the presence of the nonlinear term, as they show in SSG for all the values of the Rossby number the same slope as in SQG.

In the bottom row of Figure \ref{fig:separation effects} we show the PDFs of surface $\theta$ (left panel) and the kinetic energy spectra (right panel) for SQG (blue lines) and SQG under the application of the inverse coordinate transformation for different values of the Rossby number (magenta lines). These data represent the SSG dynamics when the nonlinear term is neglected, as usually done in the literature on the SG approximation. In this way, the difference between SQG and SSG is limited to the deformation of the flow induced by the coordinate transformation. We can see that in this case the PDFs of $\theta$ remain peaked at zero, but are characterized by mean different from zero. Slight deviations from zero of the median are one order of magnitude smaller than the value of the mean for the same Rossby number, and are probably due to numerical effects.

This is due to the fact that the coordinate transformation makes the flow divergent, and is therefore responsible for a different deformation of cyclonic and anticyclonic structures and areas. The coordinate transformation is thus the determining factor responsible to the net cooling through the horizontal divergence of the velocity field. Additionally, the kinetic energy spectra in the bottom left panel of Figure \ref{fig:separation effects} are flatter than for SQG, exactly as in Figure \ref{fig:statistics surface}, as more energy is stored at small scales due to the stretching of cyclonic areas that creates frontal structures characterized by strong horizontal gradients.
 
Remarkably,  the effects of the nonlinearity and of the coordinate transformation are thus essentially separable. In particular, the nonlinear term appears to be not negligible at all, as it is the only responsible of the fact the the PDFs of the active conserved scalar are peaked at non zero values. Both SQG and the linearized version of SSG commonly studied in the past fail to capture this distinctive qualitative feature. On the other hand, the net cooling and the change in the slope of the kinetic energy spectra is due solely to the coordinate transformation.

\subsection{Comparison with the SQG$^{+1}$ model} \label{sec: HSM}

Finite-depth SSG presents several similarities with the SQG$^{+1}$ model of \cite{Hakim&al2002}, although the two models are rather different mathematically.  It is therefore of interest to compare the results of the two models. The similarities come from the fact that both models can be see as extensions of the SQG model taking into account for the effect of ageostrophic advection at first order in Rossby number. The way the ageostrophic advection is introduced is however very different in the two models. In particular, the SSG model involves a coordinate transformation that introduces a deformation dependent on the vorticity field, that has the effect of favoring the formation of frontal structures. Further, while both SSG and SQG$^{+1}$ are second order accurate approximations to the Euler equations, SQG$^{+1}$ requires $\epsilon \sim Fr$, where $Fr$ is the Froude number, while SSG requires $\epsilon \sim Fr^2$. As a consequence of this, SQG$^{+1}$ employs a linearization of the static stability profile about a reference value, while SSG does not. It would thus be expected for SSG to perform better than SQG$^{+1}$ at describing the effect of static stability variations, but be deficient in describing vortex instabilities.

The PDFs of $\theta$ of SSG and SQG$^{+1}$ at surface present the same kind of asymmetry. In both cases the PDFs are peaked at moderately negative values, with both the median and the mean taking negative values. This net cooling is due to the restratification effect. In SSG is it possible to disentangle the effects of the nonlinear term and of the coordinate transformation, showing that the net cooling is induced solely by the coordinate transformation, an thus on the advection by the ageostrophic flow, while the shift of the peak of the distribution is due to the nonlinear term. It is not therefore possible to connect SQG$^{+1}$ to only one of these two aspects of SSG. On the other hand, SSG and SQG$^{+1}$ strongly differ in the kinetic energy spectra. In SSG the coordinate transformation leads to more energy to be stored at small scales, with a consequent flattening of the SSG spectra with respect to the SQG case. On the contrary, the SQG$^{+1}$ have exactly the same slope as the SQG spectra. Both models similarly affect the population number and morphology of the cyclones/anticyclones. \cite{Hakim&al2002} performed a detailed analysis of the vortex statistics and structure, making use of algorithms for vortex census and estimation of the vortex radius. It would be therefore of great interest to perform a similar analysis also on the SSG dynamics in a future work. 

\subsection{Interior statistics} \label{sec: interior}

\begin{figure}
  \begin{center}
  \includegraphics[height=6.5cm,width=6.5cm]{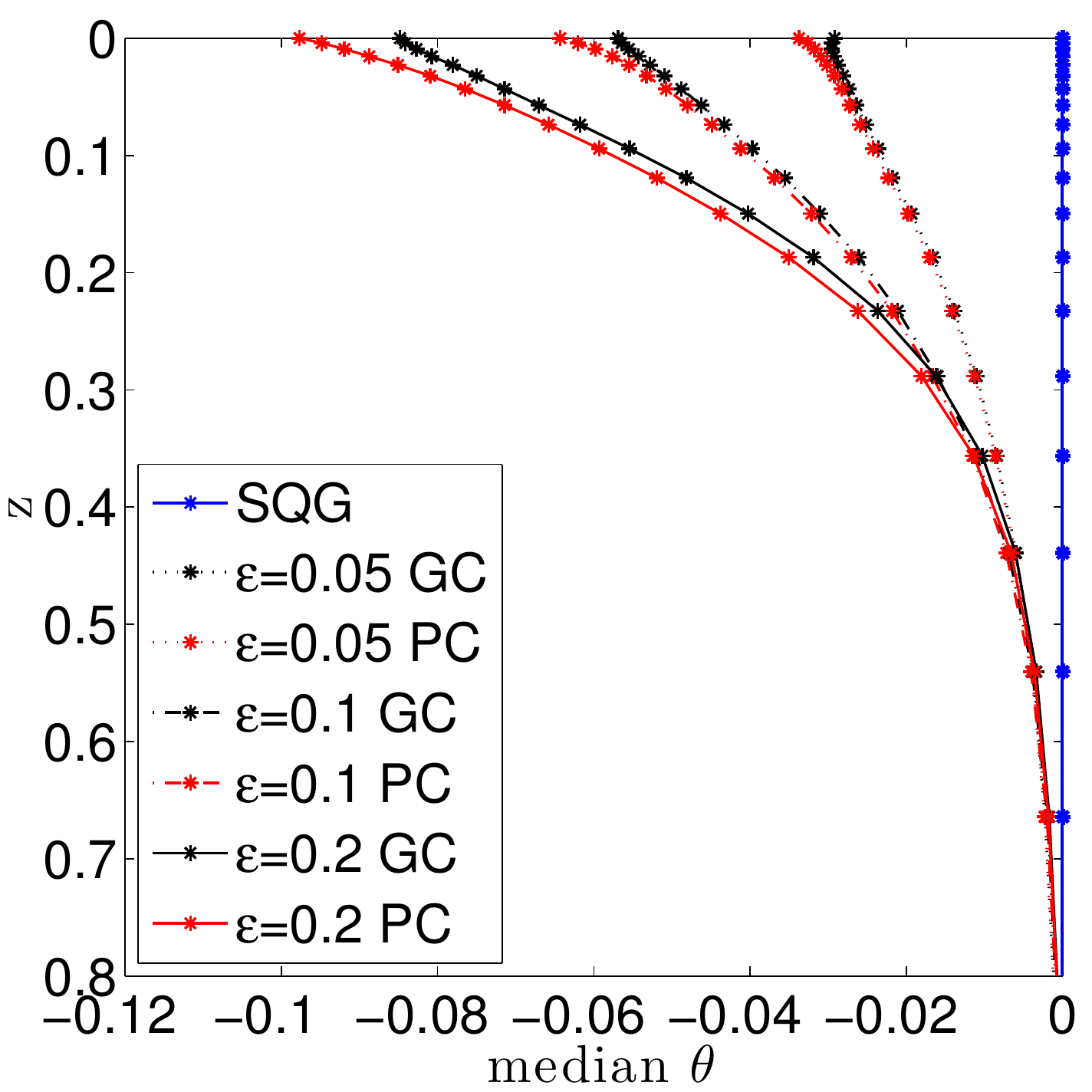}  
  \includegraphics[height=6.5cm,width=6.5cm]{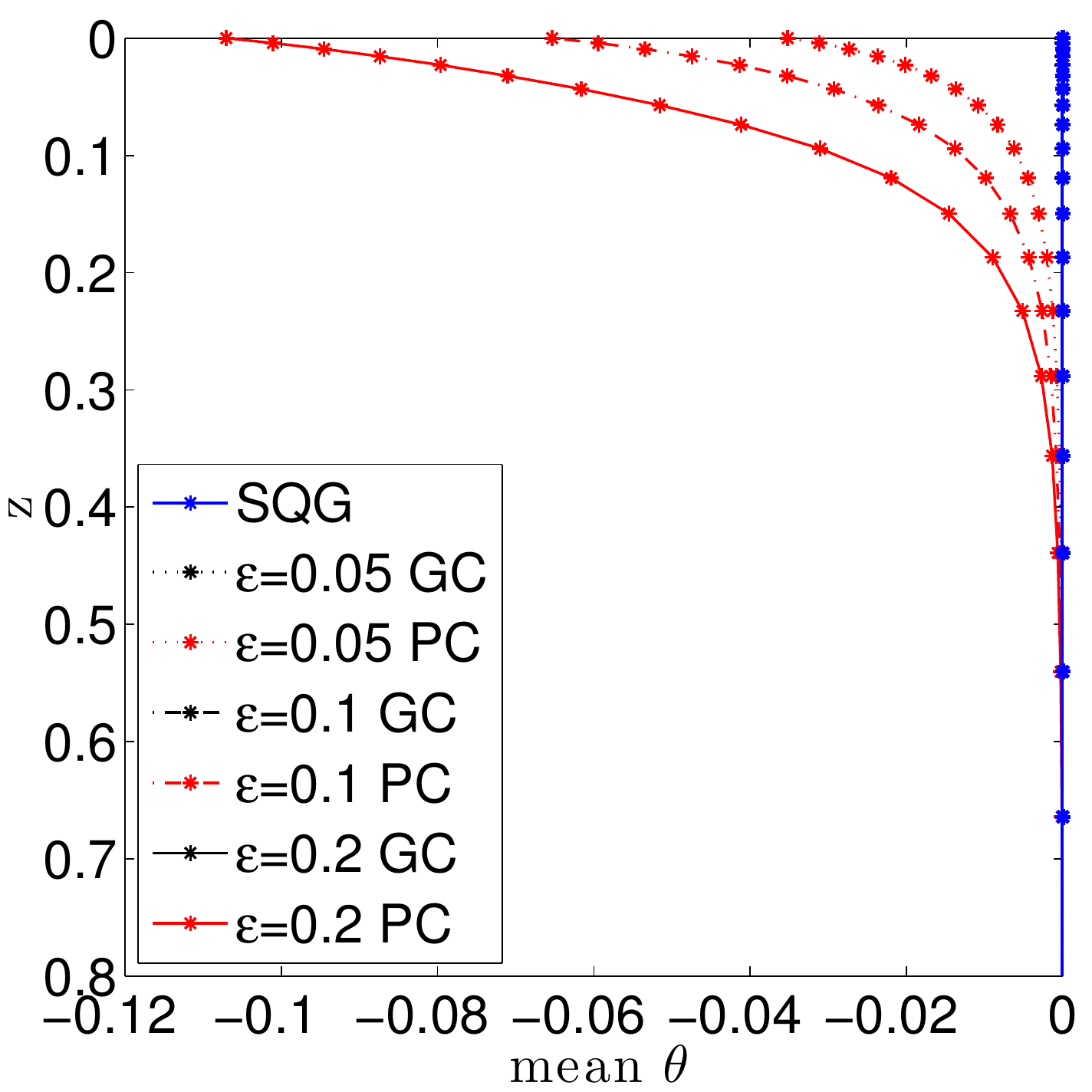}  
  \includegraphics[height=6.5cm,width=6.5cm]{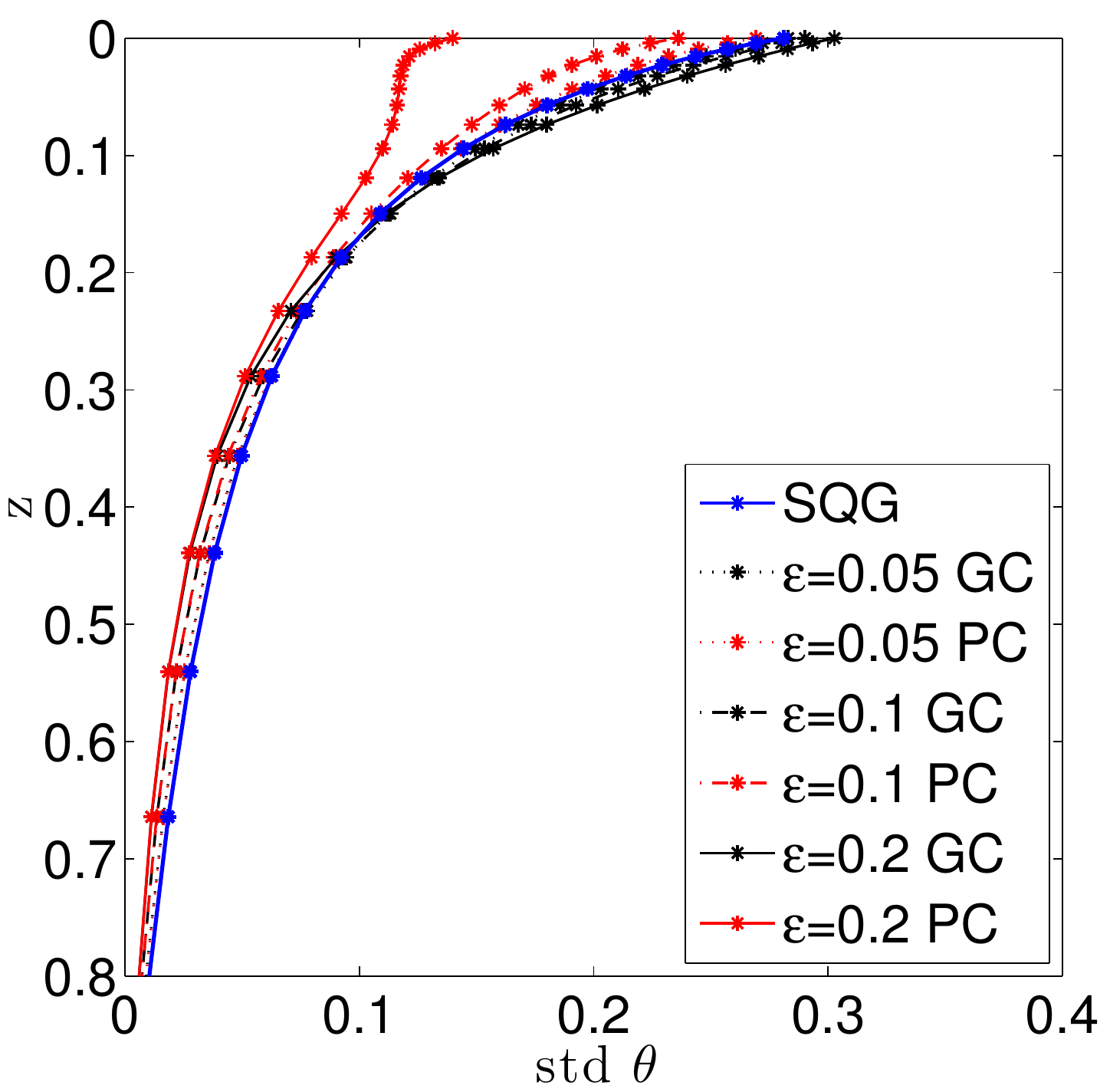}  
  \includegraphics[height=6.5cm,width=6.5cm]{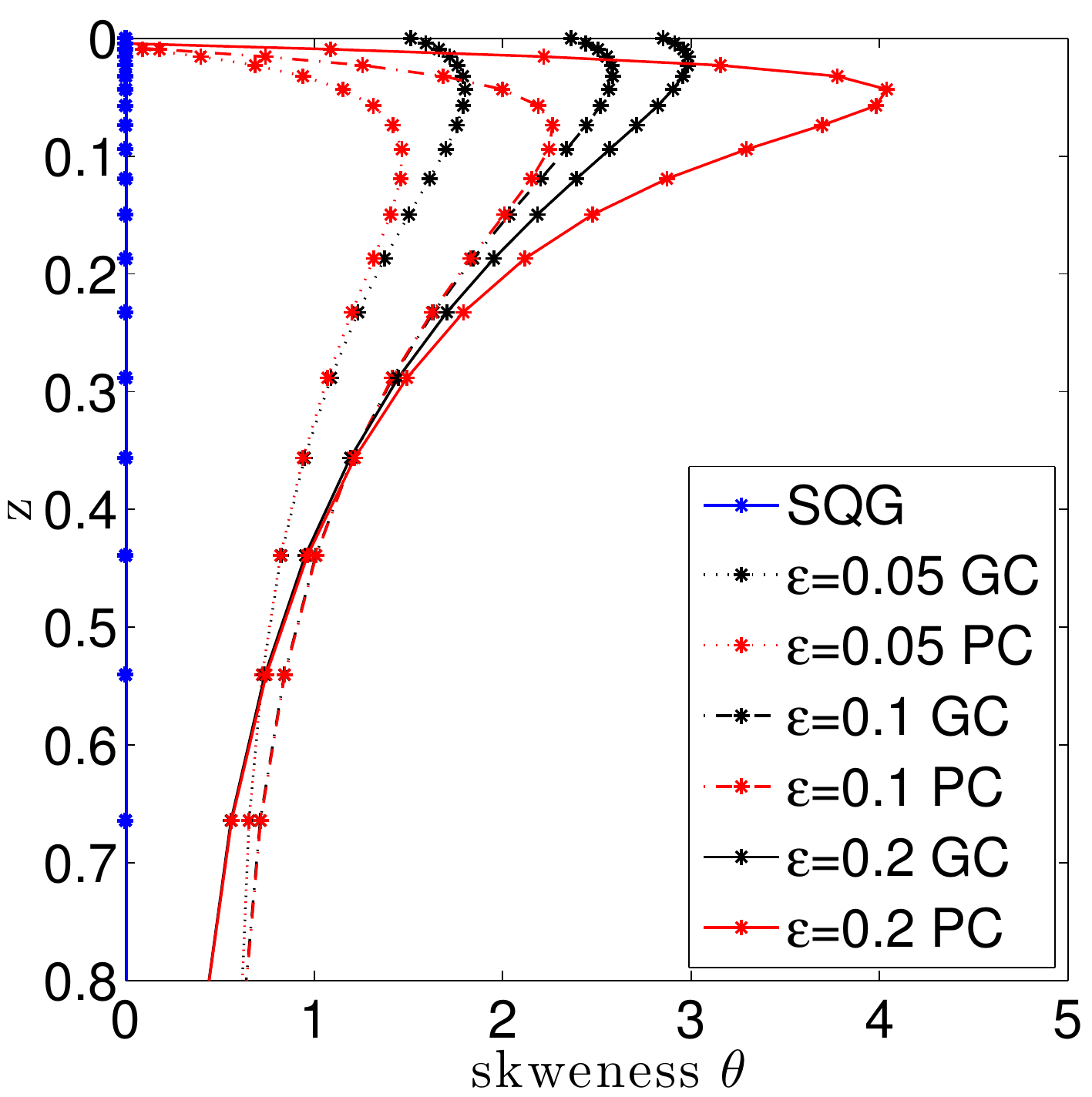}  
  \end{center}   
  \caption{Vertical profiles of moments of $\theta$ distributions for SQG (blue lines) and SSG for different values of the Rossby number in geostrophic (black lines) and physical (red lines) coordinates: median (top-left), mean (top-right), standard deviation (bottom-left), skewness (bottom-right).}
\label{fig:moments}
\end{figure}

\begin{figure}
  \begin{center}
  \includegraphics[height=6.5cm,width=6.5cm]{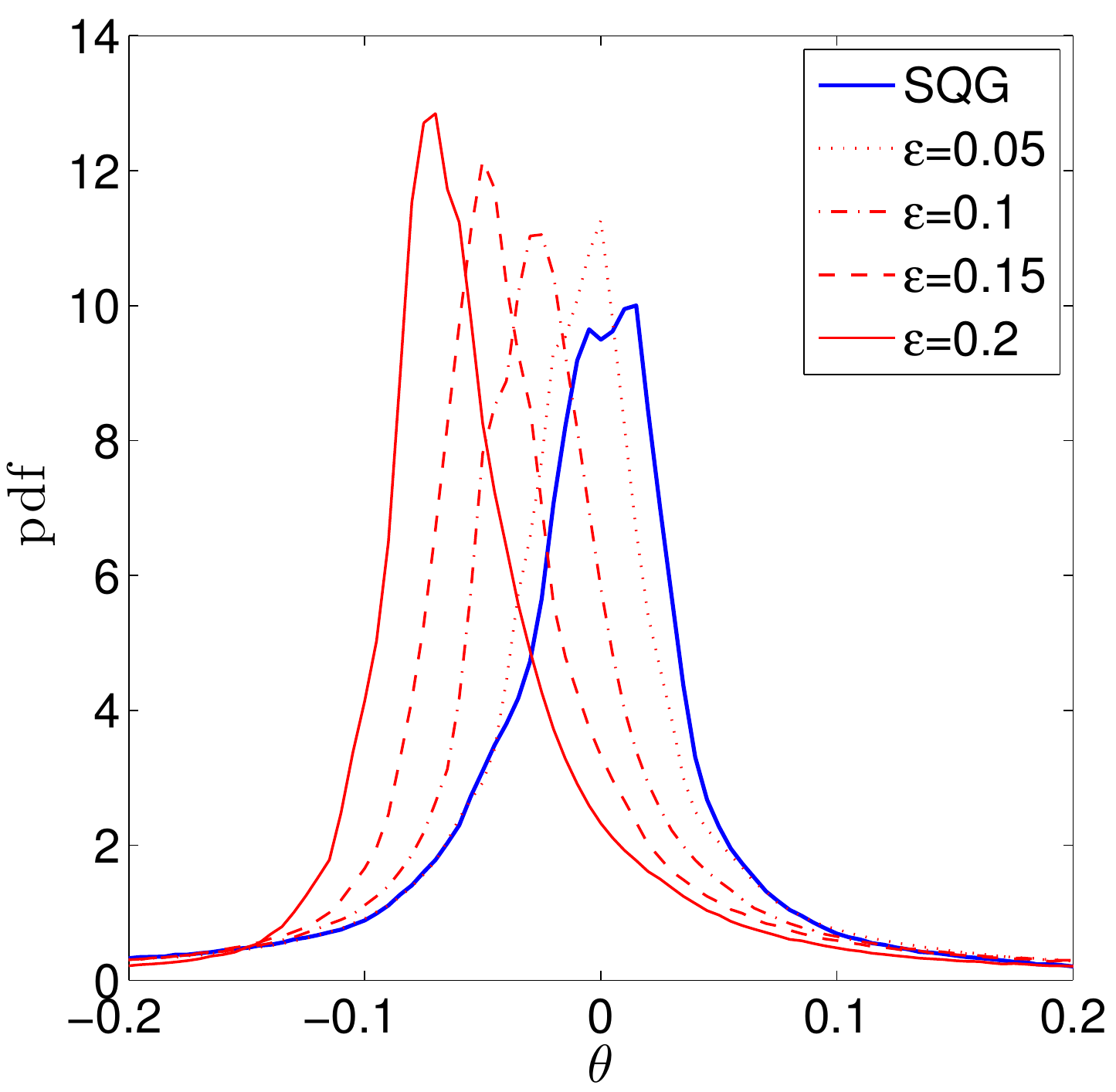}
  \includegraphics[height=6.5cm,width=6.5cm]{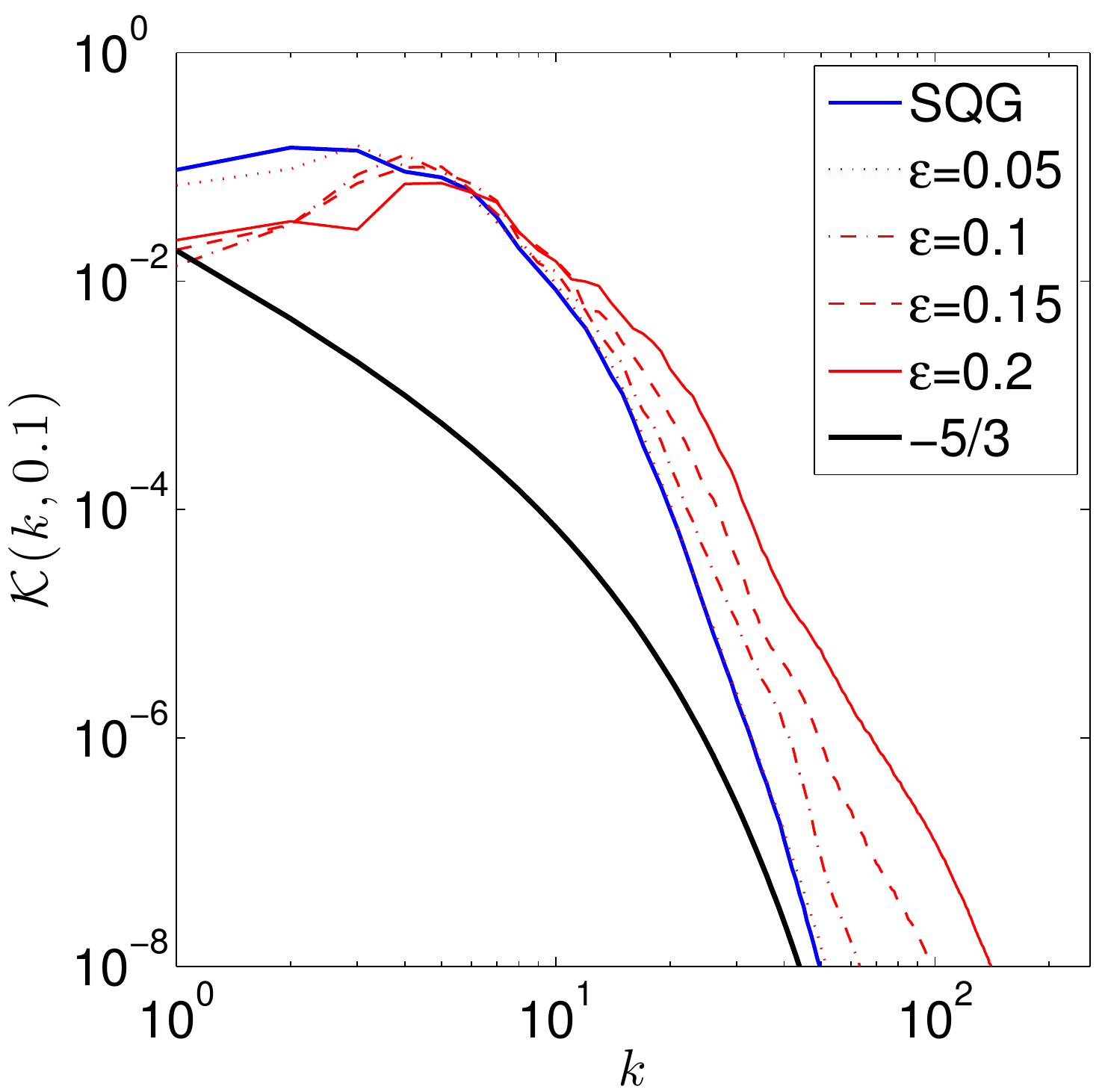}  
  \includegraphics[height=6.5cm,width=6.5cm]{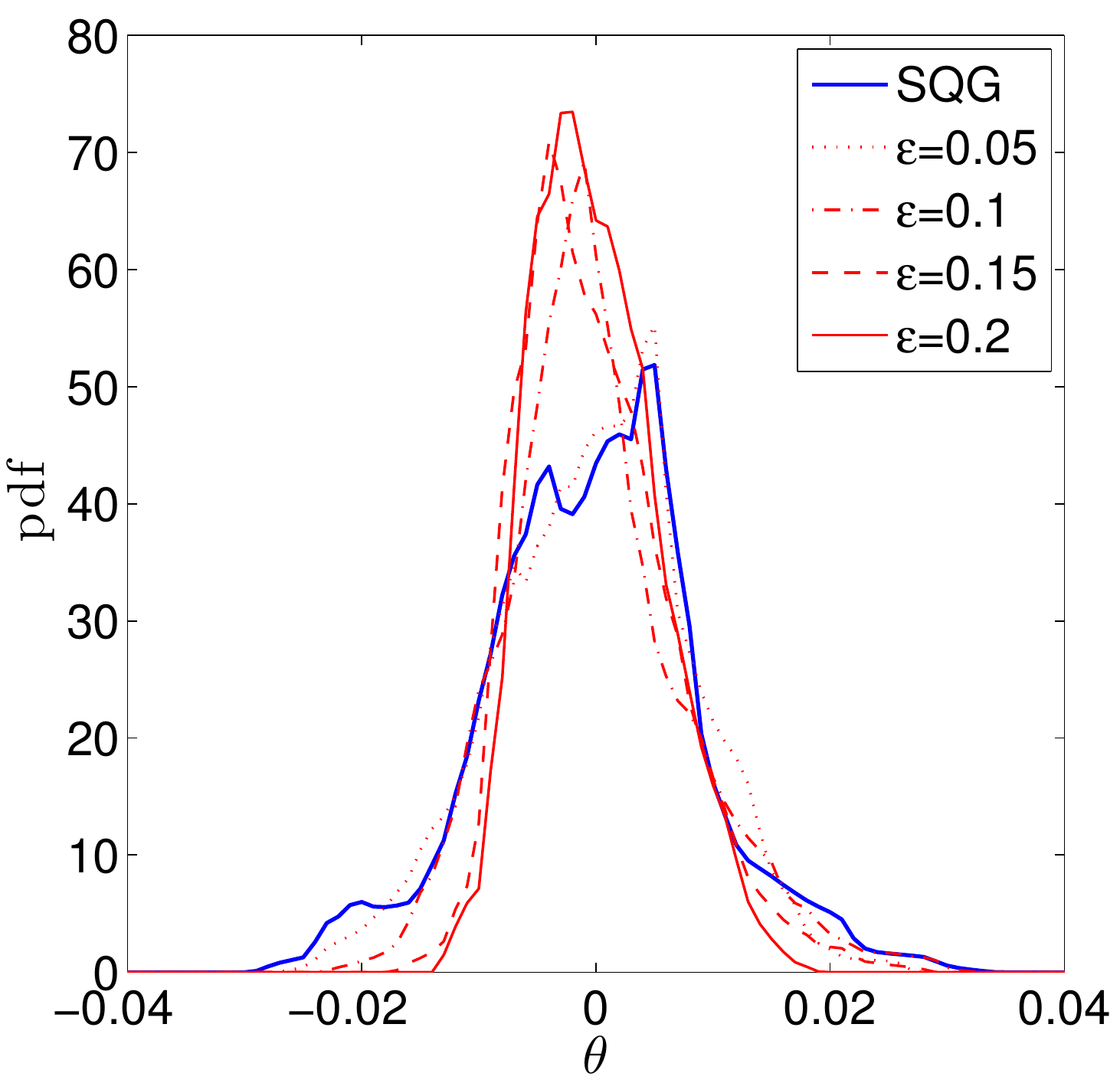}
  \includegraphics[height=6.5cm,width=6.5cm]{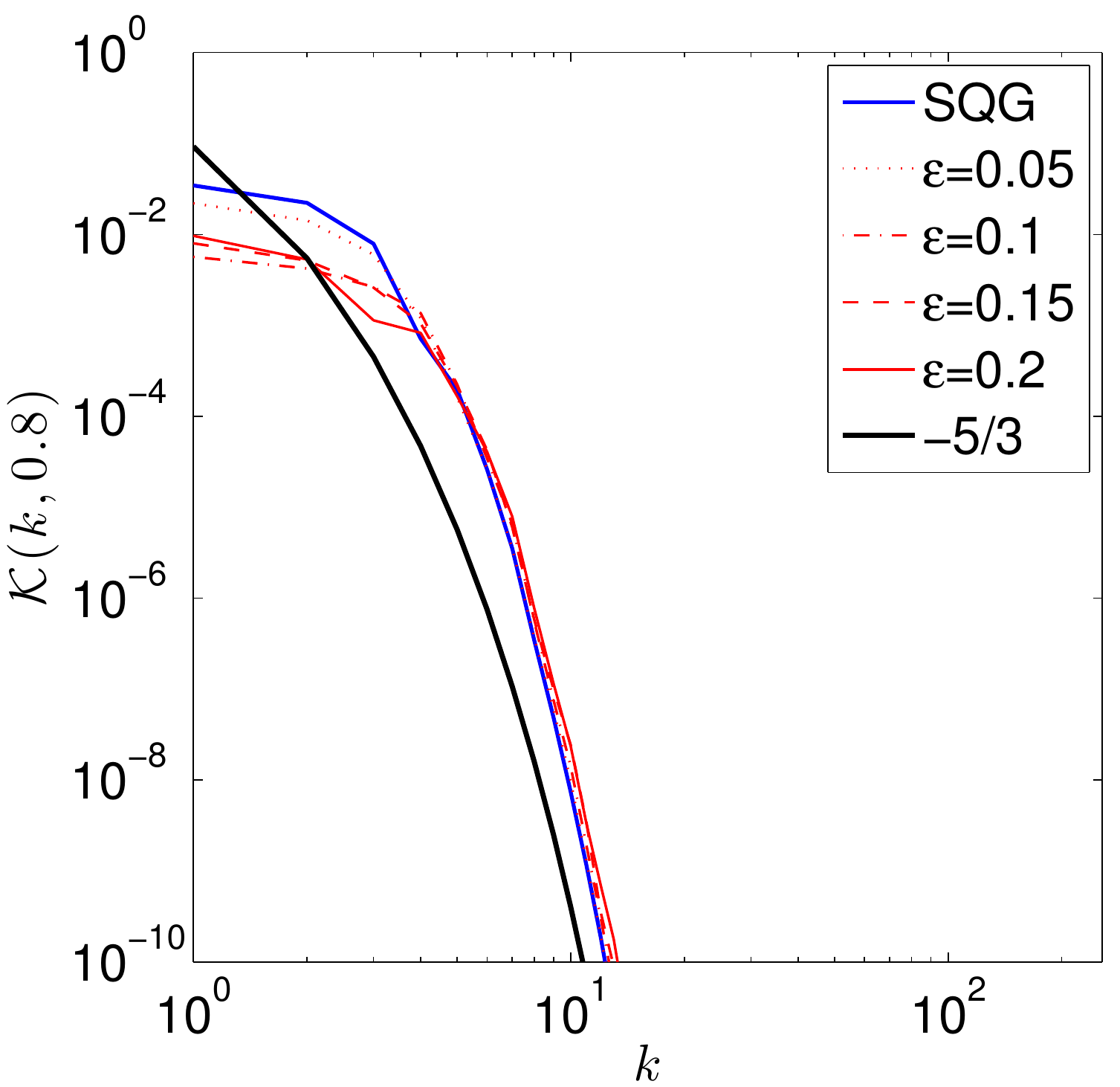} 
  \end{center}   
  \caption{(Left: PDFs of $\theta$ for SQG (blue lines) and SSG in physical coordinates (red lines) for different values of Rossby number at $Z$=0.1 (top) and $Z$=0.8 (bottom). Right: energy spectra for SQG (blue lines) and SSG in physical coordinates (red lines) for different values of Rossby number at $Z$=0.1 (top) and $Z$=0.8 (bottom).The solid black line shows the expected behavior of SQG spectra at  the corresponding depth assuming a -5/3 slope of the spectrum at the surface.}
  \label{fig:statistics interior}
\end{figure}

\begin{figure}
  \begin{center}
  \includegraphics[height=6.5cm,width=6.5cm]{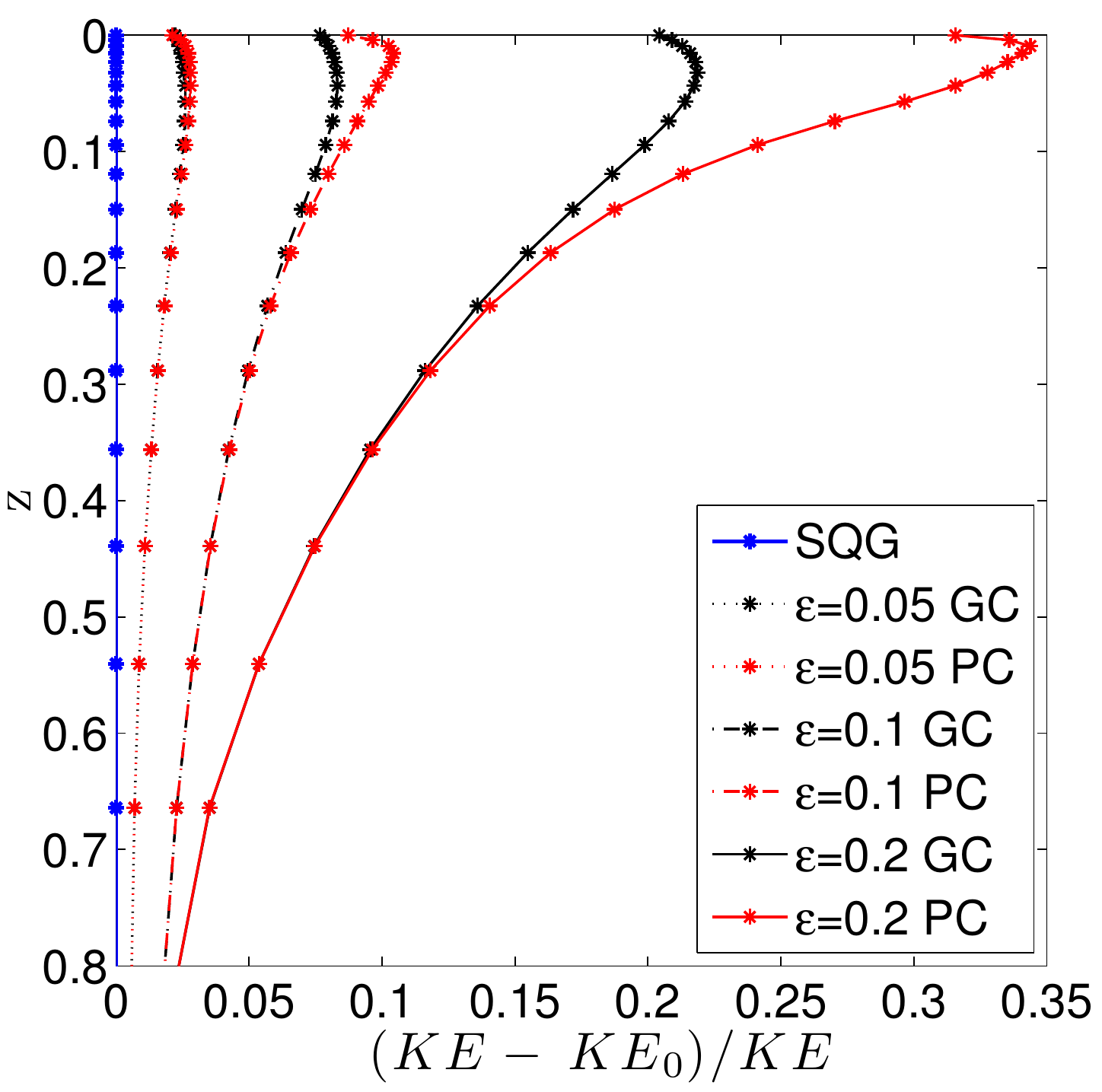}  
  \includegraphics[height=6.5cm,width=6.5cm]{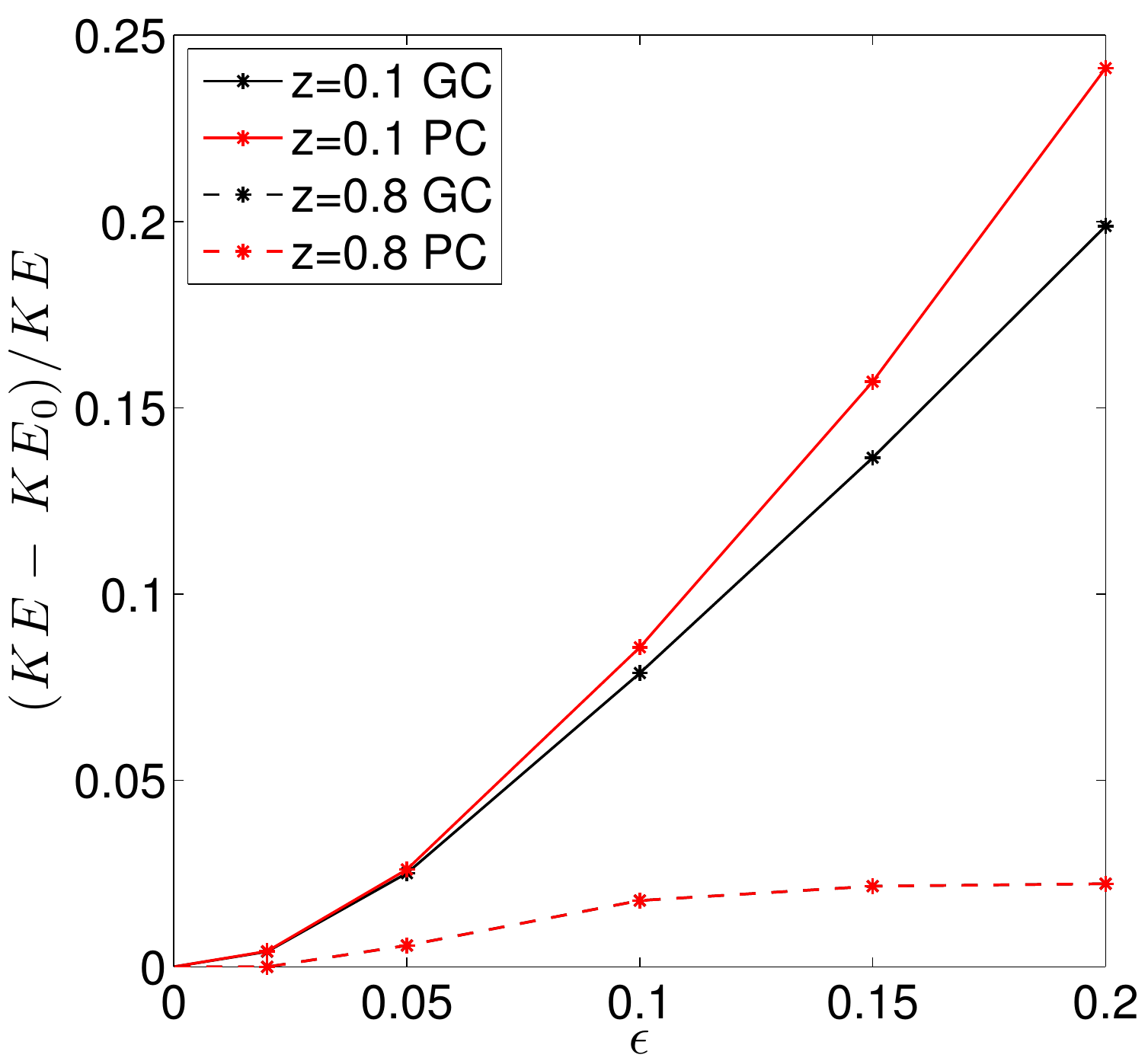}
  \end{center}   
  \caption{Left: vertical profile of fraction of horizontal kinetic energy due to the first-order correction to the SQG-like solution, in geostrophic (black lines) and  physical (red lines) coordinates. The blue line correspond to the SQG case. Right: fraction of horizontal kinetic energy due to the first-order correction to the SQG-like solution as function of $\epsilon$ for different values of $Z$, in geostrophic (black lines) and  physical (red lines) coordinates. Note that at $z=0.8$ the black and red lines are identical, as the coordinate transformation has almost no effect.}
\label{fig:KE fraction}
\end{figure}

\begin{figure}
  \begin{center}
  \includegraphics[height=6.5cm,width=6.5cm]{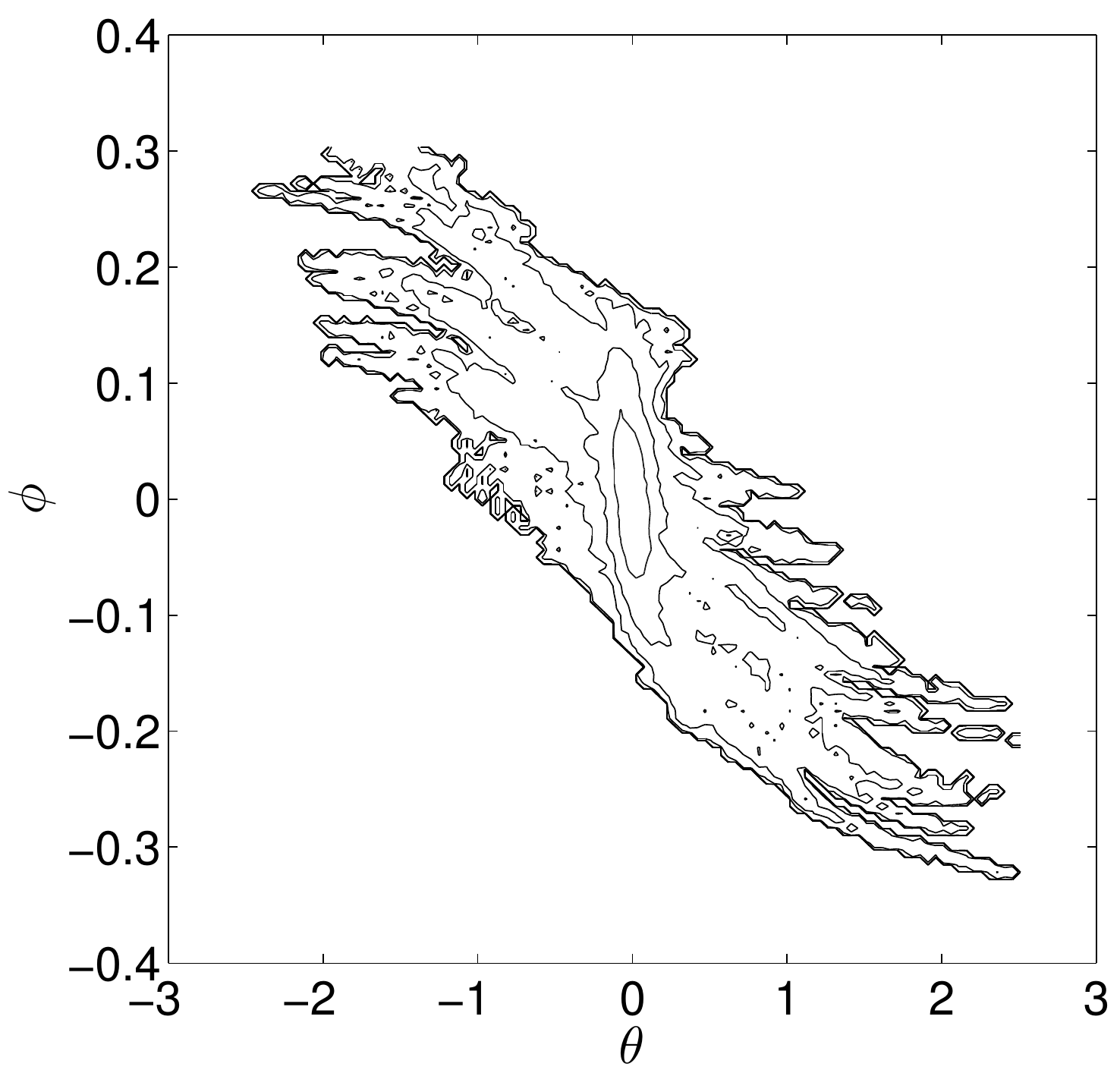}
  \includegraphics[height=6.5cm,width=6.5cm]{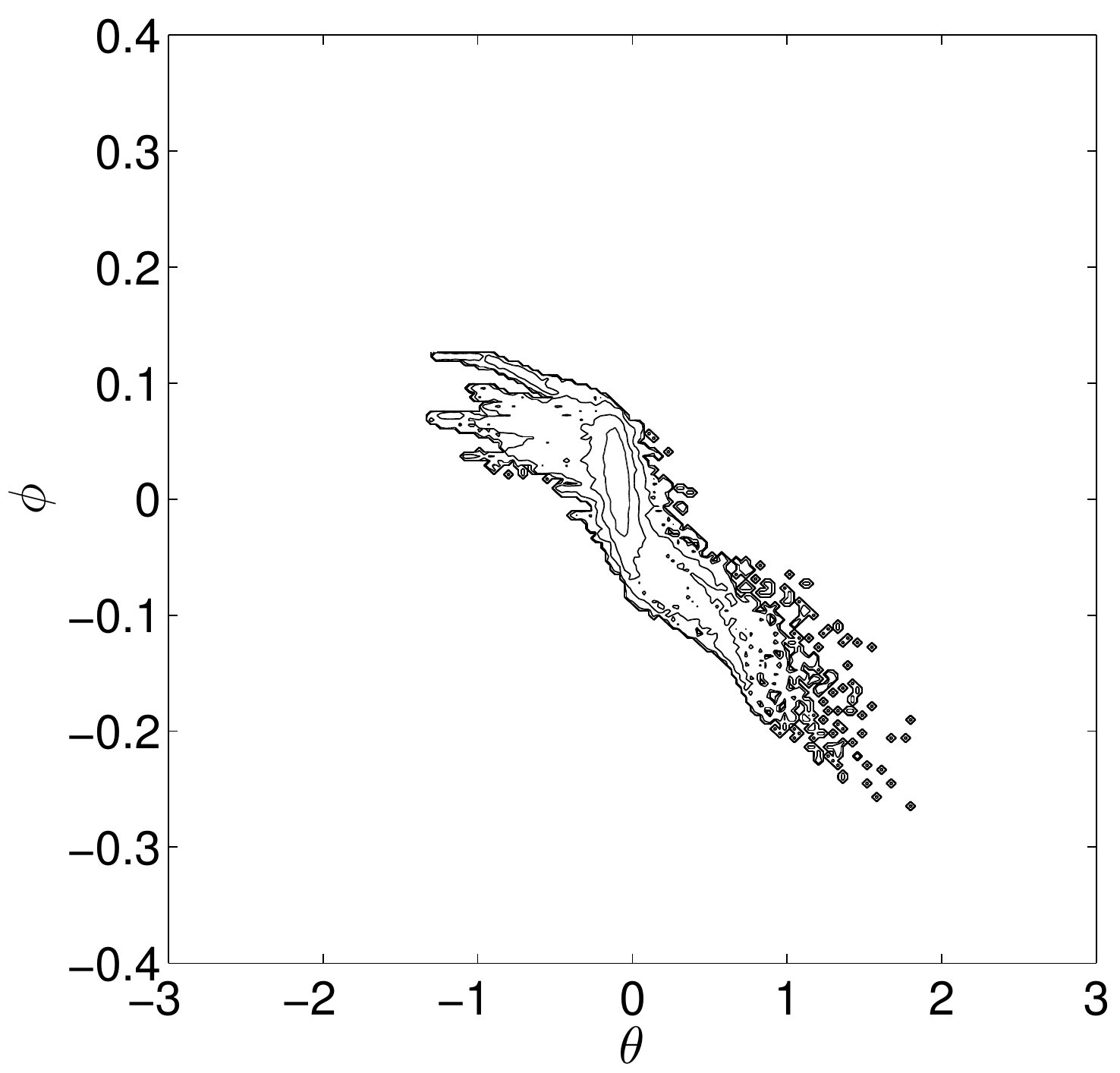}  
  \includegraphics[height=6.5cm,width=6.5cm]{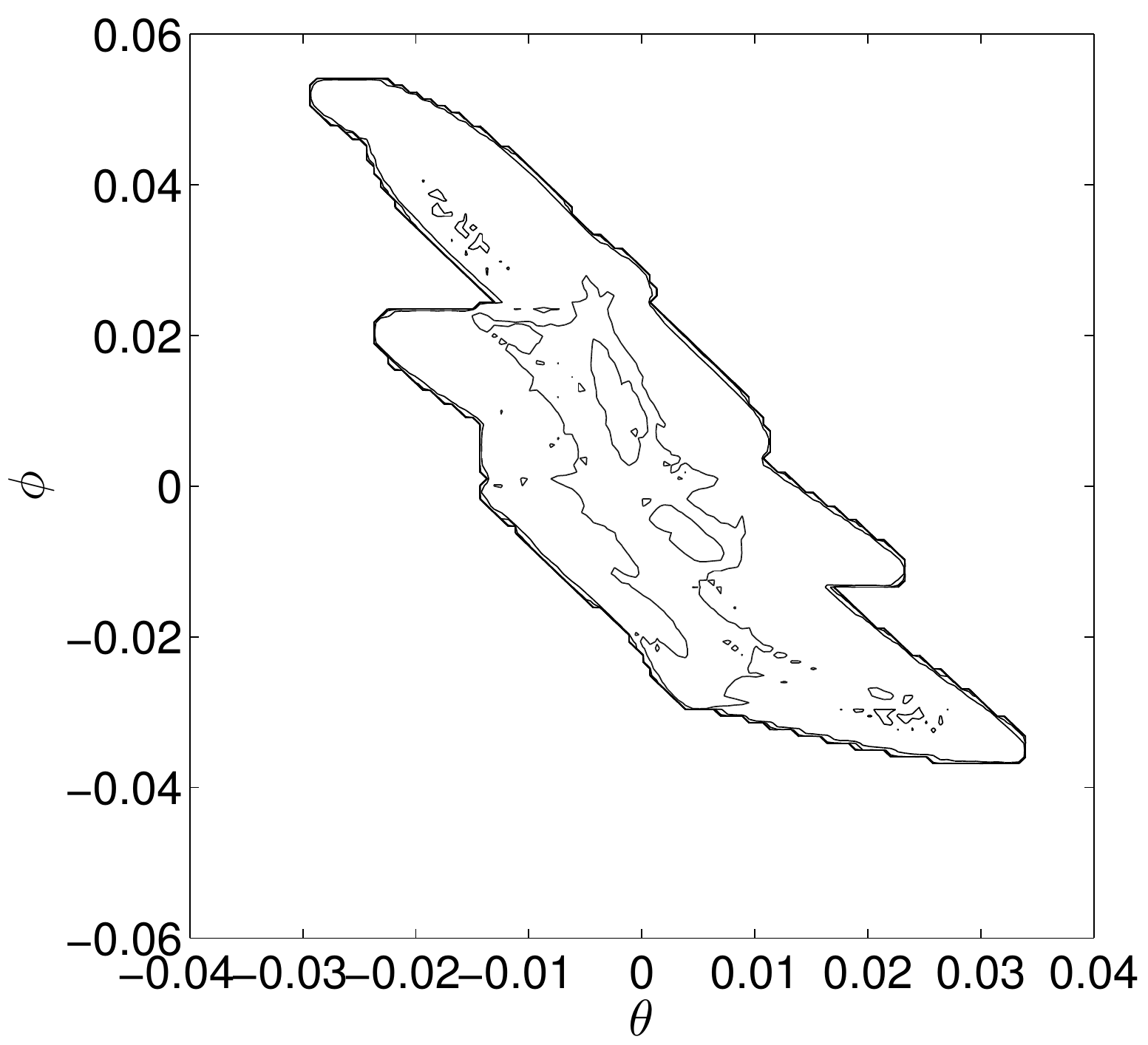}
  \includegraphics[height=6.5cm,width=6.5cm]{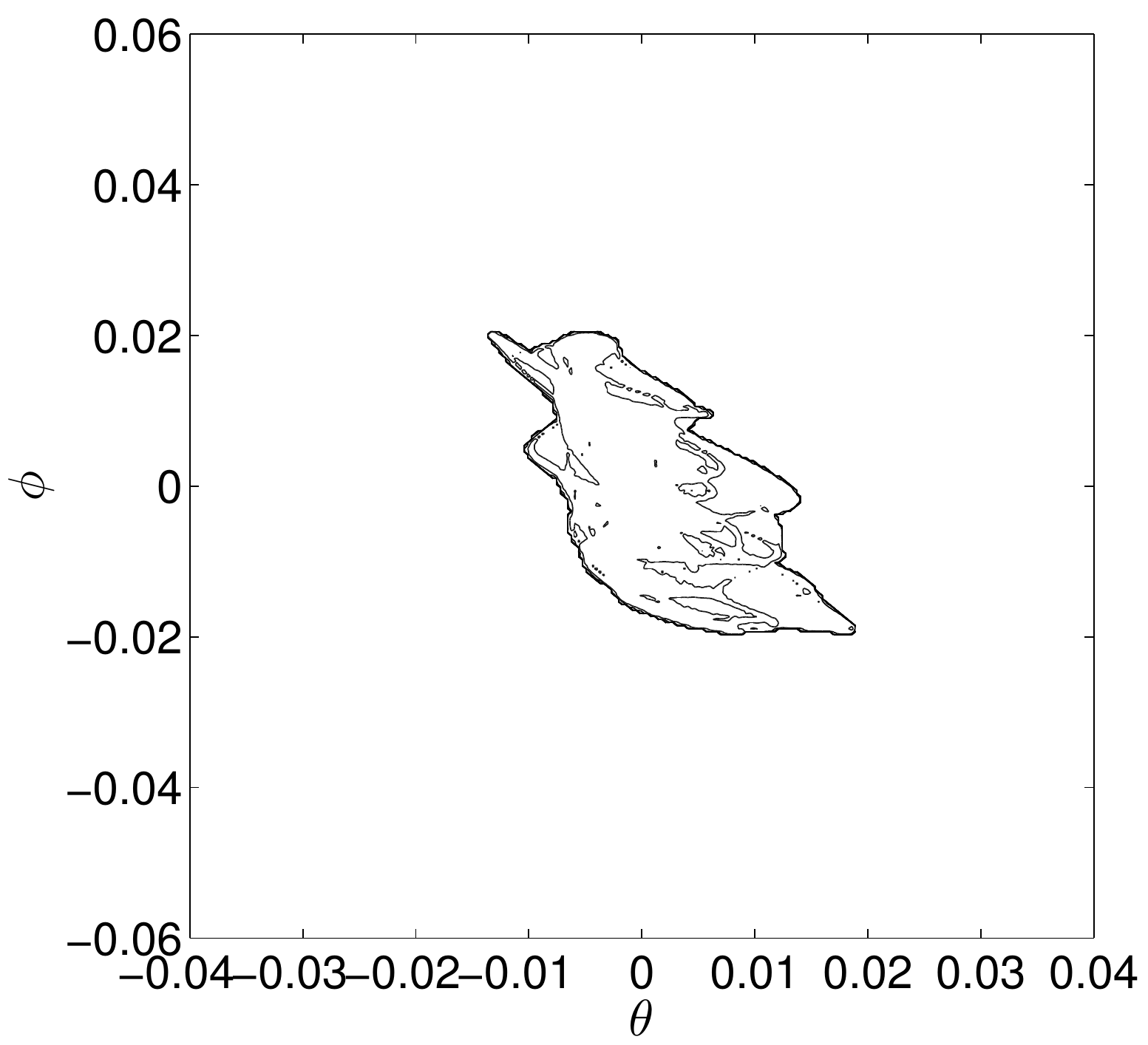}     
  \end{center}   
  \caption{Bidimensional PDFs of $\theta$ and $\phi$ for SQG (left panels) and SSG with $\epsilon=0.2$ (right panels), at surface (top row) and at Z=0.8 (bottom row) in physical coordinates. The contours are in logarithmic scale.}
\label{fig:2D pdf SQG vs SSG}
\end{figure}

\begin{figure}
  \begin{center}
  \includegraphics[height=6.5cm,width=6.5cm]{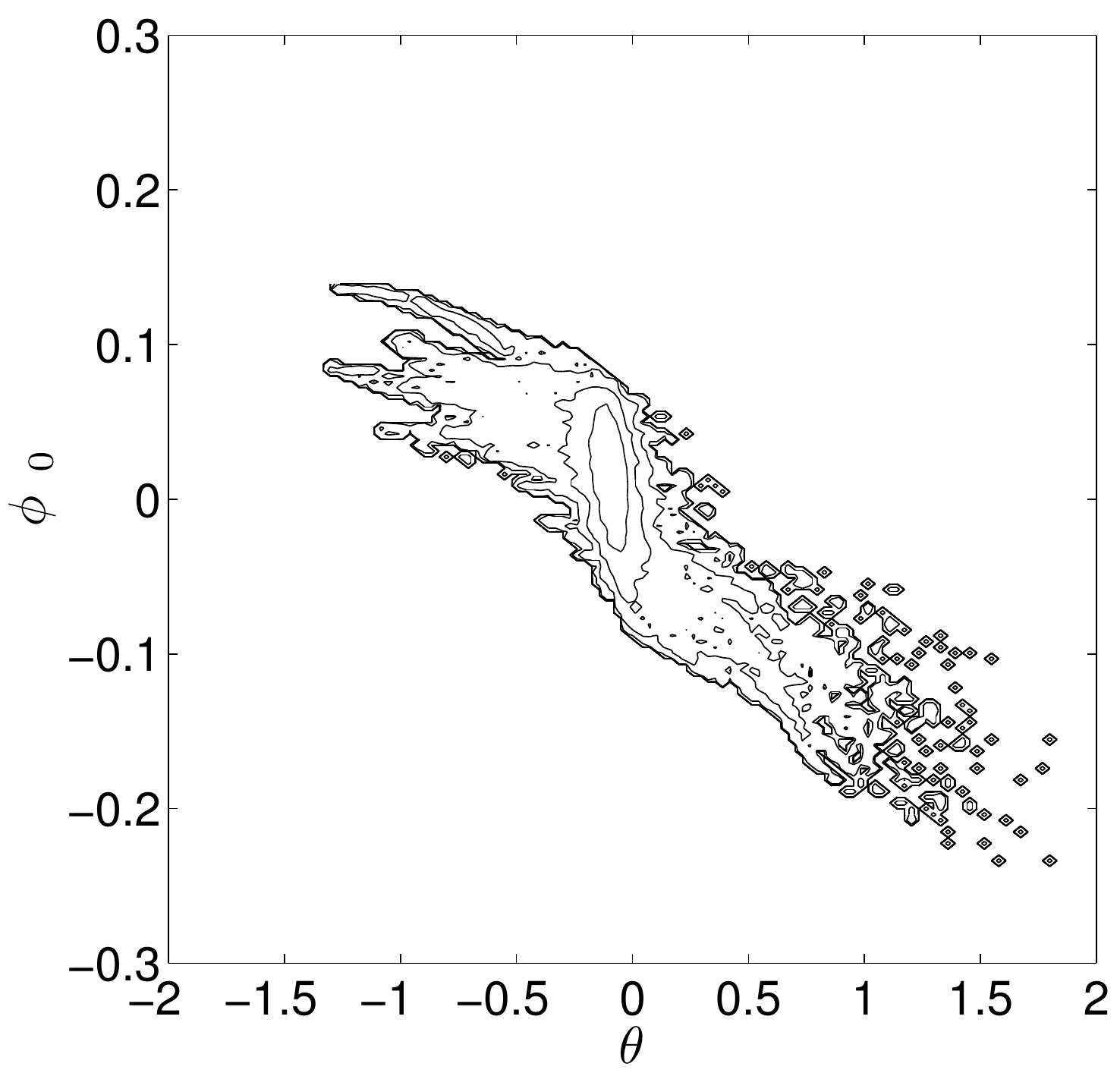}  
  \includegraphics[height=6.5cm,width=6.5cm]{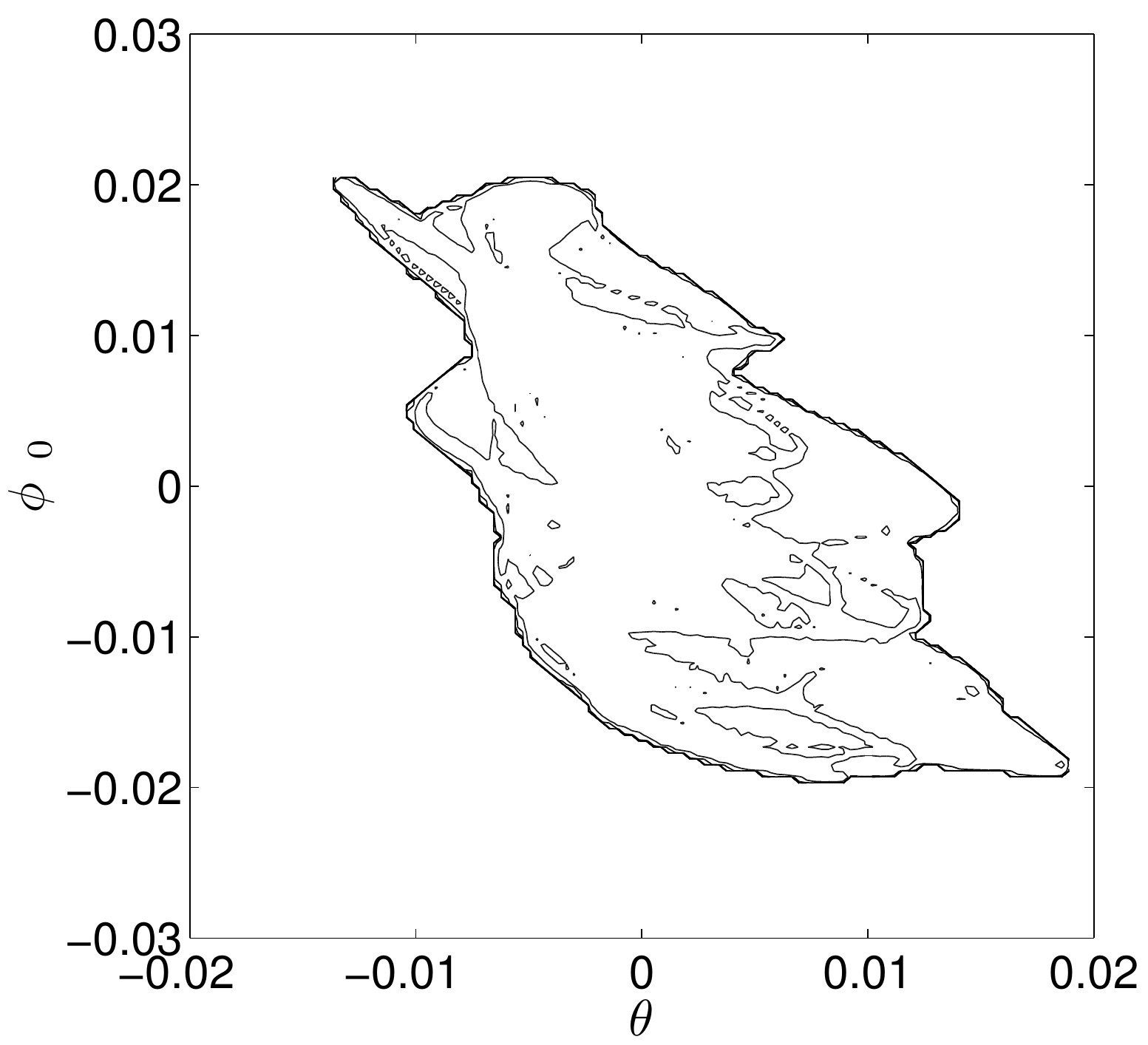}  
  \includegraphics[height=6.5cm,width=6.5cm]{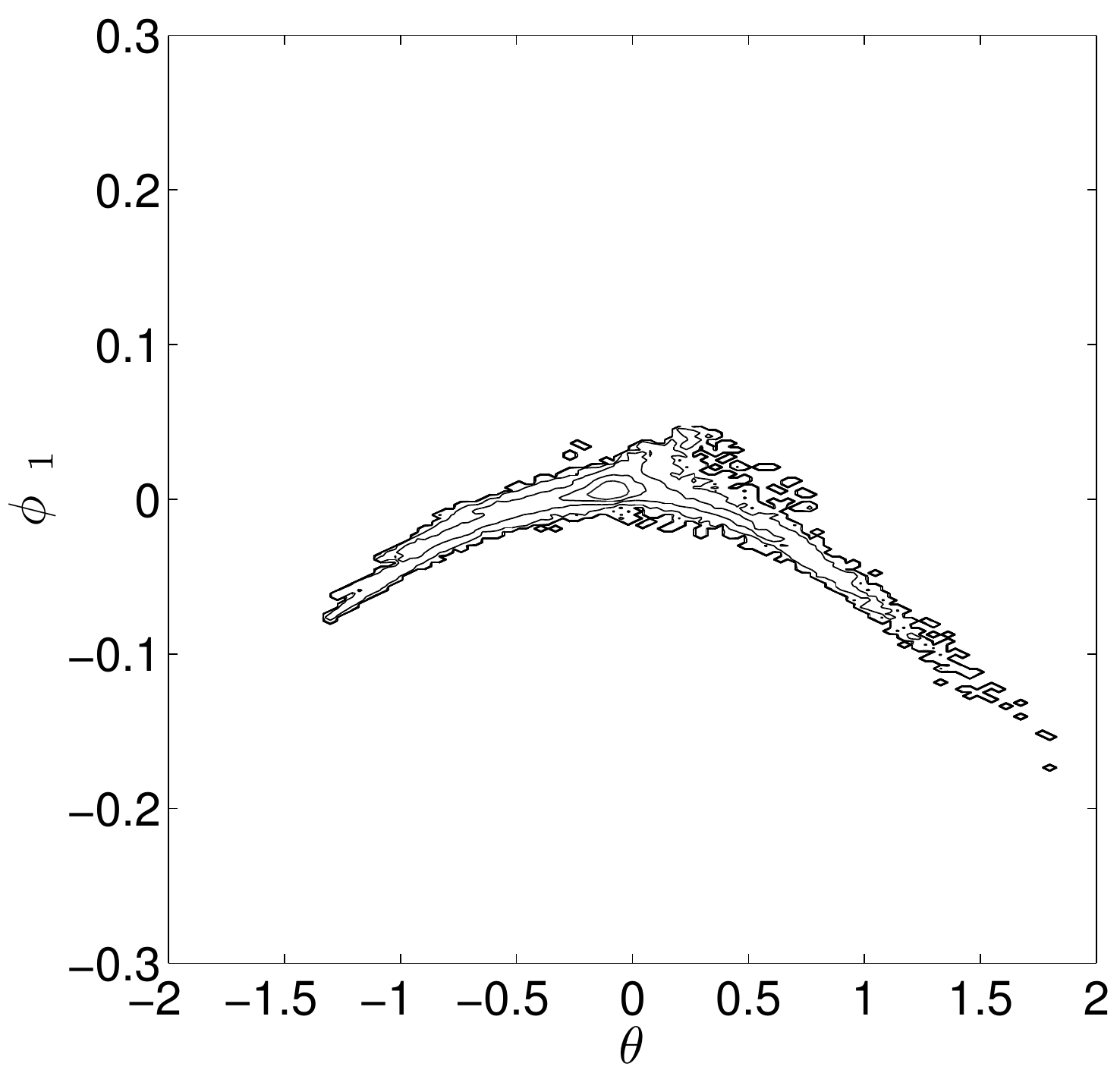}
  \includegraphics[height=6.5cm,width=6.5cm]{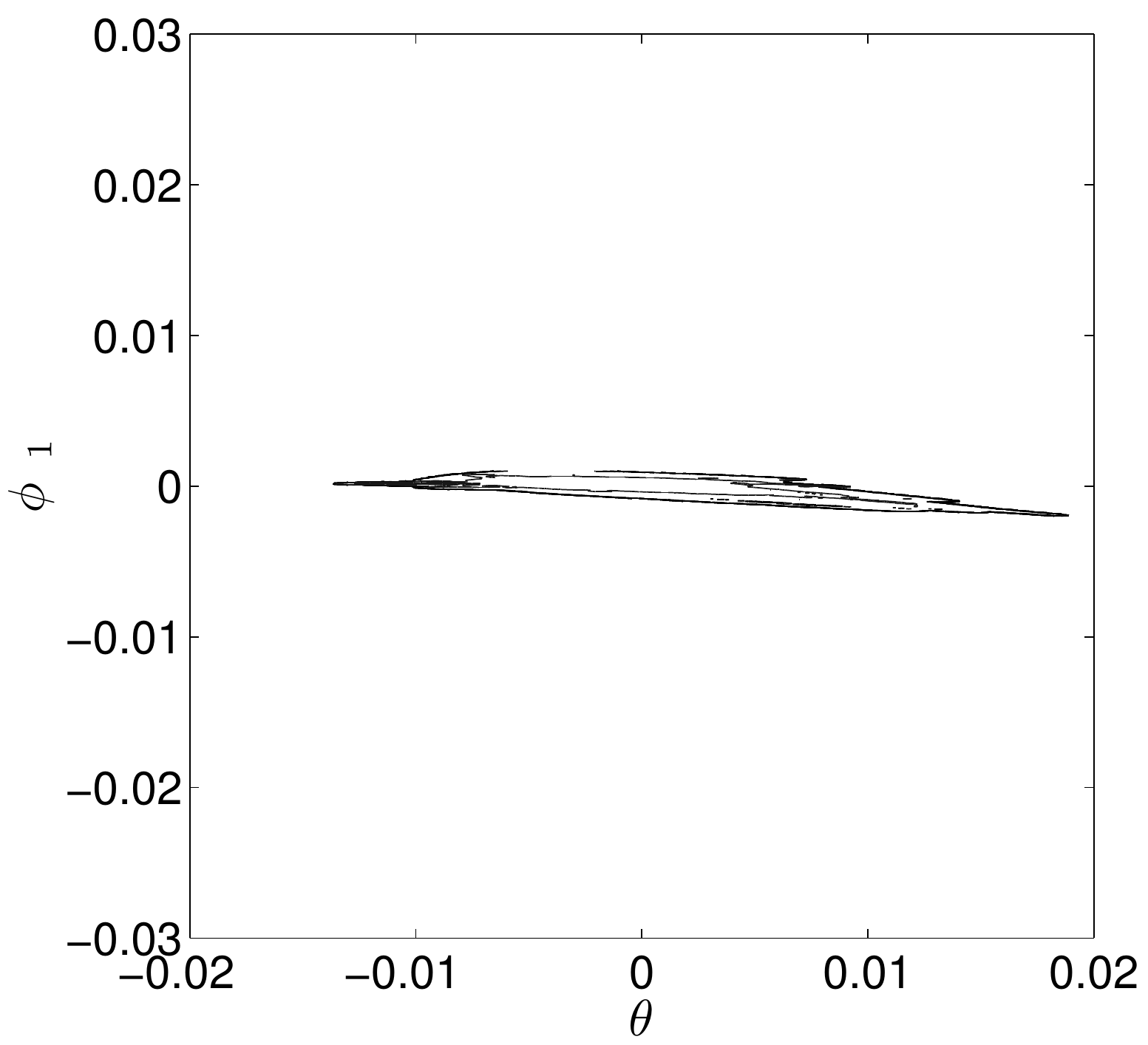}
  \includegraphics[height=6.5cm,width=6.5cm]{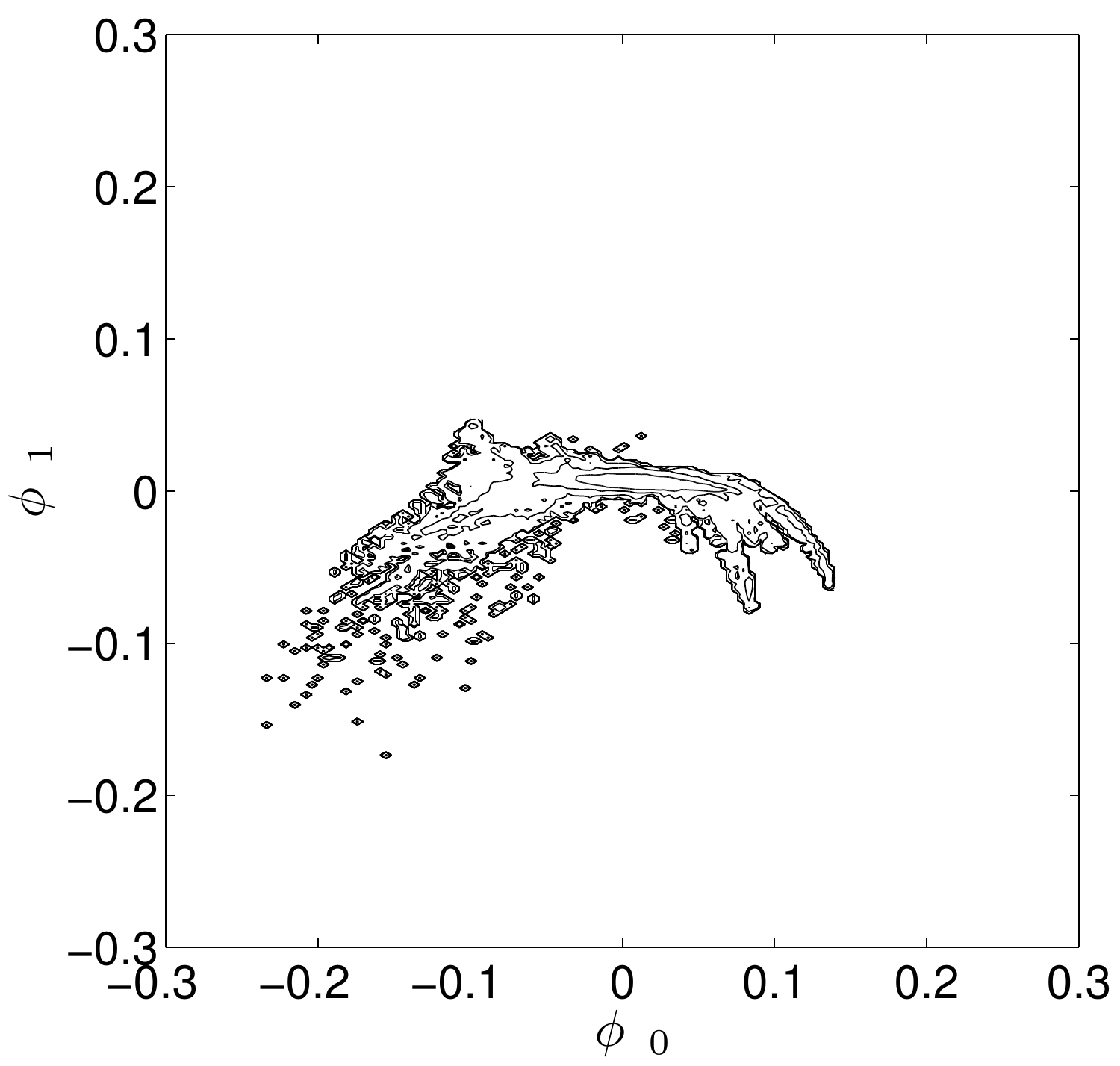}
  \includegraphics[height=6.5cm,width=6.5cm]{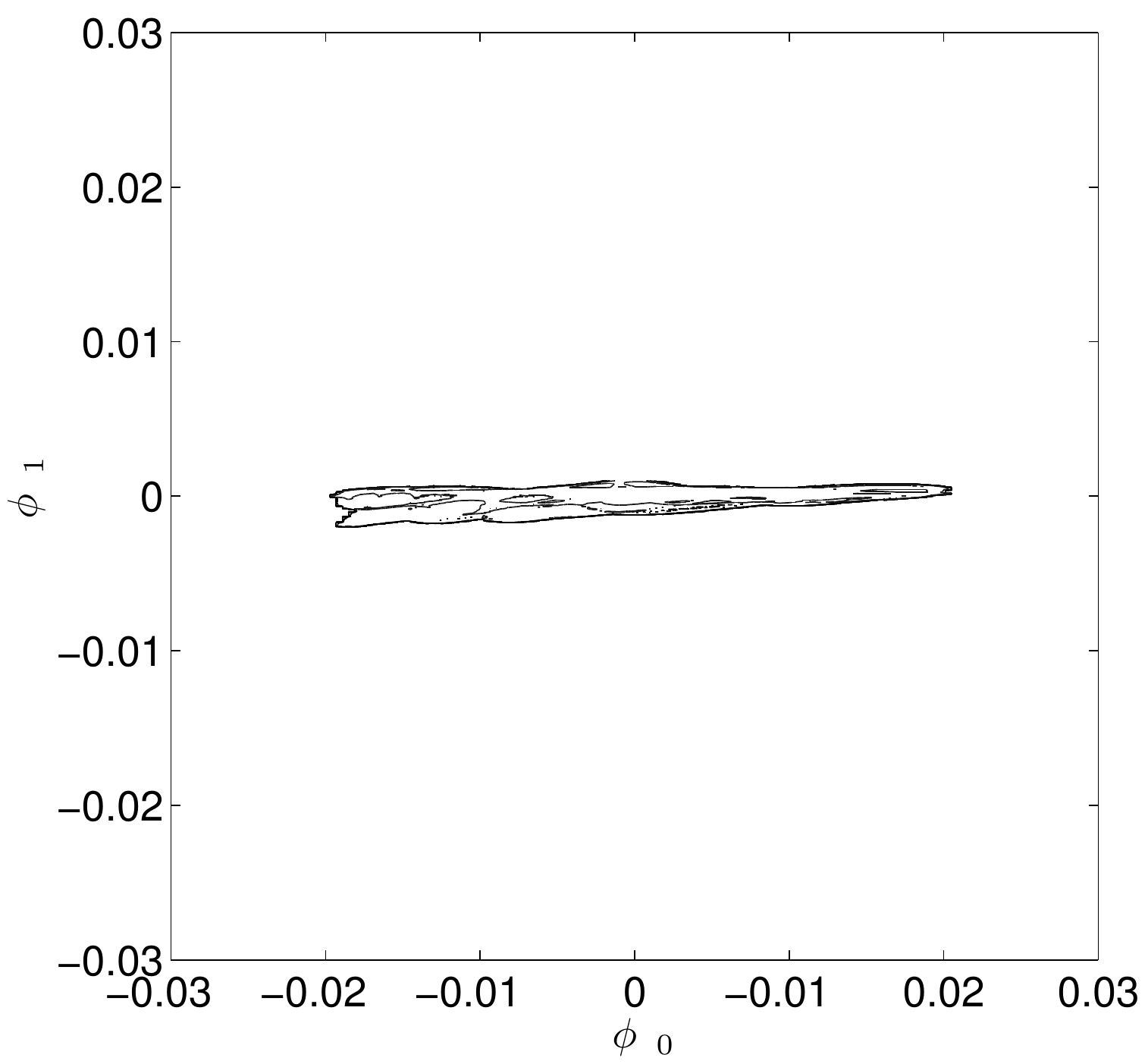}
  \end{center}   
  \caption{Top: bidimensional PDFs of $\theta$ and $\phi_0$ for SSG with $\epsilon=0.2$, at surface (left panel) and at Z=0.8 (right panel) in physical coordinates. Middle: bidimensional PDFs of $\theta$ and $\phi_1$ for SSG with $\epsilon=0.2$, at surface (left panel) and at Z=0.8 (right panel) in physical coordinates. Bottom: bidimensional PDFs of $\phi_0$ and $\phi_1$ for SSG with $\epsilon=0.2$, at surface (left panel) and at Z=0.8 (right panel) in physical coordinates. The contours are in logarithmic scale.}
\label{fig:2D pdf SSG}
\end{figure}

The analysis of the interior statistic is of particular interest, as one of the reasons of appeal of SQG-like models is the possibility to infer properties of the ocean at depth from observations of surface fields \citep{LaCasce&Mahadevan2006,Lapeyre&Klein2006,Wang2013,liuetal14}. Although rigorously SSG can not be applied to represent quantitatively processes occurring at scales smaller than the Rossby radius of deformation, the ability of reproducing the formation of frontal structures and filaments goes in the right direction to represent sub-mesoscale dynamics that are not captured by the SQG approximation. Indeed, SSG in a two dimensional vertical plane has proved to be successful in correcting the underestimation of buoyancy anomalies at depth that characterize the SQG interior profiles \citep{Badin2013}, even if the effects observed by \citet{Badin2013} are due to $\epsilon \rightarrow 1$, while this study is restricted to small values of $\epsilon$. Moreover, the SG approximation allows for $O(1)$ variations of static stability, that are physically important in determining the properties in the interior of the ocean from the surface mixed layer.

 Figure \ref{fig:moments} shows the vertical profiles of median (top-left panel), mean (top-right panel), standard deviation (bottom-left panel) and skewness (bottom-right panel) of $\theta$. The black lines refer to geostrophic coordinates, while the red lines refer to physical coordinates. Results show that the asymmetry of the PDFs strongly weakens at depth. The effect of the coordinate transformation, visible in the differences between the red and black lines, disappears at depth earlier than the effect of the nonlinearity of the Monge-Amp\'{e}re equation, visible in the difference between the black and blue lines.  From equation \eqref{eq:solution first order}, small scales (i.e. large $k$) decay faster with depth at leading order. Therefore, fronts and filaments tend to disappear at depth, restoring the symmetry of the distributions. This is in agreement with the physical interpretation on the role of enhanced ageostrophic, small-scale frontal structures in determining the differences between SSG and SQG.

Figure \ref{fig:statistics interior} shows the PDFs of $\theta$ (left panels) and the kinetic energy spectra (right panels) at $Z$= 0.1 (top row) and $Z$= 0.8 (bottom row) for SQG (blue lines) and SSG  in physical coordinates for different values of the Rossby number (red lines). In the interior of the domain, but still close to the surface ($Z$= 0.1), SSG produces PDFs of $\theta$ that are highly skewed and again peaked at negative values, while at depth ($Z$ = 0.8) the PDFs appear with zero mean and nearly symmetric even for high Rossby number. 

From Equation \eqref{eq:solution first order} the kinetic energy spectra in the interior $\mathcal{K}(k,Z)$ are linked at zeroth order to the kinetic energy spectrum at the surface $\mathcal{K}(k,0)$ by a multiplicative factor rapidly decaying for large $Z$ and large $k$ \citep{Callies&Ferrari2013}
\begin{equation}\label{eq:decay factor}
\displaystyle \mathcal{K}(k,Z) =  \mathcal{K}(k,0)\left(\frac{cosh(k(Z-1))}{cosh(k)}\right)^2,
\end{equation}
Superimposed on the spectra of Figure \ref{fig:statistics interior} is the SQG-like zeroth order expected behavior following \eqref{eq:decay factor}, and assuming a -5/3 spectral slope at surface. Results for SSG at $Z=0.1$ show higher energies at smaller scales as the Rossby number increases, with a slope approaching the -5/3 spectral slope at surface, corrected with depth. At $Z=0.8$, the SQG and SSG spectra converge to the same solution, for all values of the Rossby number. Figure \ref{fig:statistics interior} thus further confirms  that the deviation of SSG from SQG disappears with depth. Overall, it seems that SSG is successful in capturing at a qualitative level some of the distinctive properties of the statistics of sub-mesoscale processes observed in the ocean.

In classic SQG, kinetic energy and density variance spectra $\mathcal{K}(k,Z)$ and $\mathcal{T}(k,Z)$ are identical \citep{Blumen1978,Held&al1995}. In finite-depth SQG \citep{Tulloch&Smith2006} $\mathcal{K}(k,Z)$ and $\mathcal{T}(k,Z)$ are equal at small scales, with a transition occurring at scales larger than a critical wavenumber from a classical SQG-like to a QG/2D-like behavior. In our case, as discussed above, only the -5/3 regime is properly represented. In finite-depth SSG things differ due to the presence of the nonlinear term and of the coordinate transformation. From \eqref{eq:solution first order} we have
\begin{equation}\label{eq:KE components}
\displaystyle \mathcal{K}(k,Z) =  k^2 \left[ | \hat{\Phi}_0 |^2 + \epsilon \left( \hat{\Phi}_0\hat{\Phi}_1^* + \hat{\Phi}_0^* \hat{\Phi}_1 \right) + \epsilon^2 | \hat{\Phi}_1 |^2 \right] = \mathcal{K}_0(k,Z) + \epsilon \mathcal{K}_{01}(k,Z) +\epsilon^2\mathcal{K}_1(k,Z),
\end{equation}
Additional contributions to the SQG-like energy spectrum $\mathcal{K}_0$  appear due to the presence of the order-$\epsilon$ part of the SSG solution in \eqref{eq:solution first order}. We have verified that the spectra of the different components of \eqref{eq:KE components} all follow the same slope in physical coordinates (not shown), so that the change of the slope with $\epsilon$ is not due to a a flatter slope of the part of the spectrum due to the deviations from the SQG-like solution. The left panel of Figure \ref{fig:KE fraction} shows the vertical profile of the fraction of kinetic energy connected to the additional terms $\epsilon \mathcal{K}_{01} +\epsilon^2\mathcal{K}_1$ for different values of the Rossby number in geostrophic (black lines) and physical (red lines) coordinates. Again, the relative importance of the SSG first-order correction term weakens at depth, and the contribution of the coordinate transformation is negligible for $Z>0.2$. The maximum of the effect is however reached not at the surface but at a intermediate depth between $Z=0$ and $Z=0.1$. The right panel of Figure \ref{fig:KE fraction} shows how the fraction of kinetic energy connected to the additional terms $\epsilon \mathcal{K}_{01} +\epsilon^2\mathcal{K}_1$ increases with $\epsilon$ for $Z=0.1$ (solid lines) and $Z=0.8$ (dashed lines), in geostrophic (black) and physical (red) coordinates. At $Z=0.1$ the increase with $\epsilon$ is clearly nonlinear, however the values remain always on the order of $\epsilon$, thus confirming the robustness of our partition of the solution for small $\epsilon$ in a zeroth order part and a first order correction term. At $Z=0.8$ the increase is instead much slower, due to the minor importance of ageostrophic processes at depth, and the coordinate transformation does not introduce any contribution.

This has some interesting implications regarding the computation of the flow in the interior of the oceanic mixed layer from the knowledge of the surface data, one of the applications of SQG. Let us suppose that we observe a current and buoyancy field at the surface. We can deduce an estimate of the currents and buoyancy in the interior of the domain by using the SQG inversion. This is known to lead to underestimate the reconstructed fields. If we use instead the SSG inversion (equation (\ref{eq:solution first order}) at first order), the estimate of the streamfunction consists in the zeroth-order term, that is the same as if we had used the SQG inversion, plus an additional contribution. Figure \ref{fig:KE fraction}  shows that this additional contribution leads to a positive correction to the KE energy of the estimated current field corresponding to the reconstructed streamfunction. The additional KE predicted by the SSG inversion is up to $20-30\%$ in the upper layers and decays with depth, therefore it is connected with filaments and other small scale structures not captured by SQG. This could qualitatively explain why the SQG inversion underestimates the fields at depth when applied to observations at the surface. 

\subsection{$\phi - \theta$ relationship} 

It is interesting to see the relation between $\theta$ and the different parts of the solution $\Phi_1$. Figure \ref{fig:2D pdf SQG vs SSG} shows the bidimensional PDFs of potential temperature and streamfunction for SQG (left panels) and SSG with $\epsilon=0.2$ in physical coordinates (right panels), at surface (top row) and at $Z=0.8$ (bottom row). The contours are in logarithmic scale. The asymmetry at surface and the different properties highlighted above are well visible also in these plots. The PDFs of SQG are symmetric both in $\theta$ and $\phi$, while in SSG the cyclonic tails, characterized by large positive values of $\theta$ associated to large negative values of $\phi$, are longer than the anticyclonic tails, characterized by large negative values of $\theta$ associated to large positive values of $\phi$. Still, negative values of $\theta$ dominate in the mean, due to the strong asymmetries of the distributions. An additional feature visible in Figure \ref{fig:2D pdf SQG vs SSG} is the clear signature of coherent structures (vortices) in the PDFs at the surface, showing up as "fingers" in the tails of the distributions, as coherent structures are characterized by a functional relation between streamfunction and advected scalar $\theta=\theta(\phi)$. While in SQG cyclonic and anticyclonic coherent structures have the same properties, in SSG a clear asymmetry emerges, due to the different deformation of strong cyclonic and anticyclonic areas induced by the coordinate transformation that changes the morphology of the corresponding coherent structures. 

Figure \ref{fig:2D pdf SSG} shows for the case $\epsilon=0.2$ the relation between $\theta$ and the two parts of the SSG solution for the streamfunction $\phi_0$ (top row) and $\phi_1$ (middle row). At the surface (left panels), while $\phi_0$ behaves basically like $\phi$, $\phi_1$ has a very clear functional form. The values of $\phi_1$ are almost always negative, and larger in magnitude for positive $\theta$. To better understand the role of $\phi_1$ in correcting the SQG-like solution, the bottom row of Figure \ref{fig:2D pdf SSG} shows the bidimensional PDFs of $\phi_0$ and $\phi_1$ at surface (left panel) and at $Z=0.8$ (right panel). We can see that at surface the effect of  $\phi_1$ is to strongly enhance  regions of negative $\phi_0$ and weakly damp regions of positive $\phi_0$. It is then clear how this additional term systematically modifies the streamfunction, and therefore the velocity field, resulting in the observed asymmetric distributions of $\theta$ and $\phi$. Consistently with what seen previously, at $Z=0.8$ (right panels) the correction term $\phi_1$ is totally negligible with respect to $\phi_0$, as it is associated with small scale structures that do not penetrate at depth, and SQG-like symmetry is restored.

\begin{figure}
  \begin{center}
  \includegraphics[height=6.5cm,width=6.5cm]{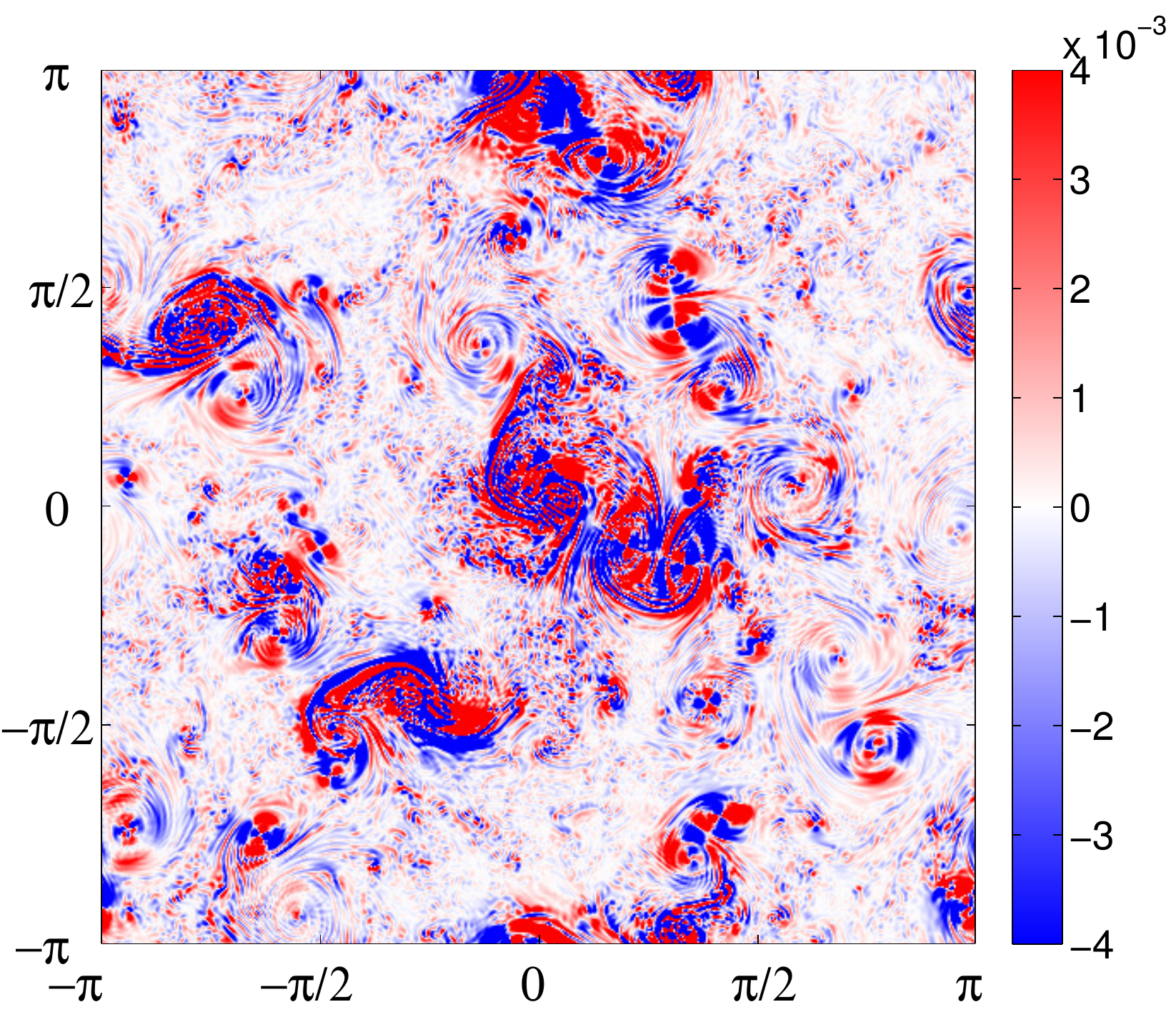}    
  \includegraphics[height=6.5cm,width=6.5cm]{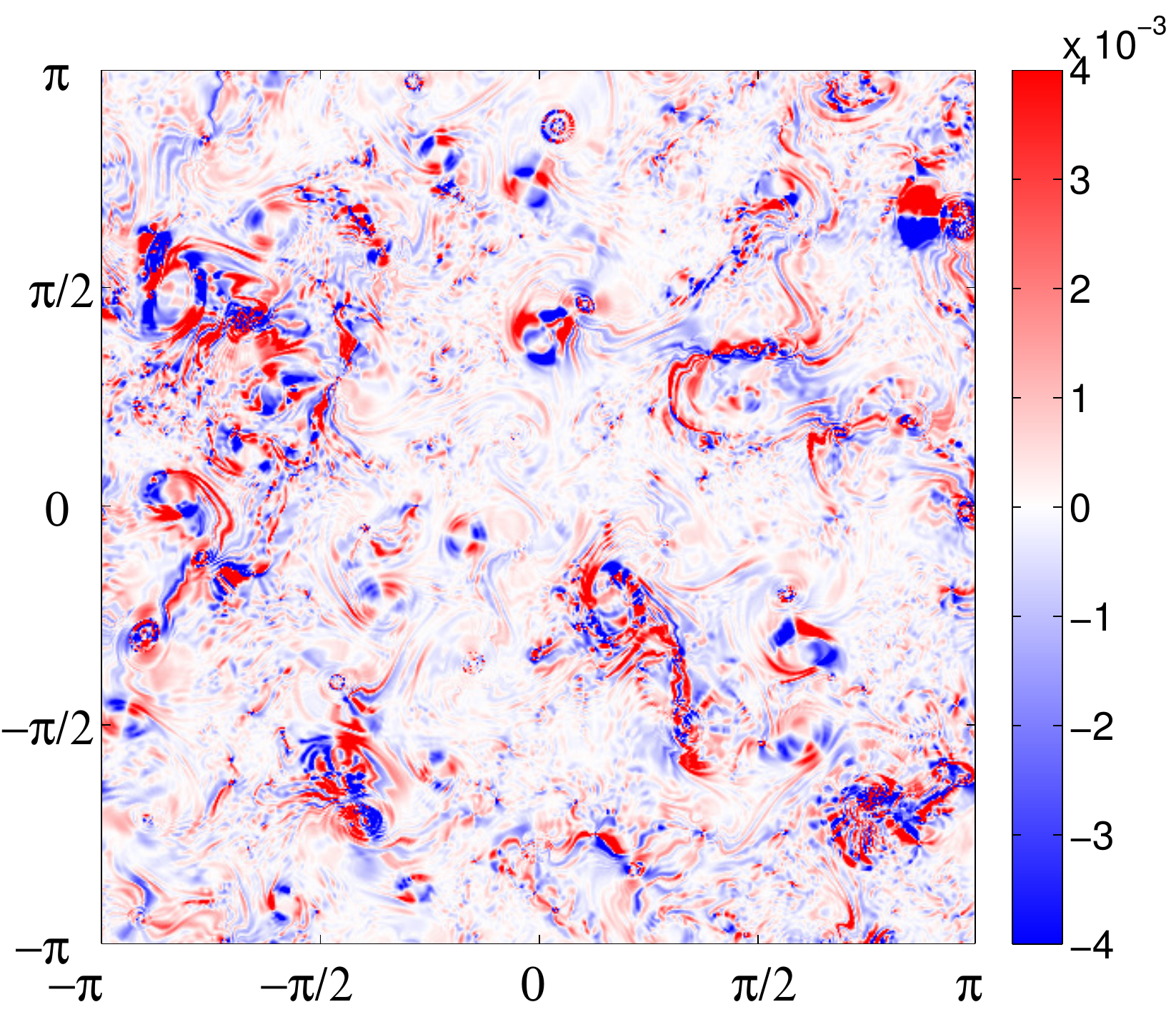}           
  \end{center}   
  \caption{Snapshot at time T=100 of vertical velocity at z=0.015 for SQG (left panel) and SSG in physical coordinates (right panel), with $\epsilon=0.2$}
\label{fig:vertical snapshot}
\end{figure}

\subsection{Vertical velocity, ageostrophic divergence and lateral strain rate}

Following \citet{Hoskins&Draghici1977}, the diagnostic equation for the vertical velocity in nondimensional variables can be written as
\begin{equation}\label{eq: omega equation}
\displaystyle \frac{\p^2 w^*}{\p X^2}+\frac{\p^2 w^*}{\p Y^2}+\frac{\p^2 w^*}{\p Z^2} = -2 \epsilon \nabla \cdot \mathbf{Q} ,\\
\end{equation}
with boundary condition $w^*=0$ at $Z=0$ and $Z=1$, where $ w=   J w^*$, and the forcing vector $Q$ in nondimensional form is
\begin{equation}\label{eq: Q vector}
\displaystyle \mathbf{Q} = \left(Q_1,Q_2\right) = \left( \frac{\p u_g}{\p X} \frac{\p \theta}{\p X} + \frac{\p v_g}{\p X} \frac{\p \theta}{\p Y}, \frac{\p u_g}{\p Y} \frac{\p \theta}{\p X} + \frac{\p v_g}{\p Y} \frac{\p \theta}{\p Y} \right).
\end{equation}

Note that SG equation for the vertical velocity in geostrophic coordinates is formally identical to the equation one derives in physical coordinates in the QG case, with two differences: the use of geostrophic coordinates and the fact that the equation is formulated for a vertical velocity $w^*$ rescaled by $  J$. 

Note also that in the derivation of \eqref{eq: omega equation}, $J$ is approximated neglecting the nonlinear terms \citep{Hoskins&Draghici1977, Hoskins&Draghici&Davies1978}. Equation \eqref{eq: omega equation} is not therefore entirely compatible with the full form of the SSG equations that we are here investigating. In order to be consistent with the derivation of equation \eqref{eq: omega equation}, we have computed $w$ using the approximated version of $J$, although there is no substantial qualitative difference between the results that we would have obtained using the full form of $J$. For further studies of semi-geostrophic vertical velocities, see e.g. \citet{Hoskins&Draghici&Davies1978,Hoskins&West1979,Pinot&al96,Pedder&Thorpe1999,Thorpe&Pedder1999,Viudez&Dritschel04}.

Figure \ref{fig:vertical snapshot} shows snapshots at time $T=100$ of the vertical velocity $w$ at $Z=0.015$ for SQG (left panel) and SSG with $\epsilon=0.2$ (right panel) in physical coordinates. Comparing Figure \ref{fig:vertical snapshot} with Figure \ref{fig:snapshots surface}, we can see that while in SQG vertical velocity is larger inside the vortices with a typical quadrupole structure, in SSG the largest values are obtained most often on the edges of the vortices and of the frontal structures, as captured in primitive equations simulations and that can be significant for the transfer of nutrients to the surface euphotic layer \citep{Levy&al2001,Mahadevan&Tandon2006}. This is due to the fact that $w$ is given by the product of $w^*$ and $J$, and the latter is largest exactly on the edges of vortices and of frontal structures.

\begin{figure}
  \begin{center}  
   \includegraphics[height=6.5cm,width=6.5cm]{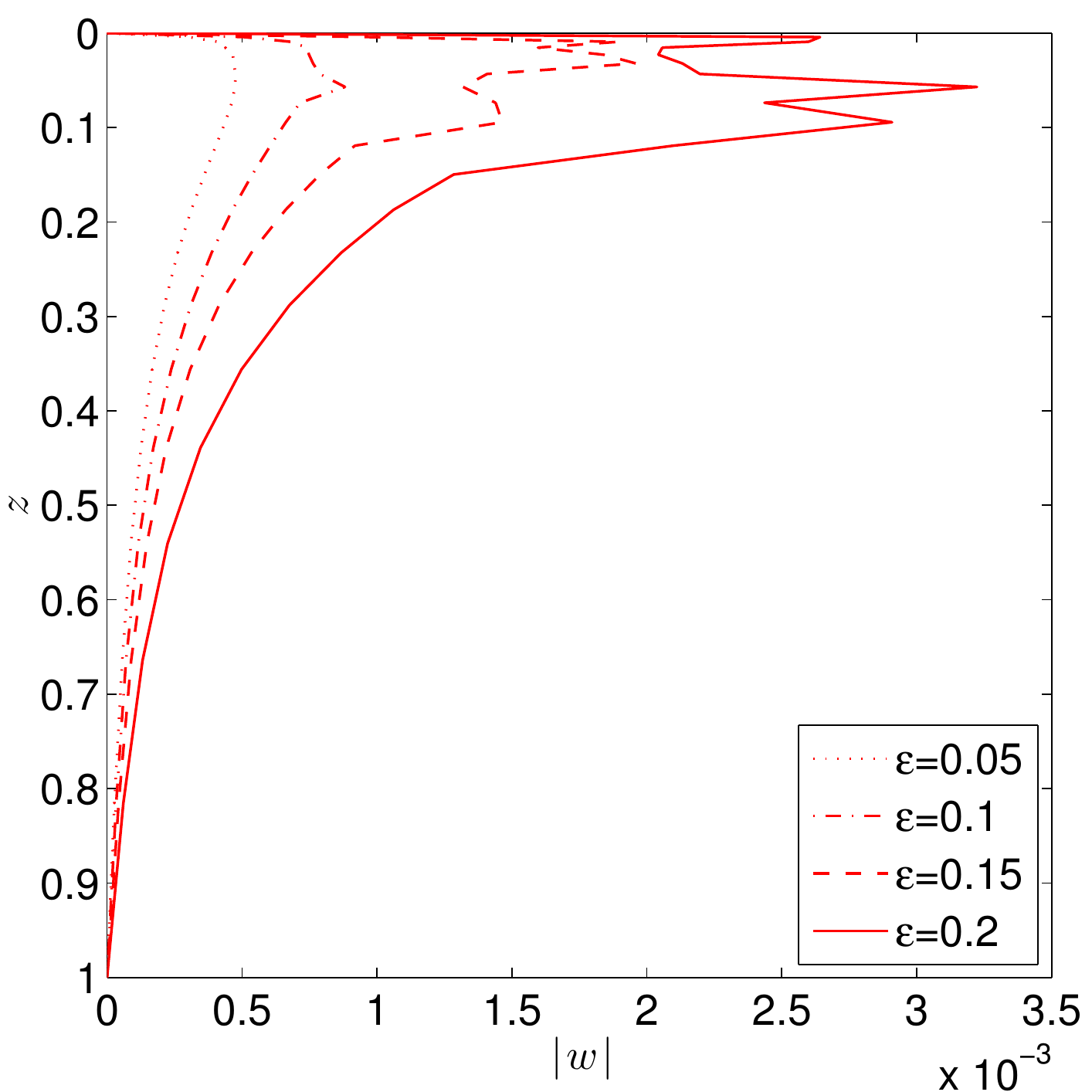}             
  \includegraphics[height=6.5cm,width=6.5cm]{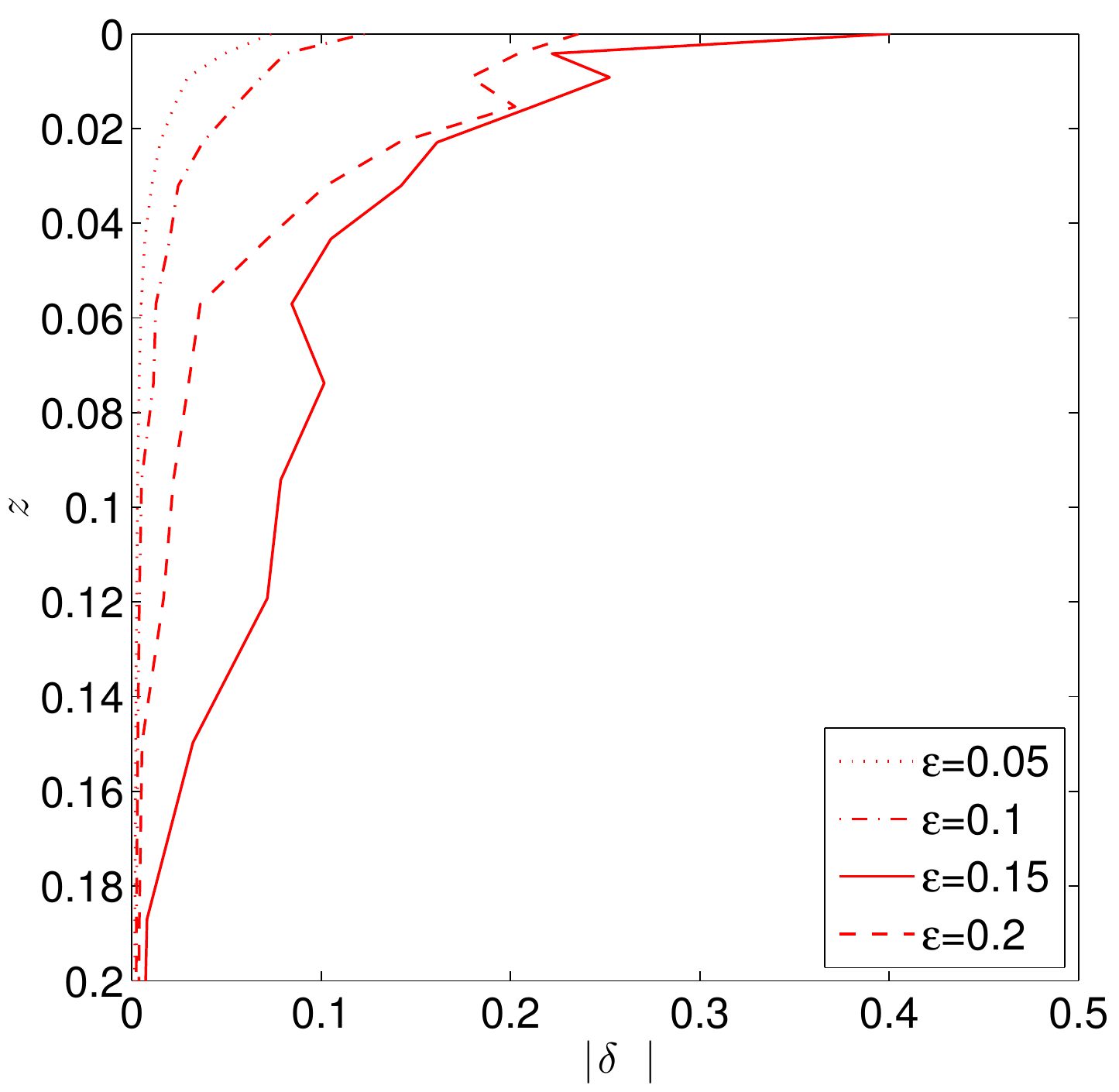}        
  \end{center}   
  \caption{Left: vertical profile of average absolute value of vertical velocity $w$ for SSG for different values of $\epsilon$ in physical  coordinates. Right: vertical profile of average absolute value of horizontal divergence $\delta$ for SSG for different values of $\epsilon$ in physical  coordinates. Note the different vertical scale of the two plots.}
\label{fig:vertical rms}
\end{figure}

Vertical velocity decays in magnitude rather fast with depth. The left panel of Figure \ref{fig:vertical rms} shows the vertical profile of the average of the absolute value of $w$ for SSG for different values of the Rossby number in physical coordinates. We can see that for increasing Rossby number the vertical velocity becomes larger close to the surface, and penetrates deeper at depth. The largest values of $w$ in the upper layers are dominated by large values of $J$ (not shown), so that stirring dominated regions become of great relevance for what concerns the vertical velocity. In comparison, vertical velocities in SQG (not shown) show larger values in the interior, due to the larger features that are present in SQG and due to the fact that the vertical decay is proportional to the wavenumber.


The largest values of vertical velocities found in the interior as $\epsilon$ increases is in agreement with the analytical results found by \citet{Badin2013}. This effect would be even more enhanced if, following \citet{Badin2013}, the potential vorticity would be expressed as a function of the Rossby number. Further, as stated in the paragraph about the comparison between SSG and SQG$^{+1}$, the increase in the vertical velocities in the interior with $\epsilon$ might be limited by the use of height coordinated, with solutions forced to be confined to the boundary. Further work is required to understand this behavior of SSG.

\begin{figure}
  \begin{center}
 \includegraphics[height=6.5cm,width=6.5cm]{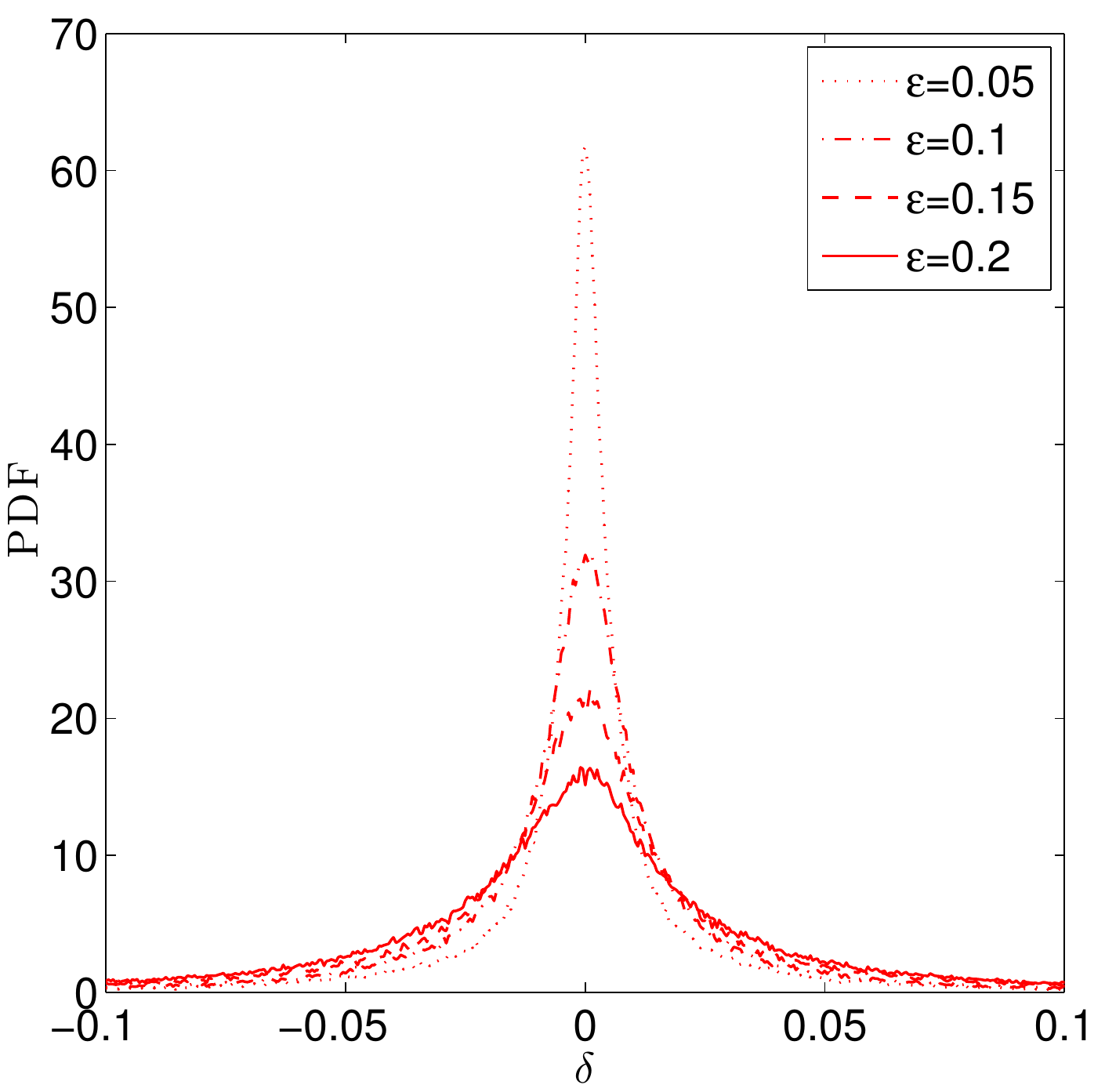}
 \includegraphics[height=6.5cm,width=6.5cm]{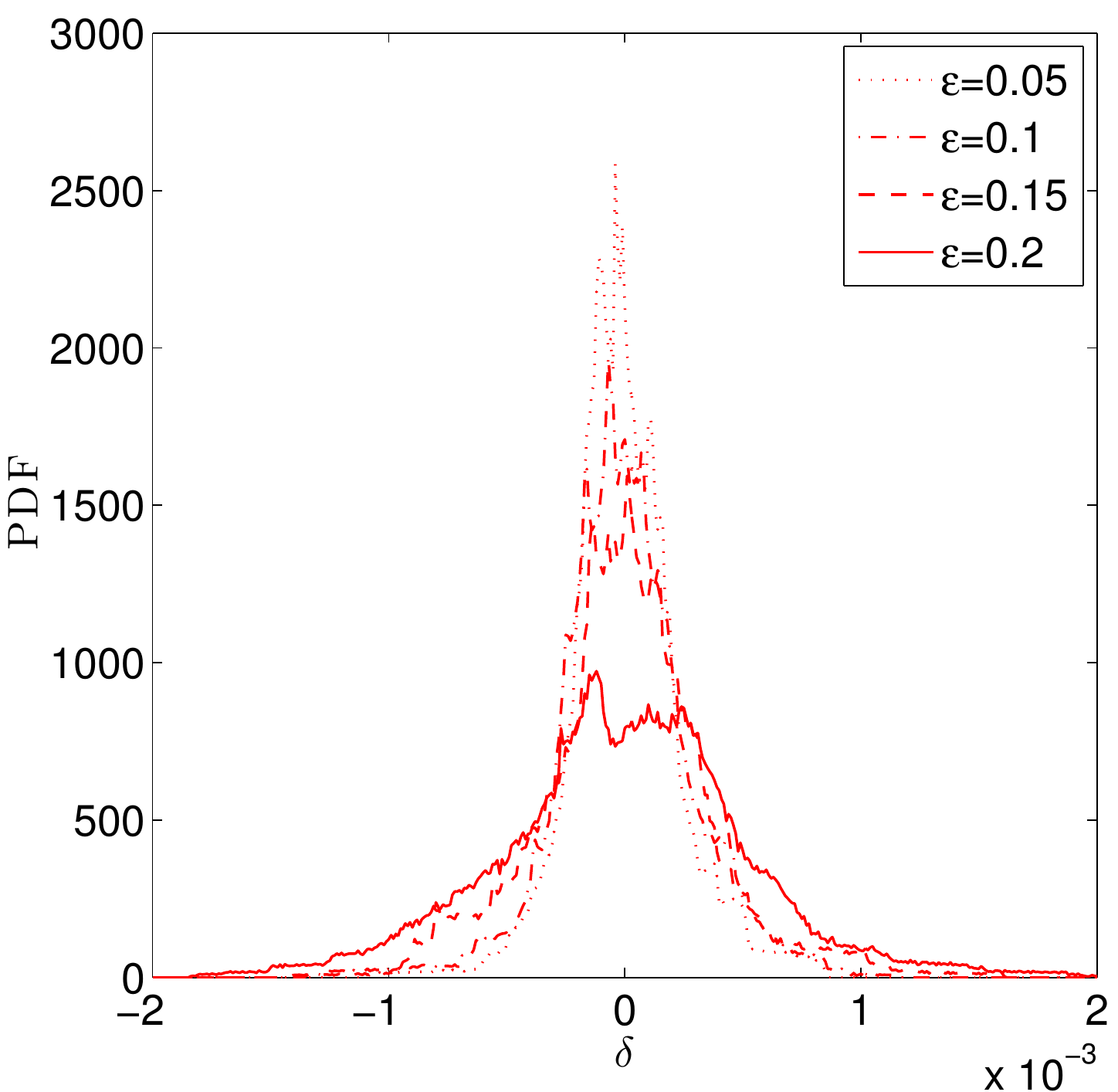}  
    \includegraphics[height=6.5cm,width=6.5cm]{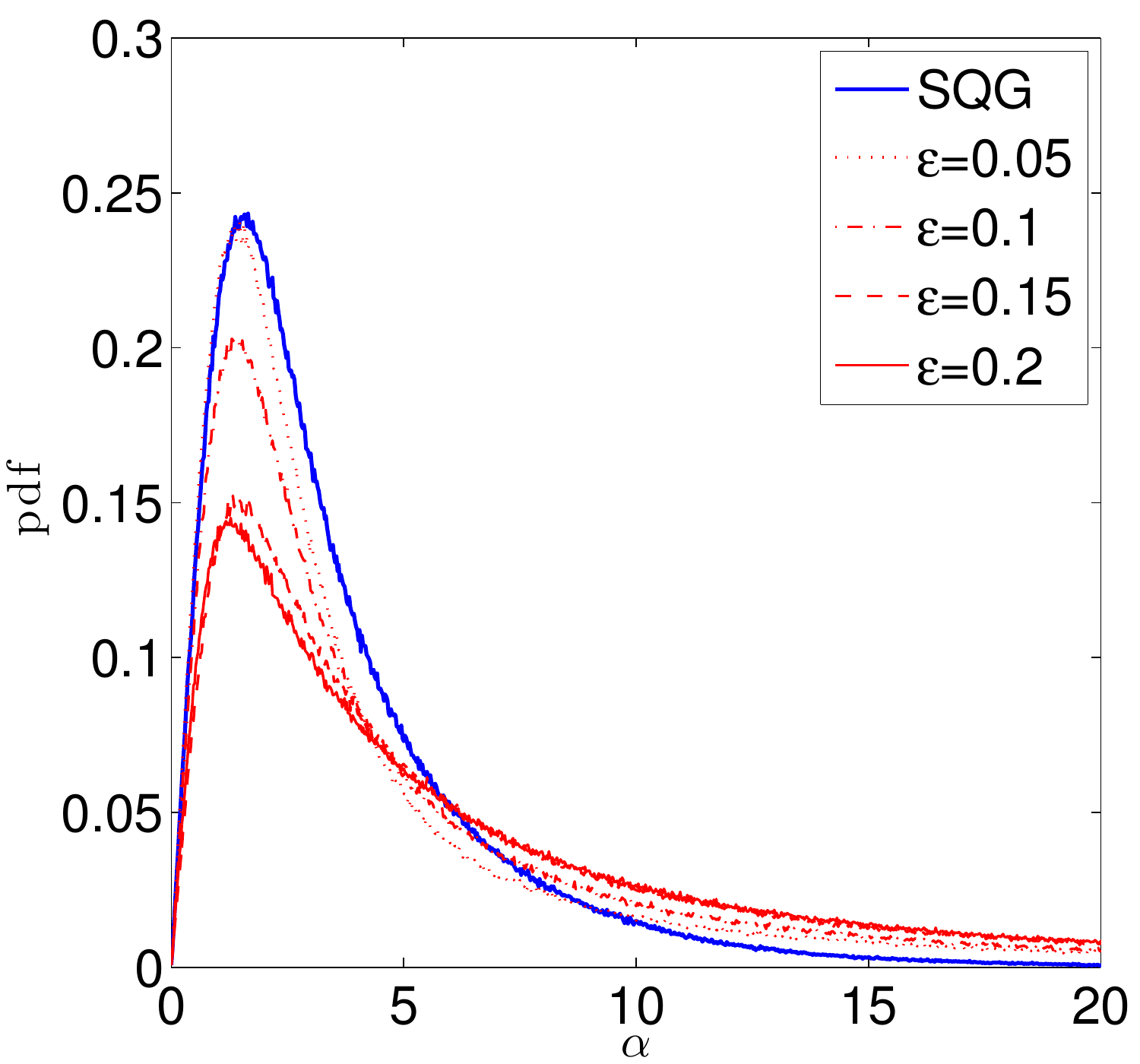}
  \includegraphics[height=6.5cm,width=6.5cm]{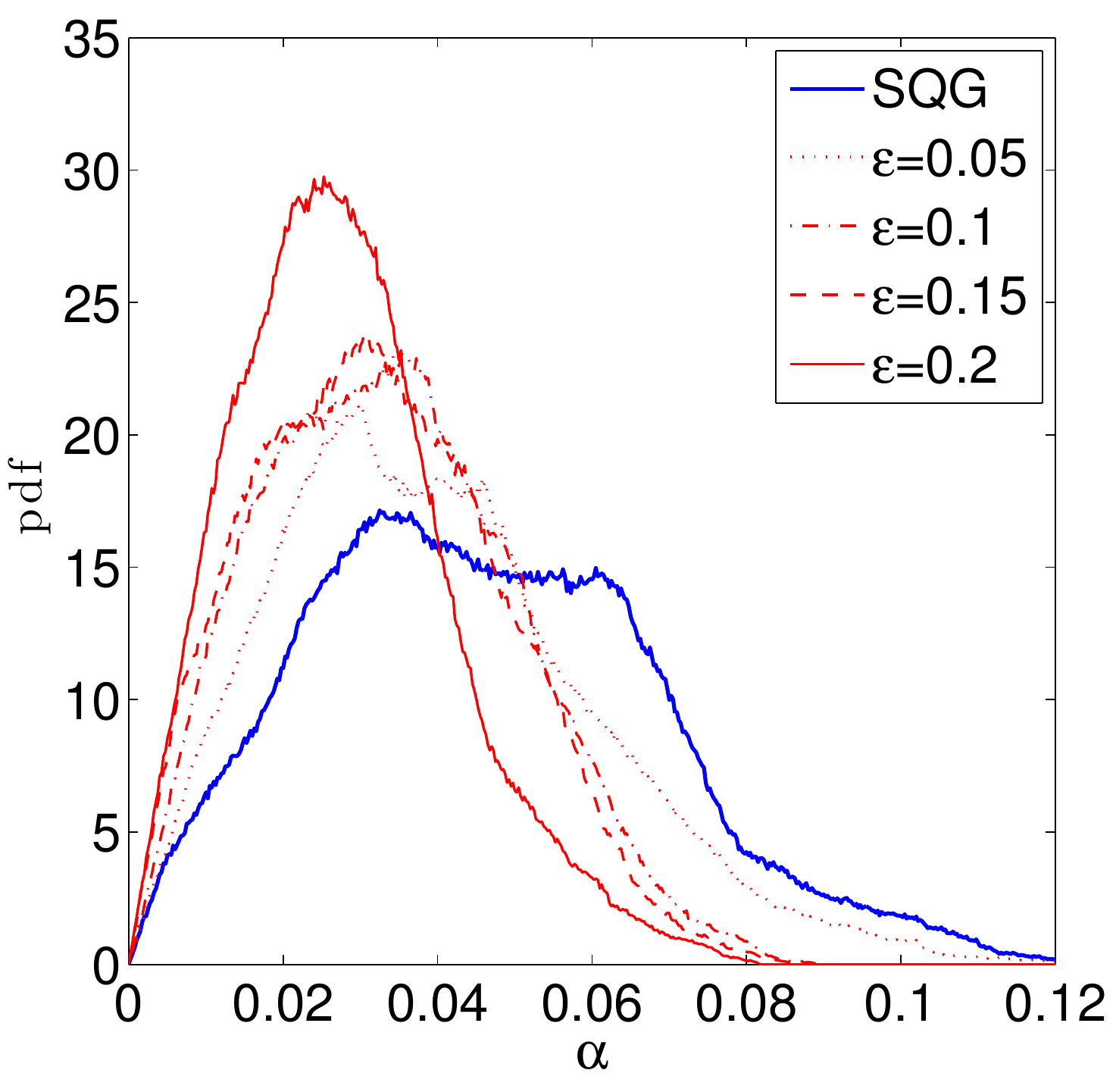} 
  \end{center}   
  \caption{Top: PDFs of ageostrophic horizontal divergence at $Z=0.015$ (left panel) and at $Z=0.8$ (right panel), for SQG (blue lines) and SSG for different values of $\epsilon$ in physical coordinates (red lines). Bottom: PDFs of lateral strain rate at $Z=0.015$ (left panel) and at $Z=0.8$ (right panel), for SQG (blue lines) and SSG for different values of $\epsilon$ in physical coordinates (red lines).}
\label{fig:vertical pdf}
\end{figure}

Once computed the vertical profile of $w$, one can easily reconstruct the horizontal divergence. The flow divergence is an important signal for frontogenetic dynamics, as frontogenesis is associated to a divergent flow of warm and cold water/air rising or sinking at the front through the secondary ageostrophic circulation. Being the geostrophic velocity field divergence free, the horizontal divergence is purely ageostrophic. From the continuity equation one can compute the horizontal divergence $\delta$ in physical coordinates as 
\begin{equation}\label{eq: divergence}
\displaystyle \delta = \frac{\p u_{ag}}{\p x} + \frac{\p v_{ag}}{\p y} = - \frac{\p w}{\p z}
\end{equation}
where $u_{ag}=u-u_{g}$ and $v_{ag}=v-v_{g}$ are the ageostrophic components of the velocity field. The right panel of Figure \ref{fig:vertical rms} shows the vertical profile of the absolute value of $\delta$ for SSG for different values of the Rossby number in physical coordinates. Values of horizontal divergence substantially different from zero are found only very close to the surface, so that the vertical scale is limited between $Z=0$ and $Z=0.2$. The magnitude of $\delta$ increases for increasing Rossby number, similarly to $w$. The top row of Figure \ref{fig:vertical pdf} shows the PDFs of $\delta$ at  $Z=0.015$ (left panel) and $Z=0.8$ (right panel) for SSG for different values of the Rossby number in physical coordinates. We can see that close to the surface ($Z=0.015$) the PDFs become flatter for increasing Rossby numbers, with larger values in the far tails of the distributions. At depth ($Z=0.8$) the same behavior occurs, but given the extremely small values of $\delta$ below $Z=0.2$ it is of relatively low interest.

Sub-mesoscale filaments are also characterized by large values of the lateral strain rate. For simplicity  we compute the lateral strain rate $\alpha$ on geostrophic velocity field only as \citep{Shcherbina&al2013}
\begin{equation}\label{eq: alpha equation}
\displaystyle \alpha = \left[ \left( \frac{\p u_g}{\p x} - \frac{\p v_g}{\p y} \right)^2 + \left( \frac{\p v_g}{\p x} + \frac{\p u_g}{\p y} \right)^2 \right]^{1/2} \\
\end{equation}
The bottom row of Figure \ref{fig:vertical pdf} shows the PDFs of $\alpha$ at  $Z=0.015$ (left panel) and $Z=0.8$ (right panel) for SQG (blue lines) and SSG for different values of the Rossby number in physical coordinates (red lines). As expected, close to the surface ($Z=0.015$) the lateral strain rate in SSG  takes larger values than in SQG, increasing as $\epsilon$ increases, with PDFs characterized by longer tails and less concentrated around small values of $\alpha$. This is again a signature of the enhanced role of filaments in the dynamics. The lateral strain rate at depth ($Z=0.8$) shows instead the opposite behavior, with smaller values of $\alpha$ at $Z=0.8$ for SSG for increasing Rossby number. Values of $\alpha$ at  $Z=0.8$ are however smaller by one order of magnitude than close to the surface, as small scale structures characterized by large values of the lateral strain rate disappear at depth. In general, it is close to the surface, where small scale ageostrophic structures are active, that we expect to see a strong impact on the mixing of tracers. 

\section{Conclusions}\label{sec: conclusions}

In this study, we have performed numerical simulations of freely decaying turbulent flows of the SSG model
in the small $\epsilon$ regime, and we have compared the results with SQG simulations.  Strong asymmetries emerge in the SSG statistics for increasing Rossby numbers at and close to the active boundary. This asymmetry is caused by the enhanced role of ageostrophic processes due to the inclusion of ageostrophic advection in the SG equations, confirming previous results on emerging vorticity asymmetries in geophysical fluid dynamics \citep{Hakim&al2002,Roullet&Klein2010}, where here the role of vorticity is taken by potential temperature as the active conserved scalar of the dynamics. Phenomenologically this appears as SSG dynamics characterized by smaller cyclones and larger anticyclones, and by a predominant role of non-local structures like elongated fronts and filaments with respect to the vortex dominated SQG phenomenology.

Kinetic energy spectra at and close to the active boundary are less steep in SSG than in SQG, with more energy stored at high wavenumbers for increasing Rossby number. We have verified that the effects due to the two aspects on which SSG differs from SQG are, in this range of Rossby numbers, almost separable. The nonlinearity of the inversion equation determines the shift in the peak of the statistics, as signature of the emerging role of filaments as the predominant structures of the dynamics. On the other hand, the coordinate transformation deforms the flow in such a way that small-scale structures are allowed to play a more important role in the dynamics, resulting in flatter energy spectra. Both affects, in different ways, the symmetry of the distribution.

These deviations from the SQG behavior tend to disappear in the interior of the domain, where SQG-like statistics and spectra are recovered, as small-scale structures decay faster with depth. The signature of the geostrophic processes on the dynamics is mostly visible at an intermediate level between $Z=0$ and $Z=0.1$, where the fraction of horizontal kinetic energy associated to the order $\epsilon$ part of the solution of the inversion equation attains its maximum value. 

Vertical velocities have been found to be larger and penetrating more at depth in SSG for increasing Rossby numbers. The horizontal divergence follows the same behavior, although in general values substantially different from zero are found only close to the active boundary. Larger values of the strain rate are found in SSG than in SQG at the layers close to the active boundary. Overall, the emerging of ageostrophic filaments  as dominant features in the SSG dynamics in this part of the domain determines enhanced vertical velocities, horizontal divergence and lateral strain rate.

SSG presents similarities and differences with the SQG$^{+1}$ model of \cite{Hakim&al2002}, similarly proposed in order to include at first order in Rossby number the ageostrophic advection. In both models the PDFs of potential temperature are strongly skewed and peaked at nonzero negative values, confirming that the relation between the inclusion of ageostrophic advection and the development of asymmetric statistics is physically robust, as it does not depend on the specific way ageostrophic advection is formally introduced in the model. Both models show a distinctive net cooling at the active boundary. In a oceanic application, with the domain inverted in the vertical, this would be a net warming of the ocean surface. Physically this is well understood as the restratification effect induced by the asymmetry in the divergence field associated with fronts and filaments created by the ageostrophic advection, and it has been observed and studied also in primitive equations simulations \citep{Lapeyre&al2006,Klein&al2008}. The change of the slope of the kinetic energy spectra, on the other hand, is a unique feature of SSG, and it is directly linked to the deformation of the flow induced by the coordinate transformation.

Besides being a new form of geophysical turbulence, the interest in the SSG model is that it shows many features that are compatible with a qualitative description of sub-mesoscale processes in the oceanic mixed-layer. Observations \citep{Shcherbina&al2013,Shcherbina&al2015} and high-resolution numerical simulations \citep{Capet&al2008a,Capet&al2008b,Capet&al2008c,Klein&al2008}  show the important presence of dynamics at scales smaller than the mesoscale in the ocean. Sub-mesoscale processes are characterized by spatial scales $O(10^2-10^4\,m)$, and by local Rossby and Richardson numbers that can be $O(1)$. These dynamics take often the shape of fronts and filaments, with enhanced vertical vorticity, strain rate and vertical velocities \citep{Mahadevan&Tandon2006}, and are thus important in determining the properties of the dynamics of the oceanic mixed-layer and can be important also for the transformation of water masses \citep{Thomas_Joyce_2010,badinetal10,badin_2013}. 
 
While studies of barotropic instabilities show that SG underestimates the growth rate of instabilities at scales smaller than the deformation radius \citep{Malardel_et_al97}, SSG produces dynamical structures like fronts and filaments that can develop ageostrophic instabilities. SSG could thus be used as an idealized laboratory to study some aspects of sub-mesoscale turbulence in the ocean. In particular, qualitative comparison between the results presented in this paper and the data from observations \citep{Shcherbina&al2013,Shcherbina&al2015} and high-resolution numerical simulations \citep{Roullet&Klein2010} show the same kind of asymmetry at surface, and the same restoring of SQG-like symmetry with depth. Qualitative comparison of the distribution of the ageostrophic horizontal divergence and the lateral strain rate with the results from observations \citep{Shcherbina&al2013} shows also striking agreement. Note that \cite{Shcherbina&al2013} report a value of the Lagrangian Rossby number $\sim$ 0.3. The observations by \cite{Shcherbina&al2015} were instead conducted in a weak strain region, where values are expected to be even lower. We stress in any case that the comparison is meant purely at a qualitative level, and a quantitative comparison with data or primitive equations simulations is left for future studies.

As a first direct follow up of this work, it would be of great interest  to study the dispersion of a passive tracer in SSG dynamics. Being SSG turbulence in many ways qualitatively consistent with observed sub-mesoscale turbulence, it could represent an idealized laboratory where to study the effects of instabilities at frontal scales in the lateral mixing of passive tracers at sub-mesoscale, and how the interaction of horizontal stirring at frontal scales and the vertical mixing affects the total mixing at different depths \citep{badinetal11}. Comparison with results obtained with the SQG and two dimensional Euler equations will be facilitated by the formal similarity between SSG and SQG. In this regard, a more abstract application of great interest could be to study the development of singular solutions in SSG (see, e.g., \citet{CullenPurser84, PurserCullen87, Cullenal91,CullenBook}), building upon earlier works based on the SQG approximation \citep{Constantin&al1994,Constantin&al2012}. Further work is required to understand this behavior of SSG. Finally, for further comparison with submesoscale ocean observations, it would be interesting to compare the results of SSG with a full 3D SG model and with observations. 


\acknowledgements 
The authors would like to thank two anonymous referees for comments that helped to improve the manuscript.

\appendix

\section{Iterative solution of the Monge-Amp\'{e}re equation} \label{sec: A}
\begin{figure}
  \begin{center}
  \includegraphics[height=6cm,width=6cm]{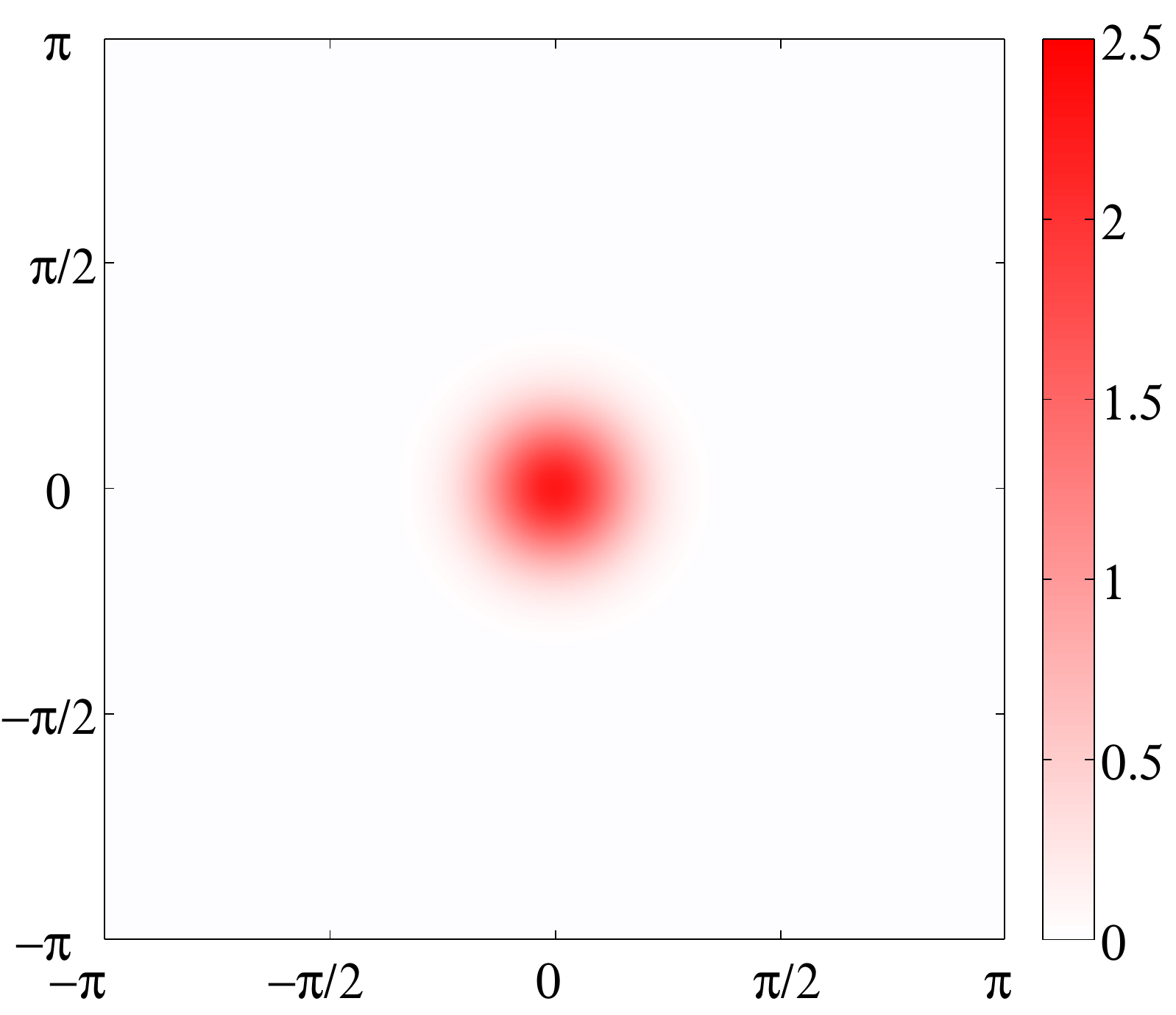}
  \includegraphics[height=6cm,width=6cm]{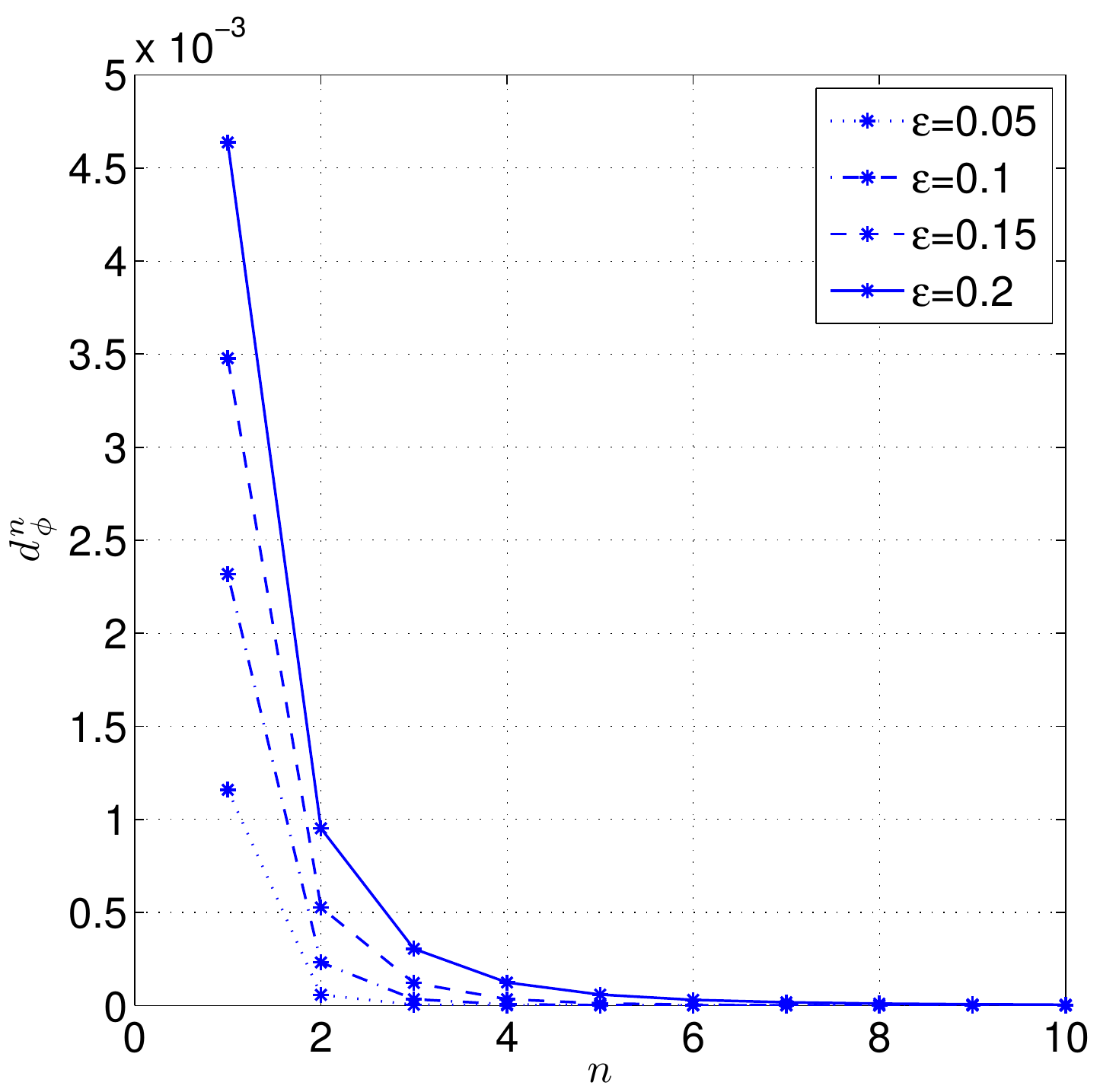}
  \end{center}   
  \caption{Left: boundary condition of surface $\theta$ for the test of the iterative solution procedure. Right: distance $d^n_{\Phi}$  between successive iterations $\Phi^{(n)}$ of the solution as a function of step $n$, for different values of $\epsilon$.}
\label{fig:convergence_MA}
\end{figure}

We test the convergence of the iterative method described in section \ref{sec: methods} to find solutions of the Monge-Amp\'{e}re inversion equation. We take as boundary condition a vortex defined by a surface potential temperature
\begin{equation}\label{eq: iterative problem new} 
\theta = \alpha \exp[-r(X^2+Y^2)]
 \end{equation} 
 where $r=4$ and $\alpha$ is chosen in order to normalize the total surface kinetic energy. The boundary condition is shown in the left panel of Figure \ref{fig:convergence_MA}.
In order to quantify the convergence of the iterative procedure, we define a distance between successive iterations $\Phi^{(n)}$ as
\begin{equation}\label{eq: iterative problem new} 
d^n_{\Phi} = \sqrt{\frac{1}{4\pi^2}\int\displaylimits_0^{2\pi} \int\displaylimits_0^{2\pi} \left(\Phi^{(n)}-\Phi^{(n)}\right)^2dxdy}
\end{equation} 
The right panel of Figure \ref{fig:convergence_MA} shows the evolution of $d^n_{\Phi}$ for different values of $\epsilon$. We can see that convergence is obtained rather fast for any value of $\epsilon$. For structures characterized by very strong local gradients, unless very small values of $\epsilon$ are considered the convergence may fail for large values of $n$ due to the accumulation of numerical errors. In this work we limit ourselves to the first iteration of the method, consistently with the choice of small $\epsilon$ and with \citet{Hoskins1975}.

\section{Coordinate transformation} \label{sec: B}
In most works on the SG approximation the analysis was limited to geostrophic space, and the transformation back to physical space was performed only for visualization purposes, with the aid of graphical methods. \cite{Hoskins1975} proposed a simple method to perform the coordinate transformation that is accurate only for small values of $\epsilon$. \cite{Schaer&Davis1990} developed an iterative algorithm for the inverse coordinate transformation that can be applied for any value of $\epsilon$, and whose first step corresponds to the method of \cite{Hoskins1975}. We briefly summarize here the general iterative method. 

We want to know how the fields defined on a regular grid in geostrophic coordinates $(X^*,Y^*)$ transform to a regular grid in physical coordinates $(x^*,y^*)$. In order to do so, we need to find the geostrophic coordinates $(X(x^*,y^*),Y(x^*,y^*))$ that correspond to the nodes of the regular grid in physical coordinates. We then interpolate the fields defined on the regular geostrophic grid $(X^*,Y^*)$ of the model to the irregular grid $(X(x^*,y^*),Y(x^*,y^*))$. In this way we find, correspondingly, the values of the fields defined on the regular physical grid $(x^*,y^*)$. Following \cite{Schaer&Davis1990}, we consider the following iteration: 
\begin{equation} \label{eq: GT iterative} 
\left \{
\begin{array}{l}
\displaystyle X^{n+1} = x^* + \epsilon \frac{\p \Phi}{\p X} |_{(X^n,Y^n)} ,\\[8pt]
\displaystyle Y^{n+1} = y^*+ \epsilon \frac{\p \Phi}{\p Y}|_{(X^n,Y^n)},
\end{array}
\right .
\end{equation}
with $(X^0,Y^0)=(X^*,Y^*)$. Using \eqref{eq:property_GT} one can see that, given a node $(x^*,y^*)$, a fixed point of the iteration satisfies \eqref{eq: nondimensional transform}, thus realizing the coordinate transformation. Invertibility requires the Jacobian of the coordinate transformation to be positive \eqref{eq:Jacobian}. This implies that the inverse transformation is single-valued and that \eqref{eq: GT iterative} has one single fixed point. Moreover, \eqref{eq: GT iterative} has to be contractive in a neighborhood of the single fixed point \citep{Schaer&Davis1990}.

Note that since the  domain is doubly periodic in the horizontal, no deformation of the horizontal boundaries has to be taken into account in passing from one coordinate system to the other. Periodicity at the boundaries is ensured by attaching to the sides of the domain identical copies of the field and performing the interpolation on the extended system. We have used a linear interpolation method, after testing in a few representative cases that using higher order methods (cubic, spline) did not change the results.   

In our experiments acceptable convergence was obtained in a few iterations. For large $n$ the map tends to oscillate between two slightly different states, probably due to local violations of the invertibility condition that are unavoidable in a turbulent simulation. \cite{Pedder&Thorpe1999} introduced a modification to the algorithm to enforce the stable convergence of the map. However, for large $n$ the fields tend to accumulate noise at small scales, due to the multiple applications of the interpolation procedure. Consistently with limiting the iteration of the solution procedure of the Monge-Amp\'{e}re equation to first order, we have limited the inverse coordinate transformation to the first iteration, thus in practice using the simpler method of \cite{Hoskins1975}. 

\bibliographystyle{jfm}


\end{document}